\documentclass[twocolumn,usenatbib]{mn2e}
\usepackage{graphicx}
\usepackage{epsfig}
\usepackage{psfig}

\title[Narrow absorption lines with two observations of SDSS]{Narrow absorption lines with two observations of Sloan Digital Sky Survey}
\author[Zhi-Fu Chen et al.]{Zhi-Fu Chen$^{1, 2, 3, 4}$\thanks{E-mail:zhichenfu@126.com}, Qiu-Sheng Gu$^{1, 3, 4,}$, Yan-Mei Chen$^{1, 3, 4}$, and Yue Cao$^{5}$\\
$^{1}$ School of Astronomy and Space Science, Nanjing University, Nanjing 210093, P. R. China \\
$^{2}$ Department of Physics and Telecommunication Engineering, Baise University, Baise 533000, China\\
$^{3}$ Key Laboratory of Modern Astronomy and Astrophysics, Nanjing University, Nanjing 210093, P. R. China \\
$^{4}$ Collaborative Innovation Center of Modern Astronomy and Space Exploration£¬ Nanjing, 210093, P.R. China\\
$^{5}$ Nanjing Foreign Language School, Nanjing 210008, P. R. China\\}
\begin{document}
\maketitle

\begin{abstract}
We assemble 3524 quasars from Sloan Digital Sky Survey (SDSS) with repeated observations to search for variations of narrow $\rm C~ IV\lambda\lambda1548,1551$ and $\rm Mg~II\lambda\lambda2796,2803$ absorption doublets in spectral regions shortward of 7000 \AA~ at the observed frame, which corresponds to time-scales of about 150 $\sim$ 2643 days at quasar rest frame. In these quasar spectra, we detect 3580 $\rm C~IV$ absorption systems with $z_{\rm abs} = 1.5188 \sim 3.5212$, and 1809 $\rm Mg~II$ absorption systems with $z_{\rm abs} = 0.3948 \sim 1.7167$. In term of the absorber velocity ($\beta$) distribution at quasar rest frame, we find a substantial number of $\rm C~IV$ absorbers with $\beta<0.06$, which might be connected to the absorptions of quasar outflows. The outflow absorptions peak at $\upsilon\approx2000~\rm km~s^{-1}$ and drop rapidly below the peak value. Among 3580 $\rm C~IV$ absorption systems, 52 systems ($\sim$ 1.5\%) show obvious variations in equivalent widths at the absorber rest frame ($W_{\rm r}$): 16 enhanced, 16 emerged, 12 weaken, and 8 disappeared systems, respectively. We find that changes in $W_{\rm r}\lambda1548$ are neither related to time-scales of the two SDSS observations, nor to absorber velocities at the quasar rest frame. Variable absorptions in low-ionization species are important to constraint the physical conditions of absorbing gas. There are two variable $\rm Mg~II$ absorption systems measured from SDSS spectra detected by Hacker et al. However, in our $\rm Mg~II$ absorption sample, we find that neither shows variable absorption with confident levels of $>4\sigma$ for $\lambda2796$ lines and $>3\sigma$ for $\lambda2803$ lines.
\end{abstract}


\section{Introduction}
Absorption lines in quasar spectra are a powerful signature to investigate gaseous conditions (e.g., density distribution, temperature, kinematics, metallicity, ionization structure) of objects that may be invisible. The absorptions might be originated from the gas located at the inner region of the accretion disc (e.g., Tombesi et al. 2010, 2011, 2013), accretion disc wind/outflow (e.g., Murray et al. 1995; Proga 2000; Filiz Ak et al. 2013), the host galaxy the quasar belongs to (e.g., Heckman et al. 1990; Kurosawa \& Proga 2009), the interstellar medium (ISM), the intergalactic medium (IGM) of galaxy cluster around the quasar (e.g., Weymann et al. 1979; Wild et al. 2008), cosmologically intervening galaxies (e.g., Bergeron 1986; Chen et al. 2010; M\'enard et al. 2011). Absorption lines caused by cosmologically intervening galaxies (known as intervening absorption lines, with $z_{\rm abs} \ll z_{\rm em}$) are shown to correlate with locations of absorbers with respect to the central region of galaxies, and with star formation rates within galaxies (Guillemin \& Bergeron 1997; Chen et al. 2010; M\'enard et al. 2011). Intervening absorption lines often show line widths of being less than a few hundreds $\rm km~s^{-1}$. Host galaxies and galaxy clusters that quasars belong to can also generate absorptions with line widths less than a few hundreds $\rm km~s^{-1}$ and $z_{\rm abs}\approx z_{\rm em}$, which are considered as associated absorptions. Absorptions produced by accretion disc wind/outflow (intrinsic to the quasar) are more complex compared to intervening and associated absorptions. Intrinsic absorption lines can show all kinds of line widths and appear at a very wide range of speed, with respect to quasars, from $10^2~\rm km~s^{-1}$ to $10^5~\rm km~s^{-1}$ (e.g., Tombesi et al. 2010, 2011, 2013; Gupta et al. 2013a,b; Chen \& Qin 2013). In term of line widths, intrinsic absorption lines can be divided into three classes: (1) narrow absorption lines (NALs) whose line widths are similar to those of intervening and associated absorption lines (e.g., Wise et al. 2004); (2) broad absorption lines (BALs) with absorption troughs exceed $2000~\rm km~s^{-1}$ below $10\%$ continuum (Weymann et al. 1991); and (3) mini-broad absorption lines (min-BALs) of which line widths are between NALs and BALs (e.g., Misawa et al. 2007, 2014).

The outflow is an important component of the quasar system, which is related to the gas lifted off the accretion disc under some mechanisms, for example, radiation pressure (e.g., Murray et al.1995; Proga et al. 2000; Proga 2007), magnetocentrifugal forces (e.g., Lery et al. 1999; Everett 2005; Fukumura et al. 2010), thermal pressure (e.g., Everett \& Murray 2007; Owen et al. 2012). The outflow can transport energy, momentum, mass, heavy elements from the accretion disc and/or the vicinity of central region to the host galaxy and IGM, which regulate the efficiency of accretion onto the central supermassive black hole (e.g., Emmering et al. 1992; Konigl \& Kartje 1994), and influence the evolution of the host galaxy and surrounding environment (e.g., Tremonti et al. 2007; Cattaneo et al. 2009; Yuan \& Narayan 2014; Ishibashi \& Fabian 2014). Intrinsic absorption lines, including BALs, mini-BALs and NALs, are often utilized to study properties of the quasar outflow. BAL, mini-BAL and NAL outflows might depict a general outflow phenomenon, and their different characteristics are related to the orientation and/or evolution effects. In the orientation scheme, sightlines going through the fast, dense, equational wind (near accretion disc) will give rise to BALs, while sightlines skimming the lower density and ragged edges of the wind (at higher latitude above the disc plane) will produce NALs and/or mini-BALs (e.g., Murray et al. 1995; Proga 2000; Ganguly et al. 2001; Chartas et al. 2009; Hamann et al. 2012). In the evolution scheme, NALs and mini-BALs might be in the initial or late phases of a powerful BAL outflow (e.g., Hamann et al. 2008; Gibson et al. 2010; Rodr\'iguez Hidalgo et al. 2013). Thus, BALs, mini-BALs and NALs may describe different phases, regions and dynamical mechanisms of the same outflow.

Profiles of intervening and associated absorption lines are similar to those of intrinsic NALs. Therefore, it is difficult to distinguish intrinsic NALs from intervening and associated absorption lines in the same spectrum, especially in low or moderate resolution spectra. Many previous studies have revealed that intrinsic absorption lines can vary in equivalent width over timescales of months to years (e.g., Wise et al. 2004; Narayanan et al. 2004; Gibson et al. 2008, 2010; Hamann et al. 2011; Hacker et al. 2013; Chen \& Qin 2013; Chen et al. 2013; Misawa et al. 2014). The line variability is a good indicator to determine whether NALs are truly intrinsic to quasars or not. This line variability can be caused by (1) bulk motion of the absorbing gas perpendicular to our sightline; (2) changes in ionization states and column densities of the gas. Over the timescale of months to years, the above two scenarios are unfeasible to intervening absorbers due to their low densities and large sizes.

Several works have focused on systematically analyzing intervening metal absorption lines in single epoch spectra of Sloan Digital Sky Survey (SDSS, York et al. 2000; Quider et al. 2011; Qin et al. 2013; Zhu \& M\'enard 2013; Cooksey et al. 2013; Seyffert et al. 2013; Chen et al. 2014a, b, c). Recently, Hacker et al. (2013) presented a catalog of 2522 narrow absorption line systems with multi-observations from SDSS Data Release 7 (DR7), and found that 33 systems with significantly variable absorption strengths ($>4\sigma$ for one line or $>3\sigma$ for two lines) over timescales of one day to several years in the quasar-frame. SDSS Data Release 9 (DR9) is the first spectroscopy release of Baryon Oscillation Spectroscopic Survey (BOSS; Eisenstein et al. 2011) to the public, which contains 7932 quasars (Ross et al. 2012; P\^aris et al. 2012) that were also included in the quasar spectroscopic catalog of SDSS DR7 (Schneider et al. 2010). In this paper, we present narrow absorption line systems which are analyzed in quasar spectra with both DR7 and BOSS observations. In section 2, we explain the quasar sample selection and the spectral analysis. In sections 3 and 4, we present properties of detected absorption systems and our discussion, respectively. Section 5 is the summary.

\section{Quasar sample and spectral analysis}
Baryon Oscillation Spectroscopic Survey (hereafter BOSS; Eisenstein et al. 2011; P\^aris et al. 2012) is the main dark time legacy survey of the third stage of Sloan Digital Sky Survey, which used the same 2.5m telescope (Gunn et al. 2006; Ross et al. 2012) as the first and second stages of Sloan Digital Sky Survey did (hereafter SDSS-I/II). SDSS-I/II spectra have a wavelength coverage from 3800\AA $\sim$ 9200\AA~ with spectral resolution of 1800 $\sim$ 2200 (e.g., York et al. 2000). BOSS spectra span a range from 3600\AA $\sim$ 10500\AA~ at a resolution of 1300 $\sim$ 2500 (P\^aris et al. 2012). In the first two years, BOSS has detected 87,822 quasars over an area of 3275 $\rm deg^2$, including 7932 quasars that were observed by SDSS-I/II as well. Quasars observed by both SDSS-I/II and BOSS provide a remarkable chance to study variabilities of absorption lines in a large population. In this paper, we search for narrow $\rm C~IV\lambda\lambda1548,1551$ and/or $\rm Mg~II\lambda\lambda2796,2803$ absorption doublets in quasar spectra with both SDSS-I/II and BOSS observations. Throughout this work, we directly take  emission redshifts of quasars provided by Hewett \& Wild (2010)\footnote[1]{http://das.sdss.org/va/Hewett$\_$Wild$\_$dr7qso$\_$newz/}.

Signal to noise ratio (SNR) of spectra is important to search for NALs. In SDSS-I/II spectra, the systematic sky-subtraction residual at longward of 7000\AA~ (observed-frame) is significant, and the region shortward of 3900\AA~ (observed frame) usually has a much low SNR (noisy region). The significant systematic sky-subtraction residual and the noisy region often prevent us from searching for NALs, therefore we constrain our detection of NALs in a wavelength range of 3900\AA $\sim$ 7000\AA~ at observed frame. The pair of $\rm O~I\lambda1302$ and $\rm S~II\lambda1304$ absorption lines with higher redshifts usually lead to misidentifications of $\rm C~IV\lambda\lambda1548,1551$ and $\rm Mg~II\lambda\lambda2796,2803$ absorption doublets at lower redshifts. Taking into account the confusion arising from $\rm Ly\alpha$, $\rm O~I\lambda1302$ and $\rm S~II\lambda1304$ absorptions, we reject the spectral region shortward of 1310\AA~ at rest frame. These limits reduce the quasar sample from 7932 to 7598.

We also rule out quasars with low SNR since searching for NALs in spectra with low SNR is unrealistic. We compute the median-SNR per pixel of each spectrum in the surveyed spectral region (3900\AA $\sim$ 7000\AA~ at observed frame and longward of 1310\AA~ at quasar rest frame). Figure 1 shows the distributions of the SNR, which have median values of 7.6 for SDSS-I/II spectra (see black solid vertical line) and 11.7 for BOSS spectra (see red dash vertical line). We limit our analysis to quasars with SNR $\ge8.0$ for both SDSS-I/II and BOSS spectra, which reduces the quasar sample from 7598 to 3524. The black solid line in Figure 2 shows the distribution of emission redshifts of these 3524 quasars, and the distribution of their time intervals between repeated observations are shown in Figure 3.


\begin{figure}
\centering
\includegraphics[width=7cm,height=6cm]{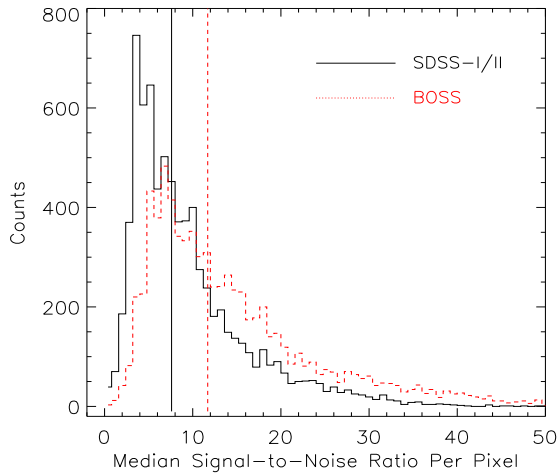}
\caption{Distributions of the median signal-to-noise ratio per pixel of SDSS-I/II or BOSS quasar spectra in the surveyed spectral region. The black solid line is for SDSS-I/II spectra, and the red dashed line is for BOSS spectra. The vertical lines mark the median values of the distributions.}
\end{figure}

\begin{figure}
\centering
\includegraphics[width=7cm,height=6cm]{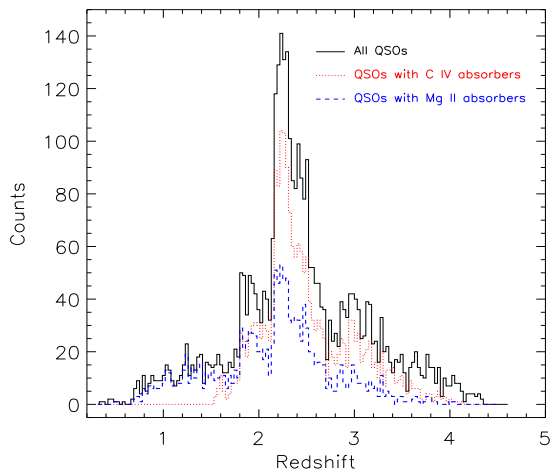}
\caption{Distributions of the emission line redshifts that are directly taken from Hewett \& Wild (2010). The black solid line is for quasars that are suitable to detect $\rm C~IV\lambda\lambda1548,1551$ or $\rm Mg~II\lambda\lambda2796,2803$ absorption doublets, the red dotted line is for quasars with $\rm C~IV\lambda\lambda1548,1551$ absorption doublets, and the blue dashed line for $\rm Mg~II\lambda\lambda2796,2803$ absorption doublets.}
\end{figure}


\begin{figure}
\centering
\includegraphics[width=8cm,height=7cm]{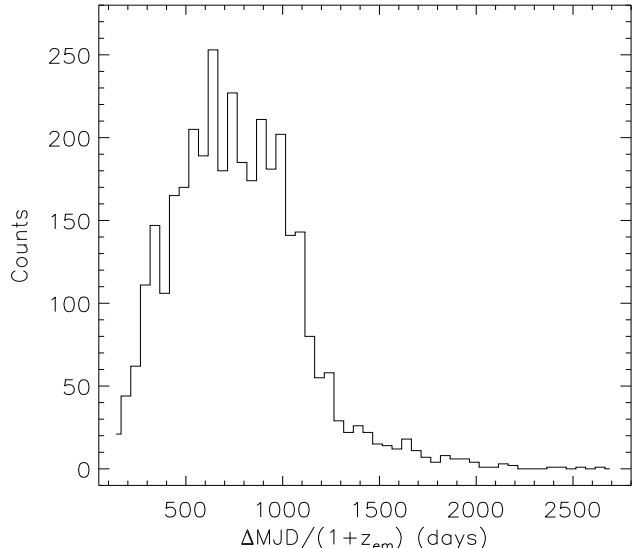}
\caption{The distribution of the time intervals between repeated observations at quasar rest frame.}
\end{figure}

We fit a pseudo-continuum (including underlying continuum and emission lines) for each spectrum of our quasar sample with a combination of cubic spline functions and Gaussian functions (e.g., Nestor et al. 2005; Quider et al. 2011; Chen et al. 2014a,b,c), which is used to normalize the flux and flux uncertainties of original spectra (the flux and flux uncertainties of original spectra divided by the pseudo-continuum). We search for narrow $\rm C~IV\lambda\lambda1548,1551$ and $\rm Mg~II\lambda\lambda2796,2803$ absorption doublets in pseudo-continuum normalized spectra. In this paper, we focus on the analysis of NALs. Therefore, as the first step, we remark continuous absorption troughs with line widths being broader than $1200~\rm km~s^{-1}$ at depths $>20\%$ below the pseudo-continuum, which won't be considered by our program of searching for narrow absorption lines. Secondly, we search for candidates of $\rm C~IV$ and $\rm Mg~II$ absorption doublets, invoke a pair of Gaussian functions to fit each absorption doublet, and measure the equivalent width at absorber rest-frame ($W_{\rm r}$) based on Gaussian fittings. The error of $W_{\rm r}$ ($\sigma_{\rm w}$) includes contributions from flux uncertainties ($\sigma_{\rm flux}$) and from uncertainties in the position of pseudo-continuum ($\sigma_{\rm cont}$), which is $\sigma_{\rm w}^2 = \sigma_{\rm flux}^2 + \sigma_{\rm cont}^2$. We evaluate $\sigma_{\rm flux}$ (e.g., Nestor et al. 2005; Quider et al. 2011) by
\begin{equation}
(1+z)\sigma_{flux}=\frac{\sqrt{\sum_i P^2(\lambda_i-\lambda_0)\sigma^2_{f_i}}}{\sum_i P^2(\lambda_i-\lambda_0)}\Delta\lambda,
\end{equation}
where, as a function of pixel, $P(\lambda_i-\lambda_0)$ is the line profile centered at $\lambda_0$, $\lambda_i$ is the wavelength, and $\sigma_{\rm f_i}$ is the normalized flux uncertainty. We adopt the method provided by Misawa et al. (2014) to calculate the value of $\sigma_{\rm cont}$ as follows. The value of $\sigma_{\rm cont}$ is related to the product of the line width near the pseudo-continuum fitting ($\lambda_{\rm max}-\lambda_{\rm min}$) and fluctuations of the pseudo-continuum fittings ($\sigma_{\rm pf}$), namely, $\sigma_{\rm cont}\propto (\lambda_{\rm max}-\lambda_{\rm min})\sigma_{\rm pf}$. Combining the SNR, in the pseudo-continuum fitting normalized spectrum, $\rm \sigma_{cont}$ can be computed by
\begin{equation}
\sigma_{cont} = \frac{A(\lambda_{max}-\lambda_{min})}{SNR}
\end{equation}
where the calibration parameter A is determined by fitting the pseudo-continuum multiple times around the absorption lines.

Thirdly, we estimate the signal-to-noise ratio of absorption line using the same method adopted by Qin et al. (2013), where $1\sigma$ noise is calculated by
\begin{equation}
\sigma_N=\sqrt{\frac{\sum\limits_{i=1}^M[\frac{F^i_{noise}}{F^i_{cont}}]^2}{M}},
\end{equation}
where $F_{\rm noise}$ and $F_{\rm cont}$ are the flux uncertainty and the flux of the pseudo-continuum in the wavelength range of
$(1+z_{\rm abs})\lambda_{\rm blue}$ - 5\AA $< \lambda_{\rm obs} < (1+z_{\rm abs})\lambda_{\rm red} +5 $\AA,
respectively. Here, the $\lambda_{\rm blue}$ and $\lambda_{\rm red}$ are vacuum wavelengths of blue and red lines of $\rm C~IV$ or $\rm Mg~II$ absorption doublets. The signal-to-noise ratio of the absorption line can be derived by
\begin{equation}
SNR_{abs}=\frac{1-S_{abs}}{\sigma_N},
\end{equation}
where $S_{\rm abs}$ is the smallest value within an absorption trough in the normalized spectrum. The same quasar usually has different SNRs from SDSS-I/II to BOSS spectra. Thus, for the same absorption line, the $SNR_{\rm abs}$ computed from SDSS-I/II spectra often differs from that of BOSS spectra. As the final step, we select the absorption doublets with $SNR_{abs}$, which can be computed in SDSS-I/II or SDSS-III spectra, being no less than 2.0 for both lines.

In term of above criterions, we collect 3580 $\rm C~IV\lambda\lambda1548,1551$ absorption doublets that are detected in 1997 quasar spectra, whose emission redshifts are shown with red dotted line in Figure 2, and 1808 $\rm Mg~II\lambda\lambda2796,2803$ absorption doublets that are detected in 1378 quasar spectra, whose emission redshifts are shown with blue dashed line in Figure 2 as well. We provide these absorption doublets in Tables A1 and A2, and display their absorption redshifts in Figure 4.

\begin{figure}
\centering
\includegraphics[width=8cm,height=7cm]{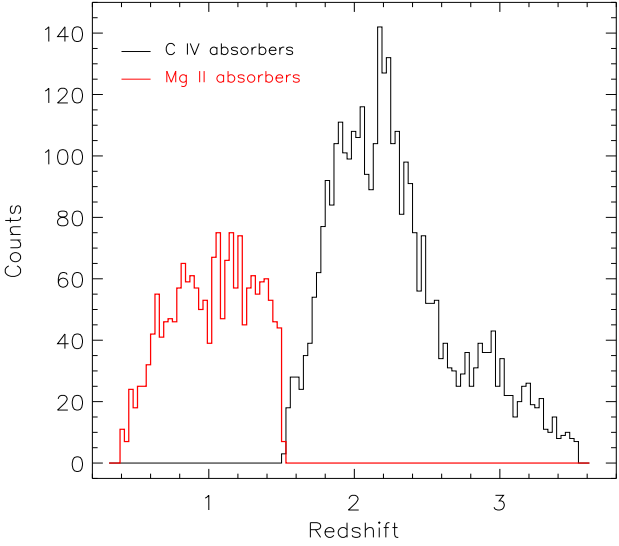}
\caption{Distributions of absorption redshifts. The black line is for $\rm C~IV$ absorbers, and the red line is for $\rm Mg~II$ absorbers.}
\end{figure}

\section{Global properties of absorbers}
We have collected 3524 quasars with both SDSS-I/II and BOSS observations, whose emission redshifts are from 0.3810 to 4.3264 (see the black solid line of Figure 2) and time intervals between the two SDSS observations are $\Delta{\rm MJD}=150 \sim 2643$ days at quasar rest frame. We find that 1997 quasars are detected to have at least one $\rm C~IV\lambda\lambda1548,1551$ absorption system, whose emission redshifts are in the range of 1.5296 to 4.3264 (see the red dotted line of Figure 2), and 1378 quasars are detected to have at least one $\rm Mg~II\lambda\lambda2796,2803$ absorption system, whose emission redshifts are from 0.4984 to 4.0688 (see the blue dashed line of Figure 2).

\subsection{The $\rm C~IV$ absorbers}
The 3580 $\rm C~IV\lambda\lambda1548,1551$ absorption systems with $z_{\rm abs}=1.5188 \sim 3.5212$ (see the black solid line of Figure 4) are detected in 1997 quasar spectra. Variabilities in equivalent widths of NALs can be happened on time intervals of months to years at rest frame (e.g., Wise et al. 2004; Narayanan et al. 2004; Hacker et al. 2013). In order to investigate variabilities of NALs, we compare $W_{\rm r}$ that are measured from SDSS-I/II and BOSS spectra, respectively, and compute normalized differences of $W_{\rm r}$ ($\Delta{W}/\sigma_{\rm w}'$) between the two SDSS observations by
\begin{equation}
\frac{\Delta{W_r}}{\sigma_w'} = \frac{W_{r2}-W_{r1}}{\sqrt{\sigma_{w1}^2+\sigma_{w2}^2}}
\end{equation}
where subscripts ``1" and ``2" represent SDSS-I/II and BOSS spectra, respectively. Distributions of $\Delta{W}/\sigma_{\rm w}'$ are displayed in Figure 5. It is clear from Figure 5 that most $W_{\rm r}$ of absorption lines are consistent with each other within $3\sigma'_{\rm w}$. It is noteworthy that a number of absorption lines show obvious variabilities ($\Delta{W_{\rm r}}>4\sigma_{\rm w}'$) between two observations. There are 52 $\rm C~IV\lambda\lambda1548,1551$ absorption doublets with $\Delta{W_{\rm r}}>4\sigma_{\rm w}'$ for $\lambda1548$ lines and $\Delta{W_{\rm r}}>3\sigma_w'$ for $\lambda1551$ lines, which are detected from 40 quasar spectra. We carefully check these 52 $\rm C~IV$ absorption systems and find that there are 32 enhanced or emerged systems, and 20 weaken or disappeared ones. We tabulate these systems in Table 1, and compare the repeated observations in the Appendix.

\begin{figure}
\centering
\includegraphics[width=8cm,height=7cm]{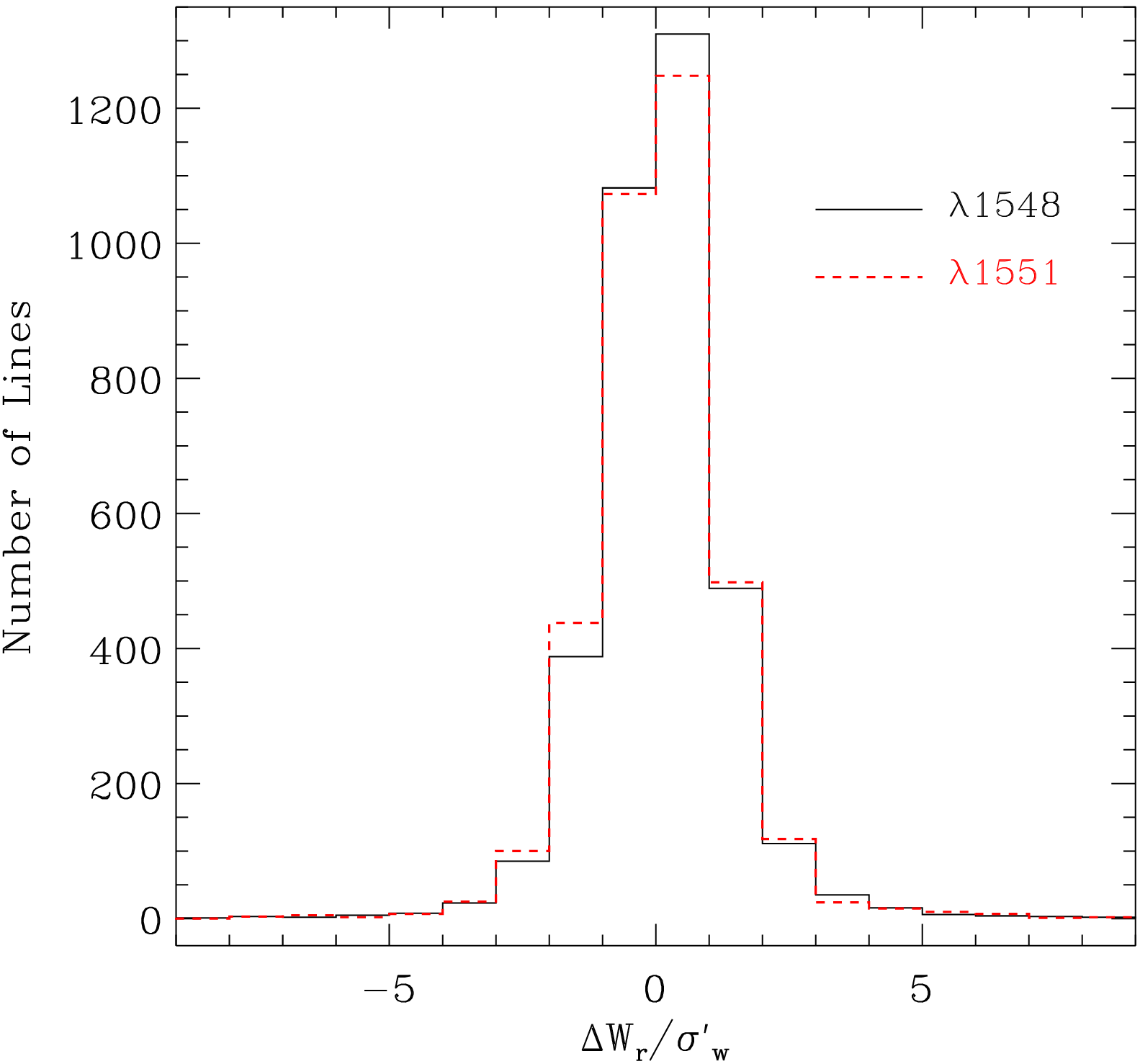}
\caption{Distributions of normalized differences of $W_{\rm r}$ between SDSS-I/II and BOSS observations. These histograms indicate that some absorption lines changed significantly between the two SDSS epochs.}
\end{figure}

\begin{table*}
\centering
\caption{The variable $\rm C~IV\lambda\lambda1548,1551$ absorption systems} \tabcolsep 0.4mm
 \begin{tabular}{ccccccccccccccccccccccc}
 \hline\hline\noalign{\smallskip}
&&&&\multicolumn{2}{c}{SDSS-I/II}&&\multicolumn{2}{c}{BOSS}\\
\cline{5-6}\cline{8-9}
SDSS NAME  & $z_{\rm em}$ &$z_{\rm abs}$ & $LogM_{\rm BH}^a$ & $W_{\rm r}\lambda1548$ & $W_{\rm r}\lambda1551$& &$W_{\rm r}\lambda1548$ & $W_{\rm r}\lambda1551$ &$\frac{\Delta{W_{\rm r}\lambda1548}}{\sigma_{w}'}$&$\frac{\Delta{W_{\rm r}\lambda1551}}{\sigma_{\rm w}'}$&$\Delta{MJD}$&$\beta$&Note\\
&&&$\rm M_\odot$&\AA&\AA&&\AA&\AA&&&days\\
\hline\noalign{\smallskip}
005157.24+000354.7	&	1.9609 	&	1.8681 	&	9.73 	&	$		<	0.04 	$	&	$		<	0.09 	$	&	&	$	0.70 	\pm	0.08 	$	&	$	0.41 	\pm	0.06 	$	&	6.9 	&	3.0 	&	3639	&	0.0318 	&	emerged	\\
015017.70+002902.4	&	3.0013 	&	2.8344 	&	8.72 	&	$		<	0.09 	$	&	$		<	0.08 	$	&	&	$	0.87 	\pm	0.15 	$	&	$	0.79 	\pm	0.15 	$	&	4.2 	&	3.9 	&	3389	&	0.0426 	&	emerged	\\
020629.33+004843.1	&	2.4988 	&	2.3624 	&	10.04 	&	$		<	0.07 	$	&	$		<	0.04 	$	&	&	$	0.71 	\pm	0.14 	$	&	$	0.66 	\pm	0.13 	$	&	4.0 	&	4.2 	&	3273	&	0.0397 	&	emerged	\\
024304.68+000005.4	&	2.0069 	&	1.9426 	&	9.38 	&	$	1.43 	\pm	0.10 	$	&	$	1.53 	\pm	0.17 	$	&	&	$	0.79 	\pm	0.06 	$	&	$	0.95 	\pm	0.07 	$	&	5.5 	&	3.2 	&	3273	&	0.0216 	&	weaken	\\
073232.79+435500.4	&	3.4618 	&	3.2569 	&	9.90 	&	$	0.09 	\pm	0.05 	$	&	$	0.19 	\pm	0.07 	$	&	&	$	0.83 	\pm	0.14 	$	&	$	0.66 	\pm	0.13 	$	&	5.0 	&	3.2 	&	1867	&	0.0470 	&	strengthen	\\
073406.75+273355.6	&	1.9239 	&	1.8609 	&	9.81 	&	$	0.51 	\pm	0.04 	$	&	$	0.42 	\pm	0.04 	$	&	&	$	0.78 	\pm	0.05 	$	&	$	0.85 	\pm	0.05 	$	&	4.2 	&	6.7 	&	3198	&	0.0218 	&	strengthen	\\
080006.59+265054.7	&	2.3438 	&	2.3033 	&	9.35 	&	$		<	0.11 	$	&	$		<	0.12 	$	&	&	$	1.17 	\pm	0.18 	$	&	$	0.94 	\pm	0.18 	$	&	4.5 	&	3.6 	&	2955	&	0.0122 	&	emerged	\\
080609.24+141146.4	&	2.2877 	&	2.0062 	&	9.53 	&	$	0.89 	\pm	0.08 	$	&	$	0.86 	\pm	0.08 	$	&	&	$		<	0.09 	$	&	$		<	0.09 	$	&	7.1 	&	6.4 	&	1921	&	0.0893 	&	disappeared	\\
080906.88+172955.1	&	2.9770 	&	2.8828 	&	9.51 	&	$		<	0.05 	$	&	$		<	0.07 	$	&	&	$	0.76 	\pm	0.07 	$	&	$	0.48 	\pm	0.07 	$	&	7.3 	&	4.4 	&	2228	&	0.0240 	&	emerged	\\
080906.88+172955.1	&	2.9770 	&	2.9384 	&	9.51 	&	$	0.32 	\pm	0.08 	$	&	$	0.32 	\pm	0.07 	$	&	&	$	1.42 	\pm	0.07 	$	&	$	1.26 	\pm	0.07 	$	&	10.3 	&	9.5 	&	2228	&	0.0098 	&	strengthen	\\
081655.49+455633.7	&	2.7168 	&	2.5745 	&	8.80 	&	$		<	0.09 	$	&	$		<	0.07 	$	&	&	$	0.74 	\pm	0.08 	$	&	$	0.68 	\pm	0.08 	$	&	5.0 	&	5.1 	&	3318	&	0.0390 	&	emerged	\\
081929.59+232237.4	&	1.8467 	&	1.7258 	&	9.87 	&	$	0.68 	\pm	0.07 	$	&	$	0.63 	\pm	0.05 	$	&	&	$	0.23 	\pm	0.08 	$	&	$	0.20 	\pm	0.08 	$	&	4.2 	&	4.6 	&	2275	&	0.0434 	&	weaken	\\
082751.78+132107.2	&	1.8289 	&	1.8019 	&	9.43 	&	$	0.93 	\pm	0.06 	$	&	$	0.61 	\pm	0.05 	$	&	&	$	1.31 	\pm	0.06 	$	&	$	0.87 	\pm	0.05 	$	&	4.5 	&	3.7 	&	1424	&	0.0096 	&	strengthen	\\
091621.46+010015.4	&	1.7410 	&	2.1264 	&	9.04 	&	$		<	0.06 	$	&	$		<	0.04 	$	&	&	$	0.80 	\pm	0.12 	$	&	$	0.57 	\pm	0.10 	$	&	5.2 	&	4.4 	&	3661	&	-0.1308 	&	emerged	\\
091621.46+010015.4	&	1.7410 	&	2.1629 	&	9.04 	&	$	0.97 	\pm	0.08 	$	&	$	1.16 	\pm	0.08 	$	&	&	$	2.04 	\pm	0.09 	$	&	$	1.99 	\pm	0.09 	$	&	8.9 	&	6.9 	&	3661	&	-0.1422 	&	strengthen	\\
091621.46+010015.4	&	1.7410 	&	2.1741 	&	9.04 	&	$	0.79 	\pm	0.11 	$	&	$	0.30 	\pm	0.06 	$	&	&	$	1.83 	\pm	0.09 	$	&	$	0.82 	\pm	0.08 	$	&	7.3 	&	5.2 	&	3661	&	-0.1457 	&	strengthen	\\
095254.10+021932.8	&	2.1526 	&	2.0056 	&	9.56 	&	$		<	0.09 	$	&	$		<	0.06 	$	&	&	$	0.72 	\pm	0.08 	$	&	$	0.56 	\pm	0.06 	$	&	4.8 	&	5.9 	&	3737	&	0.0477 	&	emerged	\\
100716.69+030438.7	&	2.1241 	&	1.9129 	&	9.72 	&	$	0.88 	\pm	0.06 	$	&	$	0.60 	\pm	0.05 	$	&	&	$	1.42 	\pm	0.10 	$	&	$	1.24 	\pm	0.10 	$	&	4.6 	&	5.7 	&	3415	&	0.0699 	&	strengthen	\\
100716.69+030438.7	&	2.1241 	&	1.9426 	&	9.72 	&	$		<	0.03 	$	&	$		<	0.06 	$	&	&	$	1.29 	\pm	0.17 	$	&	$	1.18 	\pm	0.17 	$	&	6.9 	&	5.9 	&	3415	&	0.0598 	&	emerged	\\
103115.69+374849.5	&	2.2590 	&	2.2092 	&	9.03 	&	$	0.70 	\pm	0.14 	$	&	$	0.67 	\pm	0.14 	$	&	&	$		<	0.07 	$	&	$		<	0.07 	$	&	4.1 	&	3.9 	&	2165	&	0.0154 	&	disappeared	\\
103115.69+374849.5	&	2.2590 	&	2.2253 	&	9.03 	&	$	0.67 	\pm	0.13 	$	&	$	0.62 	\pm	0.13 	$	&	&	$		\pm	0.06 	$	&	$		\pm	0.04 	$	&	4.1 	&	4.0 	&	2165	&	0.0104 	&	disappeared	\\
103842.14+350906.9	&	2.2049 	&	2.1563 	&	9.46 	&	$	0.72 	\pm	0.11 	$	&	$	0.58 	\pm	0.09 	$	&	&	$	1.50 	\pm	0.12 	$	&	$	1.17 	\pm	0.09 	$	&	4.8 	&	4.6 	&	2129	&	0.0153 	&	strengthen	\\
104841.02+000042.8	&	2.0246 	&	1.9468 	&	9.03 	&	$	1.75 	\pm	0.13 	$	&	$	1.05 	\pm	0.12 	$	&	&	$	0.53 	\pm	0.10 	$	&	$	0.33 	\pm	0.08 	$	&	7.4 	&	5.0 	&	3661	&	0.0261 	&	weaken	\\
104923.94+012224.6	&	1.9454 	&	1.9087 	&	9.73 	&	$	0.21 	\pm	0.04 	$	&	$	0.24 	\pm	0.06 	$	&	&	$	1.40 	\pm	0.08 	$	&	$	1.21 	\pm	0.08 	$	&	13.3 	&	9.7 	&	3296	&	0.0125 	&	strengthen	\\
105207.90+362219.4	&	2.3157 	&	2.2666 	&	8.91 	&	$	0.84 	\pm	0.12 	$	&	$	0.20 	\pm	0.06 	$	&	&	$	1.62 	\pm	0.09 	$	&	$	0.93 	\pm	0.06 	$	&	5.2 	&	8.6 	&	2160	&	0.0149 	&	strengthen	\\
110726.04+385158.2	&	2.6603 	&	2.6134 	&	8.99 	&	$	0.88 	\pm	0.03 	$	&	$	0.79 	\pm	0.03 	$	&	&	$	0.55 	\pm	0.06 	$	&	$	0.53 	\pm	0.07 	$	&	4.9 	&	3.4 	&	1113	&	0.0129 	&	weaken	\\
115122.14+020426.3	&	2.4085 	&	2.3269 	&	9.16 	&	$	0.47 	\pm	0.09 	$	&	$	0.50 	\pm	0.10 	$	&	&	$	1.76 	\pm	0.17 	$	&	$	1.58 	\pm	0.15 	$	&	6.7 	&	6.0 	&	3581	&	0.0242 	&	strengthen	\\
115122.14+020426.3	&	2.4085 	&	2.3742 	&	9.16 	&	$	1.58 	\pm	0.09 	$	&	$	1.71 	\pm	0.08 	$	&	&	$	2.27 	\pm	0.13 	$	&	$	2.26 	\pm	0.13 	$	&	4.4 	&	3.6 	&	3581	&	0.0101 	&	strengthen	\\
120819.29+035559.4	&	2.0213 	&	1.9500 	&	9.59 	&	$	1.84 	\pm	0.14 	$	&	$	1.31 	\pm	0.13 	$	&	&	$	2.62 	\pm	0.12 	$	&	$	2.24 	\pm	0.11 	$	&	4.2 	&	5.5 	&	3257	&	0.0239 	&	strengthen	\\
123720.85-011314.9	&	2.1620 	&	2.1283 	&	8.86 	&	$	0.24 	\pm	0.06 	$	&	$	0.24 	\pm	0.08 	$	&	&	$	1.92 	\pm	0.14 	$	&	$	1.19 	\pm	0.12 	$	&	11.0 	&	6.6 	&	3272	&	0.0107 	&	strengthen	\\
124829.46+341231.3	&	2.2285 	&	2.0621 	&	9.72 	&	$		<	0.04 	$	&	$		<	0.10 	$	&	&	$	1.12 	\pm	0.16 	$	&	$	1.39 	\pm	0.16 	$	&	6.2 	&	6.8 	&	1557	&	0.0529 	&	emerged	\\
124829.46+341231.3	&	2.2285 	&	2.0769 	&	9.72 	&	$		<	0.06 	$	&	$		<	0.11 	$	&	&	$	1.30 	\pm	0.15 	$	&	$	1.05 	\pm	0.15 	$	&	7.4 	&	5.1 	&	1557	&	0.0481 	&	emerged	\\
125216.58+052737.7	&	1.9034 	&	1.8155 	&	9.53 	&	$	0.46 	\pm	0.05 	$	&	$	0.48 	\pm	0.06 	$	&	&	$		<	0.06 	$	&	$		<	0.03 	$	&	5.2 	&	6.4 	&	2984	&	0.0307 	&	disappeared	\\
125216.58+052737.7	&	1.9034 	&	1.8638 	&	9.53 	&	$	0.75 	\pm	0.05 	$	&	$	0.70 	\pm	0.08 	$	&	&	$		<	0.02 	$	&	$		<	0.07 	$	&	13.2 	&	6.1 	&	2984	&	0.0137 	&	disappeared	\\
125216.58+052737.7	&	1.9034 	&	1.8831 	&	9.53 	&	$	2.14 	\pm	0.06 	$	&	$	1.06 	\pm	0.06 	$	&	&	$	0.59 	\pm	0.04 	$	&	$	0.33 	\pm	0.04 	$	&	21.5 	&	10.1 	&	2984	&	0.0070 	&	weaken	\\
125216.58+052737.7	&	1.9034 	&	1.8946 	&	9.53 	&	$	0.76 	\pm	0.05 	$	&	$	0.67 	\pm	0.05 	$	&	&	$	0.11 	\pm	0.02 	$	&	$	0.09 	\pm	0.04 	$	&	12.1 	&	9.1 	&	2984	&	0.0030 	&	weaken	\\
132333.03+004750.2	&	1.7785 	&	1.7340 	&	9.62 	&	$	0.79 	\pm	0.10 	$	&	$	0.60 	\pm	0.10 	$	&	&	$	0.11 	\pm	0.02 	$	&	$	0.11 	\pm	0.07 	$	&	6.7 	&	4.0 	&	3934	&	0.0161 	&	weaken	\\
134544.55+002810.7	&	2.4641 	&	2.3514 	&	9.39 	&	$	0.86 	\pm	0.13 	$	&	$	1.07 	\pm	0.15 	$	&	&	$		<	0.04 	$	&	$		<	0.05 	$	&	5.9 	&	6.4 	&	3687	&	0.0331 	&	disappeared	\\
134544.55+002810.7	&	2.4641 	&	2.3686 	&	9.39 	&	$	1.62 	\pm	0.16 	$	&	$	1.19 	\pm	0.13 	$	&	&	$		<	0.05 	$	&	$		<	0.06 	$	&	9.3 	&	7.9 	&	3687	&	0.0279 	&	disappeared	\\
134544.55+002810.7	&	2.4641 	&	2.3964 	&	9.39 	&	$	1.95 	\pm	0.13 	$	&	$	1.42 	\pm	0.11 	$	&	&	$	0.83 	\pm	0.07 	$	&	$	0.42 	\pm	0.06 	$	&	7.6 	&	8.0 	&	3687	&	0.0197 	&	weaken	\\
140815.58+060023.3	&	2.5830 	&	2.5519 	&	8.93 	&	$		<	0.12 	$	&	$		<	0.05 	$	&	&	$	1.04 	\pm	0.15 	$	&	$	0.98 	\pm	0.15 	$	&	5.0 	&	5.7 	&	2194	&	0.0087 	&	emerged	\\
150033.53+003353.6	&	2.4360 	&	2.1849 	&	10.20 	&	$		<	0.14 	$	&	$		<	0.08 	$	&	&	$	0.72 	\pm	0.09 	$	&	$	0.83 	\pm	0.09 	$	&	4.0 	&	5.6 	&	3642	&	0.0757 	&	emerged	\\
160445.92+335759.0	&	1.8776 	&	1.7709 	&	9.57 	&	$		<	0.14 	$	&	$		<	0.09 	$	&	&	$	0.97 	\pm	0.18 	$	&	$	1.28 	\pm	0.17 	$	&	4.0 	&	5.7 	&	2579	&	0.0378 	&	emerged	\\
160613.99+314143.4	&	2.0569 	&	2.0240 	&	8.80 	&	$	1.82 	\pm	0.15 	$	&	$	1.57 	\pm	0.14 	$	&	&	$	0.88 	\pm	0.11 	$	&	$	0.66 	\pm	0.11 	$	&	5.1 	&	5.1 	&	2893	&	0.0108 	&	weaken	\\
161336.81+054701.7	&	2.4855 	&	2.4152 	&	8.59 	&	$		<	0.10 	$	&	$		<	0.08 	$	&	&	$	0.87 	\pm	0.14 	$	&	$	0.80 	\pm	0.14 	$	&	4.0 	&	4.0 	&	1822	&	0.0204 	&	emerged	\\
161511.35+314728.3	&	2.0981 	&	1.9157 	&	9.89 	&	$	1.47 	\pm	0.06 	$	&	$	1.22 	\pm	0.08 	$	&	&	$	1.06 	\pm	0.08 	$	&	$	0.73 	\pm	0.06 	$	&	4.1 	&	4.9 	&	2591	&	0.0606 	&	weaken	\\
162701.94+313549.2	&	2.3263 	&	2.2785 	&	9.20 	&	$	0.96 	\pm	0.04 	$	&	$	0.70 	\pm	0.04 	$	&	&	$	1.24 	\pm	0.04 	$	&	$	1.16 	\pm	0.04 	$	&	4.9 	&	8.1 	&	2286	&	0.0145 	&	strengthen	\\
162935.68+321009.5	&	2.0364 	&	1.9345 	&	9.51 	&	$	0.62 	\pm	0.16 	$	&	$	0.63 	\pm	0.13 	$	&	&	$	1.55 	\pm	0.14 	$	&	$	1.32 	\pm	0.14 	$	&	4.4 	&	3.6 	&	2287	&	0.0341 	&	strengthen	\\
212943.25+003005.6	&	2.6802 	&	2.5726 	&	9.45 	&	$		<	0.11 	$	&	$		<	0.05 	$	&	&	$	1.09 	\pm	0.12 	$	&	$	0.85 	\pm	0.12 	$	&	5.8 	&	5.5 	&	2531	&	0.0297 	&	emerged	\\
213648.17-001546.6	&	2.1736 	&	1.8372 	&	9.63 	&	$	0.85 	\pm	0.17 	$	&	$	0.54 	\pm	0.14 	$	&	&	$		<	0.03 	$	&	$		<	0.07 	$	&	4.6 	&	3.2 	&	2984	&	0.1116 	&	disappeared	\\
222157.97-010331.0	&	2.6744 	&	2.5459 	&	10.12 	&	$	1.59 	\pm	0.06 	$	&	$	1.28 	\pm	0.06 	$	&	&	$	0.83 	\pm	0.07 	$	&	$	0.63 	\pm	0.07 	$	&	8.2 	&	7.1 	&	2472	&	0.0356 	&	weaken	\\
230034.04-004901.5	&	2.2125 	&	2.1422 	&	9.35 	&	$	1.83 	\pm	0.19 	$	&	$	1.45 	\pm	0.15 	$	&	&	$	0.43 	\pm	0.06 	$	&	$	0.82 	\pm	0.09 	$	&	7.0 	&	3.6 	&	2869	&	0.0221 	&	weaken	\\
\hline\hline\noalign{\smallskip}
\end{tabular}
\\
$^a$ The black hole mass is directly taken from the virial mass of Shen et al. (2011).
\end{table*}

In the variable $\rm C~IV$ absorption sample, there are 22 weaken or enhanced systems with $\beta>0.01$ (see equation (6) for the definition of $\beta$). Hacker et al. (2013) detected 33 variable absorption systems in quasar spectra with repeated observations from SDSS DR7. Among these 33 systems, we find that only two systems meet requirements that, between two observations, the changes of equivalent widths at rest frame are $\ge4\sigma$ for $\lambda1548$ lines and $\ge3\sigma$ for $\lambda1551$ lines. We cannot compare fractions of variable $\rm C~IV$ absorbers of our sample to that of Hacker et al. (2013), since the variable absorption systems of Hacker et al. (2013) were only limited to the system with Grade ¡°A¡± which is defined as that at least four absorption lines are detected.

Variable absorptions often imply that the system is likely intrinsic to background central object. Only a small fraction of detected $\rm C~IV$ NALs ($\sim$1.5\%) show significant variation, this indicates that at least a part of narrow $\rm C~IV$ absorption systems should be truly intrinsic to quasars. The variations in absorptions may be related to time intervals of two observations, distances of absorbers with respect to central regions of quasars, gaseous densities and sizes of absorbers (e.g., Sargent et al. 1988; Steidel \& Sargent 1991; Wild et al. 2008; Capellupo et al. 2013; Hacker et al. 2013; Misawa et al. 2014). Figure 6 displays the distribution of differences in $W_{\rm r}\lambda1548$ versus time intervals between two observations at absorber rest frame for all narrow $\rm C~IV$ absorption systems. It suggests that, at a certain timescale, only a small portion of NALs be expected to be variable ones.
\begin{figure}
\centering
\includegraphics[width=8cm,height=7cm]{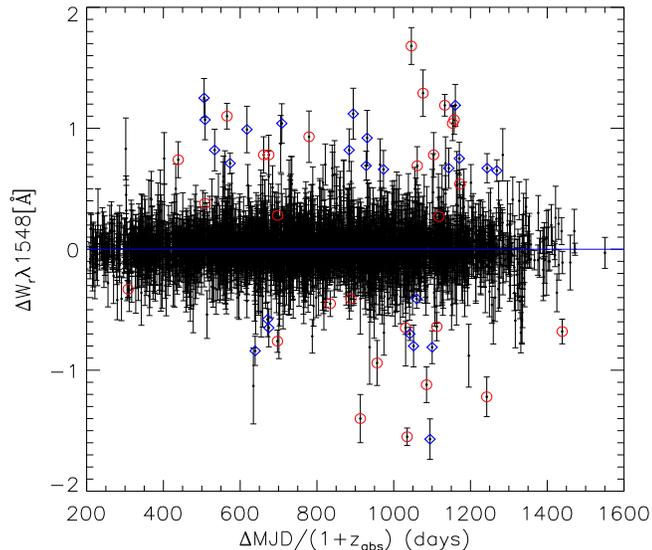}
\caption{The distribution of differences in $W_{\rm r}\lambda1548$ versus the time intervals between the two SDSS epochs at absorber rest frame for all narrow $\rm C~IV$ absorption systems. Diamonds represent absorption systems disappeared or emerged from BOSS spectra, and unfilled circles represent weaken or enhanced absorption systems.}
\end{figure}

Assuming absorptions are physically related to quasar central regions, for the same spectra, the difference between absorption redshift and emission redshift would mainly originate in the relative velocity of the absorber with respect to the quasar central region. If so, relative velocities of absorbers, $\rm \beta$, can be calculated via
\begin{equation}
\beta \equiv \frac{v}{c} = \frac{(1+z_{em})^2-(1+z_{abs})^2}{(1+z_{em})^2+(1+z_{abs})^2}
\end{equation}
We show the distribution of $\beta$ in the top panel of Figure 7. This distribution is similar to that given by Wild et al. (2008) and Nestor et al. (2008). A remarkable excess at $\beta \approx 0$ is detected, and the excessive tail extends out to $\beta \approx0.06$. In the range of $\beta > 0.06$, a platform with average count of 42 is observed as well. This velocity distribution would be mainly contributed from absorptions due to cosmologically intervening galaxies, galaxies in the vicinity of the quasar, quasar host galaxies, and quasar outflows. The platform might suggest that absorption systems with $\beta > 0.06$ would be mainly dominated by intervening absorptions. Nestor et al. (2008) claimed that a large fraction (more than 60\%) of all $\rm C~IV$ NALs with low velocities (1800 $\rm km~s^{-1}\sim$4400 $\rm km~s^{-1}$) are associated with quasar outflows. It is out of the topic of this work to determine the fraction of different type absorptions (quasar outflow, quasar host galaxy, environment around the quasar, or intervening galaxy). We take a relaxed assumption that, in the range of $\beta>0$, cosmologically intervening absorptions show uniform distribution in velocity space at quasar rest frame. This assumption may overpredict intervening absorptions with small $\beta$, but won't change the primary conclusions of this paper. The mean value of counts shown in the top panel of Figure 7 with $\beta>0.06$ is 42, which is used to account for the counts of intervening absorptions with $0<\beta<0.06$. We take out the contribution of intervening absorption (minus the mean value of 42 in the range of $\beta>0$). The result is shown in middle panel of Figure 7 with black histogram. For environmental contributions from quasar host galaxy and galaxies in the vicinity of the quasar, we invoke a Gaussian component centred at $\rm \beta=0$ to model only the data with $\rm \beta<0$, which is plotted in the middle panel of Figure 7 with red solid line.

The bottom panel of Figure 7 shows the velocity distribution after subtracting contributions from intervening galaxies, quasar host galaxies and galaxies in the vicinities of quasars. Note that there are substantially residual in the range of $\beta \approx0 \sim 0.06$, which would be mainly arisen from absorptions of quasar outflows. Similar results were found by Wild et al. (2008) and Nestor et al. (2008). Nestor et al. found that the velocity distribution of narrow $\rm C~IV$ absorption outflows shows a peak at $\upsilon\equiv\beta c \approx2000~\rm km~s^{-1}$, decreases rapidly at low velocities, and truncates at $\upsilon\approx750~\rm km~s^{-1}$. This similar shape of velocity distribution is also detected from our sample.

Here, there are some caveats for Figure 7. The assumption that all absorptions with $\beta<0$ originate from the quasar environment might be less reliable. The errors of emission redshifts may result in negative $\beta$ for low velocity outflows. Therefore, the Gaussian modelling shown in the middle panel might overpredict environmental contributions and underpredict outflow absorptions with low velocities. The distribution of cosmologically intervening absorptions may not be uniform in velocity space at quasar rest frame. Nestor et al. (2008) claimed that more than 60\% $\rm C~IV$ NALs with low velocities are related to quasar outflows. In low velocity, the uniform distribution might overpredict cosmologically intervening absorptions and thus underpredict outflow absorptions. In addition, the assumption that all absorptions with $\beta>0.06$ arise from cosmologically intervening galaxies ignores very high velocity outflows (e.g., $\beta>0.06$, Tombesi et al. 2010, 2011, 2013; Gupta et al. 2013a,b; Chen \& Qin 2013). This also underpredicts outflow absorptions. In order to check these underpredictions, we plot the velocity distribution for the 52 variable absorptions of our sample in Figure 8. It can be seen that there are 5 (9.6\%) systems with $\beta>0.06$ and a few systems locate in the range of small $\beta$. This implies that the velocity distribution shown in the bottom panel of Figure 7 is the lower limit of the outflow fraction.

The velocity distribution of variable absorptions is very interesting, which implies that the population of narrow $\rm C~IV$ absorption outflows, as shown in the bottom panel of Figure 7, does exist a peak at $\upsilon\approx2000~\rm km~s~^{-1}$ and drops dramatically at low velocities. The accurate redshifts of quasar systems are important to the velocity distribution of intrinsic absorptions. Such accurate redshifts can be estimated by narrow emission lines (e.g., $\rm [O~III]$; $\rm [O~II]$), and are not currently available for our sample. In the future, we will further investigate the velocity distribution of intrinsic absorptions when a larger sample and more accurate emission redshifts are available.

\begin{figure}
\centering
\includegraphics[width=8cm,height=10cm]{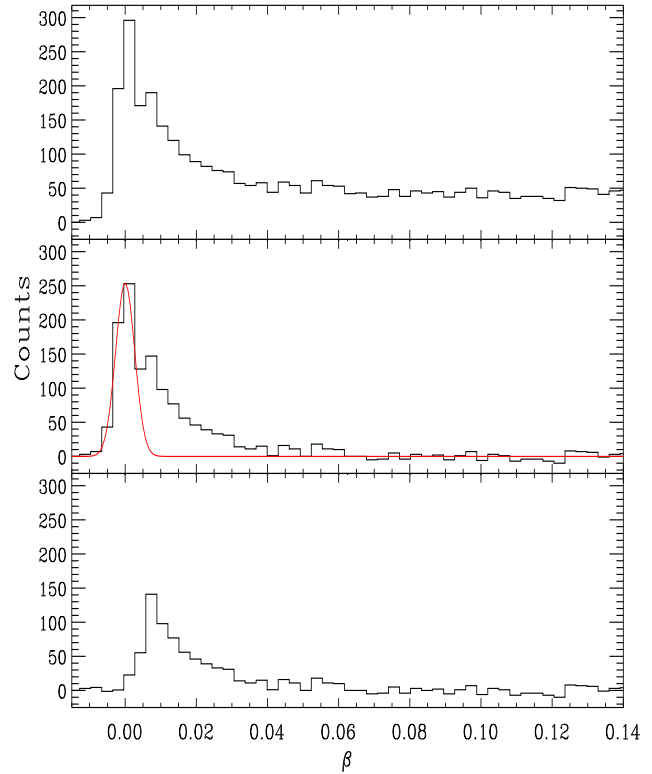}
\caption{The relative velocity distributions of absorbers. Top panel: note that a remarkable excess at $\beta \approx 0$ and a platform with average count of 42 in the range of $\beta > 0.06$. Middle panel: the counts showed in top panel minus the average count of 42 in the range of $\beta > 0$. The red solid line is the Gaussian fitting centered at $\beta=0.0$ to the data with $\beta<0.0$. Bottom panel: the counts showed in middle panel minus the Gaussian fitting (red solid line).}
\end{figure}

\begin{figure}
\centering
\includegraphics[width=8cm,height=7cm]{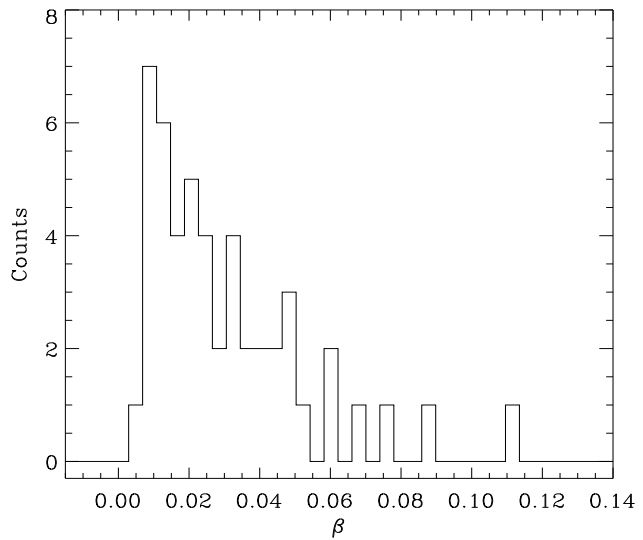}
\caption{Relative velocity distributions for variable absorptions. Note that there are 5 systems with $\beta>0.06$ and some systems locate in the range of $0<\beta<0.01$.}
\end{figure}

Figure 7 shows that there are 5 variable absorption systems with $\beta>0.06$, which are in the following.

J080609.24+141146.4 (Figure B8, $z_{\rm em}=2.2877$): A $\rm C~IV$ absorption system with $z_{\rm abs}=2.0062$ is detected in the SDSS-I/II spectrum, which disappeared from the BOSS spectrum. This system has a high velocity of $\beta=0.0893$. The FWHM of $\rm C~IV\lambda\lambda1548,1551$ doublet, measured from the SDSS-I/II spectrum, is $808~\rm km~s^{-1}$ at quasar rest frame.

J100716.69+030438.7 (Figure B15, $z_{\rm em}=2.1241$): The variable $\rm C~IV$ absorption systems are located at $z_{\rm abs}=1.9129$ and $z_{\rm abs}=1.9426$, respectively, whose relative velocities are $\beta=0.0699$ and $\beta=0.0598$, respectively. The system with $z_{\rm abs}=1.9426$ cannot be detected in the SDSS-I/II spectrum, whose FWHM of $\rm C~IV\lambda\lambda1548,1551$ doublet, measured from the BOSS spectrum, is $1108~\rm km~s^{-1}$ at quasar rest frame. The system with $z_{\rm abs}=1.9129$ changes obviously in both strengths and profiles between the two observations. An unsymmetrical profile of $\rm C~IV$ absorption is clear in the SDSS-I/II spectrum. The FWHMs of $\rm C~IV$ absorption troughs are broader than $800~\rm km~s^{-1}$ at quasar rest frame.

J150033.53+003353.6 (Figure B30, $z_{\rm em}=2.4360$): The $\rm C~IV$ absorption system with $z_{\rm abs}=2.1849$ emerged from the BOSS spectrum, whose relative velocity is $\beta=0.0757$. The FWHM of $\rm C~IV$ absorption trough is $821~\rm km~s^{-1}$ at quasar rest frame.

J161511.35+314728.3 (Figure B34, $z_{\rm em}=2.0981$): Both strengths and profiles change obviously between the two observations for $\rm C~IV$ absorption system with $z_{\rm abs}=1.9157$, whose relative velocity is $\beta=0.0757$. The FWHMs of $\rm C~IV$ absorption troughs at quasar rest frame are $840~\rm km~s^{-1}$ and $668~\rm km~s^{-1}$ for SDSS-I/II and BOSS spectra, respectively.

J213648.17-001546.6 (Figure B38, $z_{\rm em}=2.1736$): The $\rm C~IV$ absorption system with $z_{\rm abs}=1.8372$ is detected in the SDSS-I/II spectrum, which disappeared from the BOSS spectrum. This system has a very high velocity of $\beta=0.1116$. The FWHM of $\rm C~IV$ absorption trough is $557~\rm km~s^{-1}$ at quasar rest frame.

These 5 $\rm C~IV$ absorption systems with $\beta>0.06$ show large FWHMs ($>550~\rm km~s^{-1}$). Among them, three systems show fairly smooth profiles, which are in J080609.24+141146.4 (Figure B8), J100716.69+030438.7 (Figure B15, $z_{\rm abs}=1.9129$ and $\beta=0.0699$), and J150033.53+003353.6 (Figure B30), respectively, and might be potentially mini-BALs. The other two systems, namely J161511.35+314728.3 (Figure B34) and J213648.17-001546.6 (Figure B38), show spectroscopically resolved profiles.


The dependence of changes or fractional changes in $W_{\rm r}\lambda1548$ on time intervals has been observed for BALs and mini-BALs (e.g., Filiz Ak et al.2013; Misawa et al. 2014), finding that the larger changes or fractional changes in $W_{\rm r}\lambda1548$ are usually detected over the longer time intervals. For NALs, Hacker et al. (2013) observed obvious correlation between fractional changes in $W_{\rm r}\lambda1548$ and time intervals, but did not between changes in $W_{\rm r}$ and time intervals. There are 16 enhanced and 12 weaken absorption systems in our $\rm C~IV$ absorption sample. We show changes or fractional changes in $W_{\rm r}\lambda1548$ of these 28 systems in Figures 9 and 10, respectively. Both changes and fractional changes in $W_{\rm r}\lambda1548$ show large scatters with respect to time intervals.

\begin{figure}
\centering
\includegraphics[width=8cm,height=7cm]{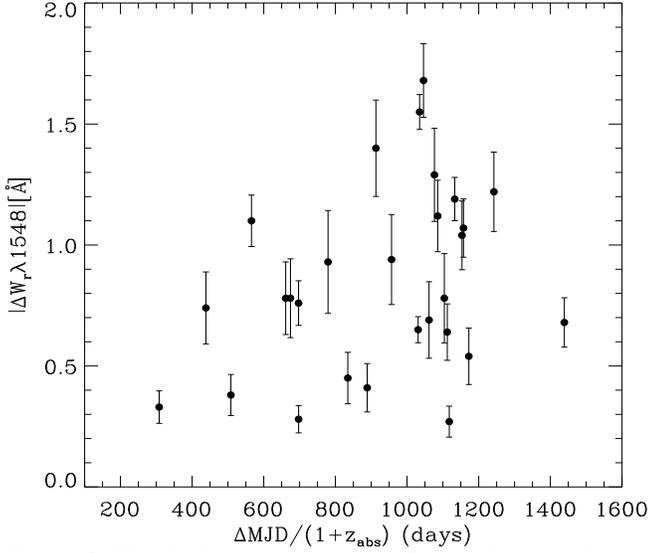}
\caption{The absolute changes in $W_{\rm r}\lambda1548$ as a function of time intervals at absorber rest frame for weaken or enhanced absorption systems.}
\end{figure}

\begin{figure}
\centering
\includegraphics[width=8cm,height=7cm]{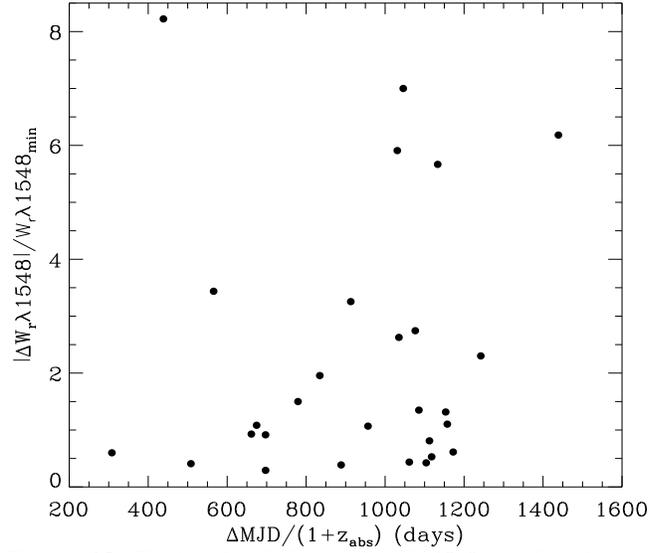}
\caption{The fractional changes in $W_{\rm r}\lambda1548$ as a function of the minimum $W_{\rm r}$ for weaken or enhanced absorption systems.}
\end{figure}

\subsection{The $\rm Mg~II$ absorbers}
We have detected 1808 $\rm Mg~II\lambda\lambda2796,2803$ absorption systems with $z_{\rm abs}=0.3948 \sim 1.7167$ (see the red solid line of Figure 4) from 1378 quasar spectra. We show differences of $W_{\rm r}\lambda2796$ between the two observations versus $W_{\rm r}\lambda2796$ measured from SDSS-I/II spectra in Figure 11, and show normalized differences of $W_{\rm r}$ in Figure 12. We find that most systems are non-variation within $3\sigma'_{\rm w}$, and none systems shows significant variation with $\Delta W_{\rm r} > 4\sigma_{\rm w}'$ for $\lambda2796$ lines and $\Delta W_{\rm r} > 3\sigma_{\rm w}'$ for $\lambda2803$ lines.

\begin{figure}
\centering
\includegraphics[width=8cm,height=7cm]{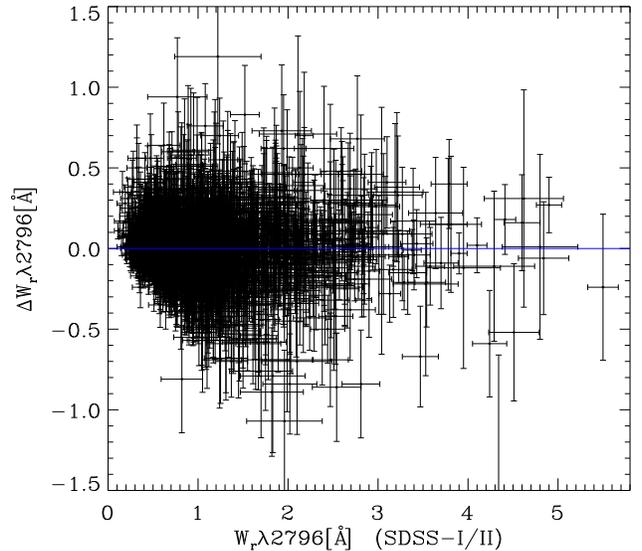}
\caption{The distribution of differences in $W_{\rm r}\lambda2796$ versus $W_{\rm r}\lambda2796$ measured from the SDSS-I/II spectra.}
\end{figure}

\begin{figure}
\centering
\includegraphics[width=8cm,height=7cm]{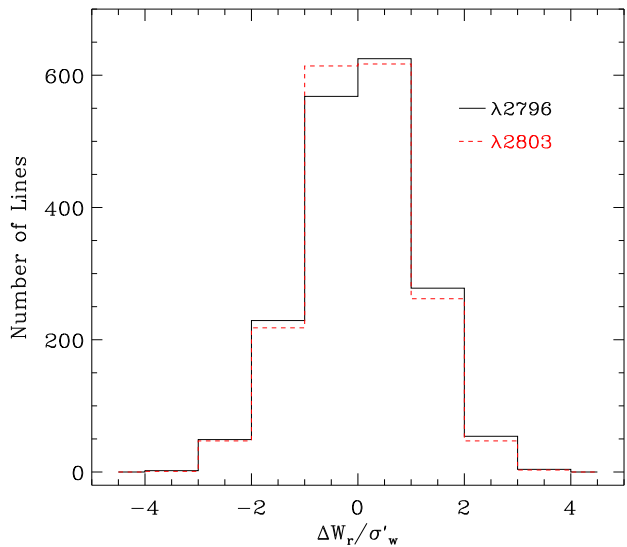}
\caption{Distributions of normalized differences of $W_{\rm r}$ between SDSS-I/II and BOSS observations. None of systems shows obvious variation with $\Delta W_{\rm r} > 4\sigma_{\rm w}'$ for $\lambda2796$ lines and $\Delta W_{\rm r} > 3\sigma_{\rm w}'$ for $\lambda2803$ lines.}
\end{figure}

Hacker et al. (2013), includes two systems with $\Delta W_{\rm r} > 4\sigma_{\rm w}'$ for $\lambda2796$ lines and $\Delta W_{\rm r} > 3\sigma_{\rm w}'$ for $\lambda2803$ lines. One system with $z_{\rm abs}=1.1140$ and $\beta=0.1518$ in J152555.81+010835.4 is blended with a separate $\rm Mg~II$ system with $z_{\rm abs}=1.1162$, which is shown in Figure 13. Two pairs of Gaussian component are invoked to fit this blended absorption troughs. The fitting results are also shown in Figure 13 with purple and green lines. We find that the absorption strength of each component is consistent within $3\sigma_{\rm w}'$ in the two SDSS epochs, although obvious changes in the profiles of the system with $z_{\rm abs}=1.1162$ are observed (see the panel (B) of Figure 13).
The other system is $z_{\rm abs}=0.9114$ and $\beta=0.1492$, which was imprinted in the spectrum of quasar J143826.73+642859.8 that is not reobserved by BOSS. Hacker et al. found that the $\rm Mg~II$ absorption of this system changes significantly in the two SDSS observations. That is $\Delta W_{\rm r}=0.32\pm0.09$\AA~ for the $\rm \lambda2796$ line and $\Delta W_{\rm r}=0.40\pm0.09$\AA~ for the $\rm \lambda2803$ line. It is difficult to determine whether variable $\rm Mg~II$ absorption systems with large $\beta$ are intrinsic to quasars or not (Hacker et al. 2013). It might be absorptions caused by high velocity outflows, and their variations can be due to changes in ionization states or bulk motion of absorbing gas. It might also be formed in cosmologically intervening galaxies, and their variations are produced by bulk motion of absorbing gas.

\begin{figure}
\centering
\includegraphics[width=8cm,height=10cm]{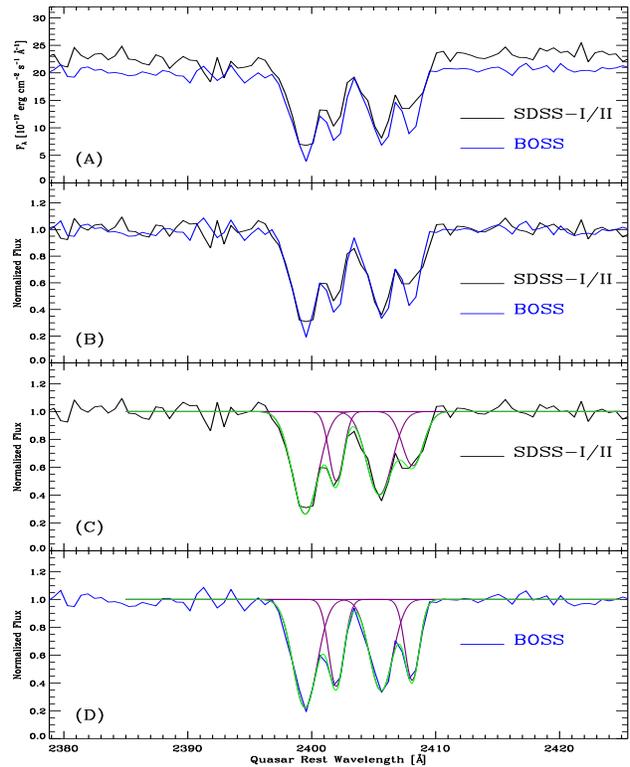}
\caption{The spectra of quasar J152555.81+010835.4. The SDSS-I/II spectra are plotted with black lines, and the BOSS spectra with blue lines. The panel (A) shows the SDSS-I/II (black line) and BOSS (blue line) spectra overplotted. The panel (B) shows the pseudo-continuum normalized spectra from the SDSS-I/II (black line) and BOSS (blue line), respectively. The panel (C) is the pseudo-continuum normalized spectra from the SDSS-I/II with Gaussian fittings, and the panel (D) is the pseudo-continuum normalized spectra from the BOSS with Gaussian fittings. In panels (C) and (D), a purple line represents a Gaussian component, and the green line is just the sum of multiple Gaussian components.}
\end{figure}

In the same single outflow, the ion with higher ionization potential is expected to be closer to the ionizing source. In other words, the intrinsic $\rm C~IV$ and $\rm Mg~II$ (the ionization potentials are 64.494eV for the $\rm C~IV$ and 15.035eV for the $\rm Mg~II$, respectively) absorptions may not arise from the same region of the outflow. In spectral regions of $20,000~\rm km~s^{-1}$ bluewards of $\rm C~IV$ emission lines, we search for accompanying $\rm C~IV$ NALs for each $\rm Mg~II$ absorption system with relative velocity $< 20,000~\rm km~s^{-1}$ with respect to the quasar system. Only 3 $\rm Mg~II$ absorption systems are found to have accompanying narrow $\rm C~IV$ absorption. We carefully check these accompanying narrow $\rm C~IV$ absorptions and find that their absorption strengths are consistent with each other within $1\sigma'_{\rm w}$ in the two SDSS epochs.

\section{Discussions}
The origin  of variable absorptions includes (1) the clumpy absorbing gas moving across our sightline, (2) variations in ionization structures of the absorbing gas, or (3) a combination of above two scenarios. Most of the UV continuum emission is believed to be emitted from the inner parts of a geometric thin and optical thick accretion disc with a radius scale of $r_{\rm cont} \sim 5R_{\rm s} $ (e.g., Wise et al. 2004; Misawa et al.2005; Chen \& Qin 2013), where $R_{\rm s} \equiv 2GM_{\rm BH}/c^2$ is the Schwarzchild radius. For supermassive black holes with $M_{\rm BH} \sim$ a few $\times10^9~M_{\rm \odot}$, one can estimate the size scales of UV continuum emission sources. Those are, $r_{\rm cont} \sim 5\times10^{\rm 15}~ \rm cm$. We take velocity scale of absorbers across UV continuum emission sources to be $v_{\rm cross} \sim 10^4~\rm km~s^{-1}$, which leads to crossing times of $t_{\rm cross} = r_{\rm cont}/v_{\rm cross} \sim 0.2$ year. Time intervals of variable $\rm C~IV$ absorption systems between two observations are over months to years, which are comparable to crossing times required to fully move across UV continuum emission sources. We find that multiple $\rm C~IV$ absorption systems detected from the same quasar spectra vary together (see Table 1 and Appendix B). It is difficult to orchestrate multiple parcels of absorbing gas to move across the continuum emission source at the same time. Therefore, the scheme that clumpy absorbing gas passes through the cylinder of sightline may not be a plausible explanation for the well coordinated variations of multiple absorption systems, although we can not rule it out.

Changes in the incident flux of absorbers can give rise to increase or decrease the ionization levels of the absorbing structure, and thus induce variable absorption. The fluctuations of the incident flux can be caused by (1) the variation of the quasar continuum emission; and (2) the change in the optical depth between the absorber and the continuum emission source. The total column density of the shielding material located between the continuum emission source and the absorber play an important role in the ionizing continuum emission reaching the absorber. The motions of the clumpy shielding material might alter the effective coverage fraction of the absorber to the continuum emission source, and ultimately bring about changes in the ionization structure of absorbing gas. Multiple parcels of absorbing gas at different velocities, which are located in the same outflow, can be affected together by the variable obscuration of ionizing continuum. This might be the most plausible origin leading to well coordinated variations of multiple absorption systems. Of course, both bulk motions and changes in the ionization levels of the absorbing gas can be applicable to a single variable absorption system.

\section{Summary}
Based on the quasar spectra population obtained by both SDSS-I/II and BOSS, we have constructed a sample of 3524 quasars with high signal-to-noise ratio to analyze narrow $\rm C~IV\lambda\lambda1548,1551$ and $\rm Mg~II\lambda\lambda2796,2803$ absorption systems. The time intervals of the two SDSS observations for these quasars are about 150 $\sim$ 2643 days at quasar rest frame.

In this work, we have detected a total of 3580 narrow $\rm C~IV$ absorption systems with $z_{\rm abs} = 1.5188 \sim 3.5212$. These $\rm C~IV$ systems can be well accounted for by cosmologically intervening absorptions with an uniform distribution at $\beta>0$, environmental absorptions with a Gaussian distribution centred at $\beta=0$ that are formed in quasar host galaxies and galaxies in the vicinities of quasars, and quasar outflow absorptions. The outflow contribution peaks at $\upsilon\approx2000~\rm km~s^{-1}$, reaches beyond $18000~\rm km~s^{-1}$, and decreases rapidly below the peak velocity. The quasar outflow absorptions might be underestimated since we might overestimate the contributions from absorptions of cosmologically intervening galaxies, quasar host galaxies and galaxies in the vicinities of quasars. Among 3580 $\rm C~IV$ absorption systems, 52 systems show variable absorptions with significant levels of $>4\sigma$ for blue lines and $>3\sigma$ for red lines. The velocity distribution of variable $\rm C~IV$ systems suggests that there might be a peak velocity distribution for quasar outflow absorptions. However, the accurate shape of the velocity distribution of quasar outflow absorptions needs accurate emission redshifts estimated from narrow emission lines.

We find that, for variable $\rm C~IV$ absorption systems, changes in $W_{\rm r}\lambda1548$ depend on neither timescales of two observations, nor relative velocities of absorbers. We do not find, for the weaken or enhanced absorption systems, remarkable correlation between fractional changes in $W_{\rm }r\lambda1548$ and time intervals as well.

We have also detected a total of 1808 narrow $\rm Mg~II$ absorption systems with $z_{\rm abs} = 0.3948 \sim 1.7167$. All detected $\rm Mg~II$ absorptions are stable between two observations. None of systems shows variation with significant levels of $> 4\sigma$ for $\lambda2796$ lines and $> 3\sigma$ for $\lambda2803$ lines.

\vspace{6mm}We thank the anonymous referee for the careful and thorough review. We also thank Zhi-Yuan Li (Nanjing University) for useful discussions and comments. This work was supported by the National Natural Science Foundation of China (NO. 11363001, grants 11273015 and 11133001), National Basic Research Program (973 program No. 2013CB834905), and the Guangxi University of Science and Technology research projects (No. KY2015YB289).

Funding for SDSS-III has been provided by the Alfred P. Sloan Foundation, the Participating Institutions, the National Science Foundation, and the U.S.  Department of Energy Office of Science. The SDSS-III web site is http://www.sdss3.org/.

SDSS-III is managed by the Astrophysical Research Consortium for the Participating Institutions of the SDSS-III Collaboration including the University of Arizona, the Brazilian Participation Group, Brookhaven National Laboratory, Carnegie Mellon University, University of Florida, the French Participation Group, the German Participation Group, Harvard University, the Instituto de Astrofisica de Canarias, the Michigan State/Notre Dame/JINA Participation Group, Johns Hopkins University, Lawrence Berkeley National Laboratory, Max Planck Institute for Astrophysics, Max Planck Institute for Extraterrestrial Physics, New Mexico State University, New York University, Ohio State University, Pennsylvania State University, University of Portsmouth, Princeton University, the Spanish Participation Group, University of Tokyo, University of Utah, Vanderbilt University, University of Virginia, University of Washington, and Yale University.

\appendix
\section{The data of absorption systems}
\begin{table*}
\centering
\caption{Catalog of $\rm C~IV\lambda\lambda1548,1551$ absorption systems} \tabcolsep 0.7mm
 \begin{tabular}{ccccccccccccccccccccc}
 \hline\hline\noalign{\smallskip}
&&&\multicolumn{5}{c}{SDSS-I/II}&&\multicolumn{5}{c}{BOSS}\\
\cline{4-8}\cline{10-14}
SDSS NAME&$z_{\rm em}$&$z_{\rm abs}$ & PLATEID & MJD & FIBERID&$W_{\rm r}\lambda1548$&$W_{\rm r}\lambda1551$&&PLATEID&MJD&FIBERID&$W_{\rm r}\lambda1548$ & $W_{\rm r}\lambda1551$\\
&&&&&&\AA&\AA&&&&&\AA&\AA\\
\hline\noalign{\smallskip}
000330.18+000813.2	&	2.5962 	&	2.5593 	&	0686	&	52519	&	0356	&	$	0.79 	\pm	0.11 	$	&	$	0.29 	\pm	0.11 	$	&	&	4217	&	55478	&	0532	&	$	0.83 	\pm	0.07 	$	&	$	0.38 	\pm	0.06 	$	\\
000330.18+000813.2	&	2.5962 	&	2.5796 	&	0686	&	52519	&	0356	&	$	0.24 	\pm	0.07 	$	&	$	0.30 	\pm	0.10 	$	&	&	4217	&	55478	&	0532	&	$	0.22 	\pm	0.04 	$	&	$	0.34 	\pm	0.06 	$	\\
001016.49+001227.6	&	2.2761 	&	2.0552 	&	1491	&	52996	&	0395	&	$	0.24 	\pm	0.08 	$	&	$	0.28 	\pm	0.06 	$	&	&	4217	&	55478	&	0933	&	$	0.19 	\pm	0.05 	$	&	$	0.30 	\pm	0.06 	$	\\
001022.14-003701.2	&	3.1478 	&	2.6214 	&	0388	&	51793	&	0114	&	$	0.56 	\pm	0.10 	$	&	$	0.30 	\pm	0.08 	$	&	&	4217	&	55478	&	0096	&	$	0.60 	\pm	0.07 	$	&	$	0.38 	\pm	0.05 	$	\\
001022.14-003701.2	&	3.1478 	&	2.6600 	&	0388	&	51793	&	0114	&	$	0.49 	\pm	0.09 	$	&	$	0.44 	\pm	0.10 	$	&	&	4217	&	55478	&	0096	&	$	0.55 	\pm	0.06 	$	&	$	0.44 	\pm	0.07 	$	\\
001022.14-003701.2	&	3.1478 	&	3.1163 	&	0388	&	51793	&	0114	&	$	0.47 	\pm	0.06 	$	&	$	0.30 	\pm	0.06 	$	&	&	4217	&	55478	&	0096	&	$	0.61 	\pm	0.06 	$	&	$	0.41 	\pm	0.06 	$	\\
001057.59+011011.5	&	3.1069 	&	2.8948 	&	0686	&	52519	&	0561	&	$	0.45 	\pm	0.13 	$	&	$	0.24 	\pm	0.12 	$	&	&	4218	&	55479	&	0582	&	$	0.62 	\pm	0.13 	$	&	$	0.39 	\pm	0.14 	$	\\
001057.59+011011.5	&	3.1069 	&	3.0292 	&	0686	&	52519	&	0561	&	$	1.29 	\pm	0.12 	$	&	$	1.31 	\pm	0.12 	$	&	&	4218	&	55479	&	0582	&	$	1.50 	\pm	0.10 	$	&	$	1.53 	\pm	0.11 	$	\\
001057.59+011011.5	&	3.1069 	&	3.0789 	&	0686	&	52519	&	0561	&	$	1.34 	\pm	0.10 	$	&	$	1.33 	\pm	0.11 	$	&	&	4218	&	55479	&	0582	&	$	1.37 	\pm	0.08 	$	&	$	1.32 	\pm	0.09 	$	\\
001240.24+002433.6	&	2.6251 	&	2.5170 	&	1490	&	52994	&	0624	&	$	0.54 	\pm	0.08 	$	&	$	0.26 	\pm	0.05 	$	&	&	4218	&	55479	&	0696	&	$	0.49 	\pm	0.05 	$	&	$	0.27 	\pm	0.04 	$	\\
\hline\hline\noalign{\smallskip}
\end{tabular}
\end{table*}

\begin{table*}
\centering
\caption{Catalog of $\rm Mg~II\lambda\lambda2796,2803$ absorption systems} \tabcolsep 0.7mm
 \begin{tabular}{cccccccccccccccccccc}
 \hline\hline\noalign{\smallskip}
&&&\multicolumn{5}{c}{SDSS-I/II}&&\multicolumn{5}{c}{BOSS}\\
\cline{4-8}\cline{10-14}
SDSS NAME&$z_{\rm em}$&$z_{\rm abs}$ & PLATEID & MJD & FIBERID&$W_{\rm r}\lambda2796$&$W_{\rm r}\lambda2803$&&PLATEID&MJD&FIBERID&$W_{\rm r}\lambda2796$ & $W_{\rm r}\lambda2803$\\
&&&&&&\AA&\AA&&&&&\AA&\AA\\
\hline\noalign{\smallskip}
000042.90+005539.5	&	0.9517 	&	0.4964 	&	0387	&	51791	&	0531	&	$	0.38 	\pm	0.07 	$	&	$	0.43 	\pm	0.10 	$	&	&	4216	&	55477	&	0758	&	$	0.51 	\pm	0.07 	$	&	$	0.43 	\pm	0.09 	$	\\
000654.10-001533.3	&	1.7209 	&	1.0937 	&	0388	&	51793	&	0234	&	$	1.29 	\pm	0.16 	$	&	$	0.86 	\pm	0.15 	$	&	&	4217	&	55478	&	0290	&	$	1.11 	\pm	0.06 	$	&	$	0.81 	\pm	0.06 	$	\\
000654.10-001533.3	&	1.7209 	&	1.1317 	&	0388	&	51793	&	0234	&	$	1.51 	\pm	0.15 	$	&	$	1.44 	\pm	0.15 	$	&	&	4217	&	55478	&	0290	&	$	1.40 	\pm	0.06 	$	&	$	1.24 	\pm	0.07 	$	\\
000654.10-001533.3	&	1.7209 	&	1.4076 	&	0388	&	51793	&	0234	&	$	0.63 	\pm	0.10 	$	&	$	0.52 	\pm	0.10 	$	&	&	4217	&	55478	&	0290	&	$	0.56 	\pm	0.06 	$	&	$	0.55 	\pm	0.06 	$	\\
000701.32+002242.3	&	0.8882 	&	0.6541 	&	0388	&	51793	&	0430	&	$	0.41 	\pm	0.07 	$	&	$	0.22 	\pm	0.07 	$	&	&	4217	&	55478	&	0728	&	$	0.43 	\pm	0.03 	$	&	$	0.28 	\pm	0.03 	$	\\
000912.69+010919.0	&	2.4623 	&	0.8999 	&	0686	&	52519	&	0523	&	$	0.88 	\pm	0.16 	$	&	$	0.95 	\pm	0.24 	$	&	&	4217	&	55478	&	0864	&	$	0.88 	\pm	0.10 	$	&	$	0.71 	\pm	0.14 	$	\\
001016.49+001227.6	&	2.2761 	&	1.0216 	&	1491	&	52996	&	0395	&	$	1.44 	\pm	0.14 	$	&	$	0.94 	\pm	0.11 	$	&	&	4217	&	55478	&	0933	&	$	1.18 	\pm	0.08 	$	&	$	0.94 	\pm	0.08 	$	\\
001022.14-003701.2	&	3.1478 	&	0.9572 	&	0388	&	51793	&	0114	&	$	2.40 	\pm	0.19 	$	&	$	2.26 	\pm	0.19 	$	&	&	4217	&	55478	&	0096	&	$	2.23 	\pm	0.11 	$	&	$	2.02 	\pm	0.12 	$	\\
001223.56+003800.1	&	1.8593 	&	1.1161 	&	0686	&	52519	&	0586	&	$	1.09 	\pm	0.29 	$	&	$	0.82 	\pm	0.30 	$	&	&	4218	&	55479	&	0682	&	$	1.08 	\pm	0.24 	$	&	$	0.94 	\pm	0.23 	$	\\
001240.24+002433.6	&	2.6251 	&	1.1511 	&	1490	&	52994	&	0624	&	$	0.90 	\pm	0.16 	$	&	$	0.67 	\pm	0.20 	$	&	&	4218	&	55479	&	0696	&	$	0.66 	\pm	0.11 	$	&	$	0.43 	\pm	0.16 	$	\\
\hline\hline\noalign{\smallskip}
\end{tabular}
\end{table*}

\section{The spectra of the variable $\rm C~IV\lambda\lambda1548,1551$ absorption systems}
\clearpage

\begin{figure}
\centering
\includegraphics[width=8.cm,height=3.cm]{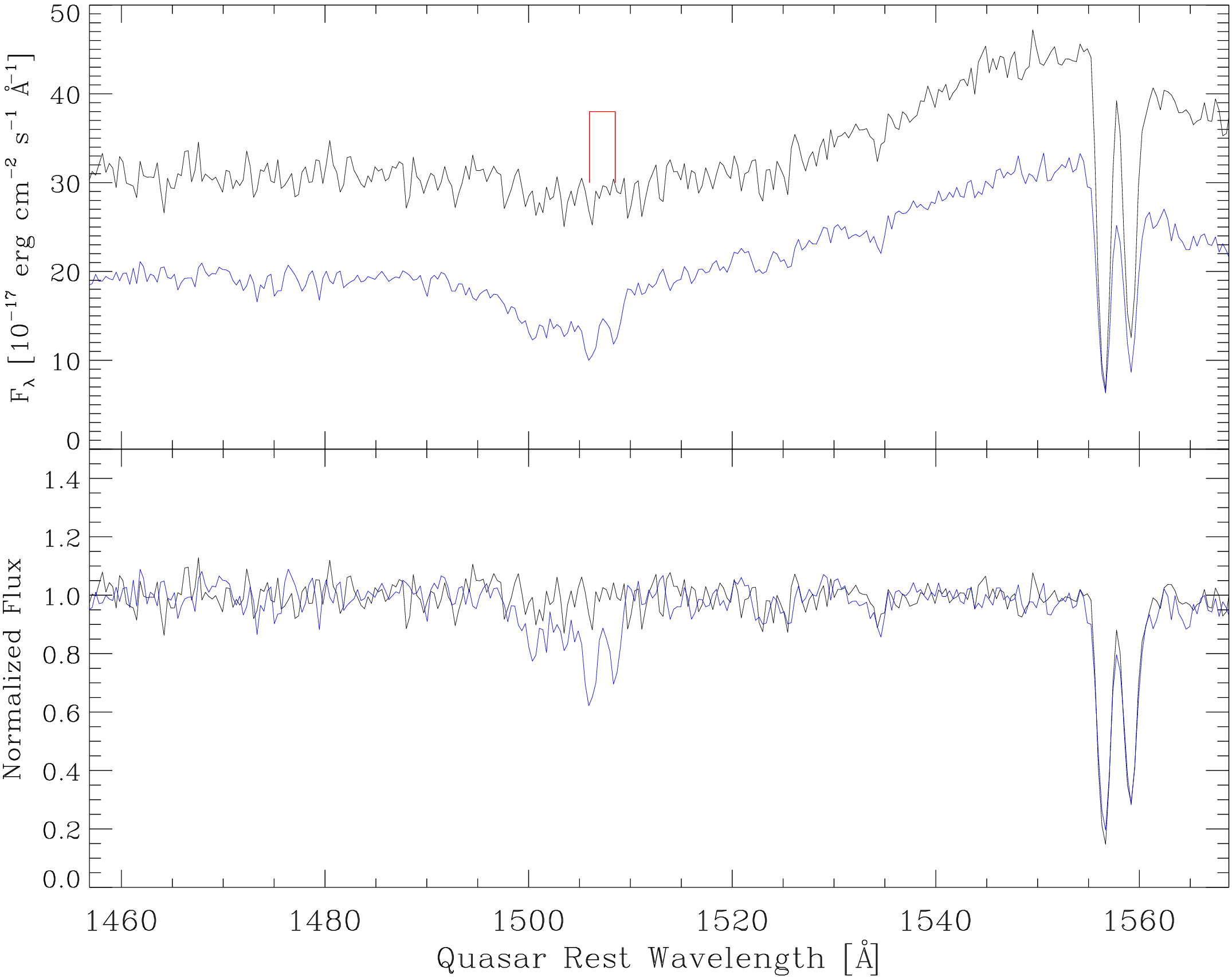}
\caption{The quasar J005157.24+000354.7 with $z_{\rm em}=1.9609$. The top panel shows the SDSS-I/II (black line) and BOSS (blue line) spectra overplotted. The bottom panel shows the pseudo-continuum normalized spectra from the SDSS-I/II (black line) and BOSS (blue line), respectively. The variable $\rm C~IV\lambda\lambda1548,1551$ absorption lines are marked by red lines. It is located at $z_{\rm abs}=1.8681$ and has a relative velocity value of $\beta=0.0318$ with respect to the emission line redshift.}
\end{figure}

\begin{figure}
\centering
\includegraphics[width=8.cm,height=3.cm]{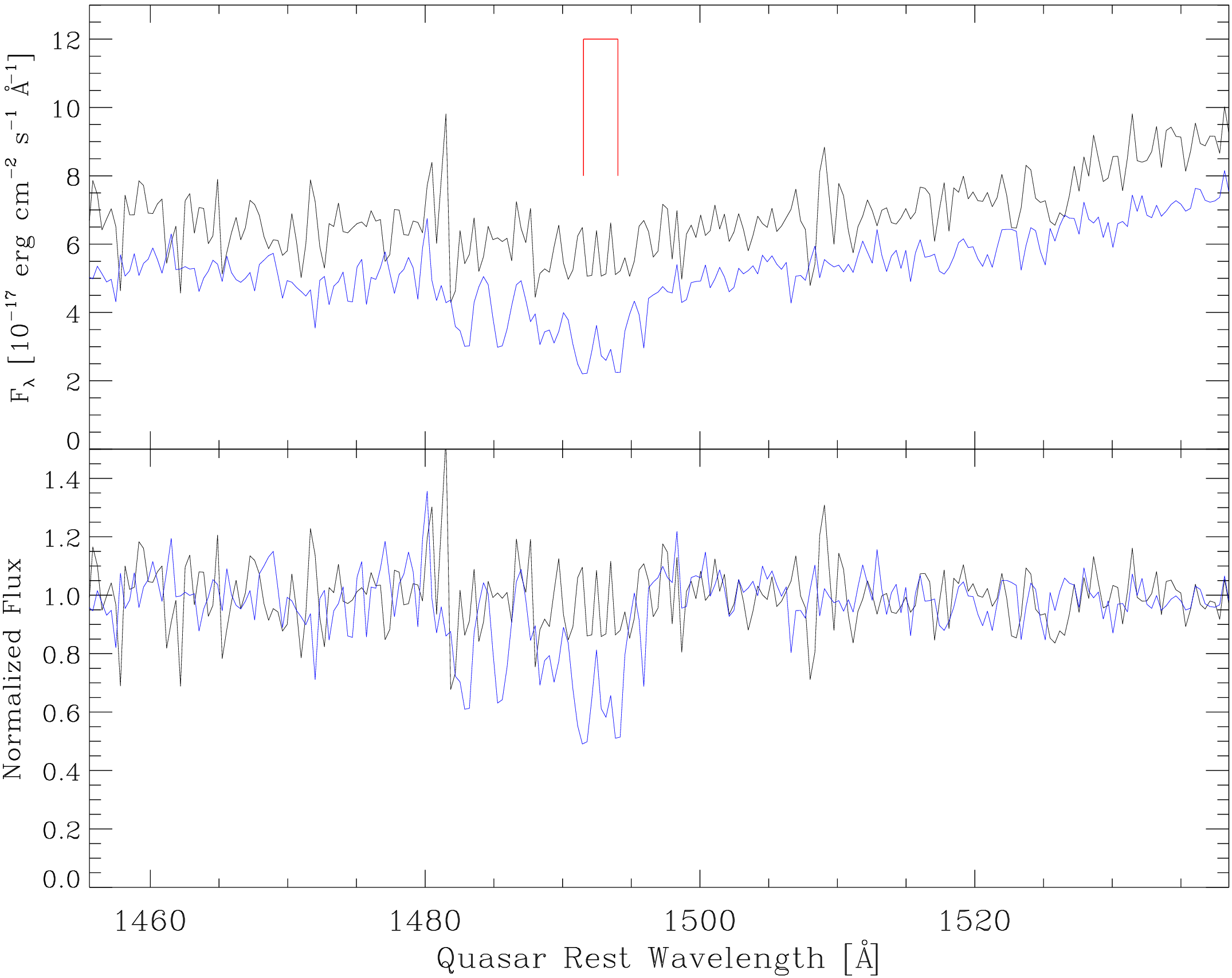}
\caption{The quasar J015017.70+002902.4 with $z_{\rm em}=3.0013$. See Figure A1 for the meanings of the color lines. The variable $\rm C~IV$ absorption system is located at $z_{\rm abs}=2.8344$, and has a relative velocity value of $\beta=0.0426$ with respect to the emission line redshift.}
\end{figure}

\begin{figure}
\centering
\includegraphics[width=8.cm,height=3.cm]{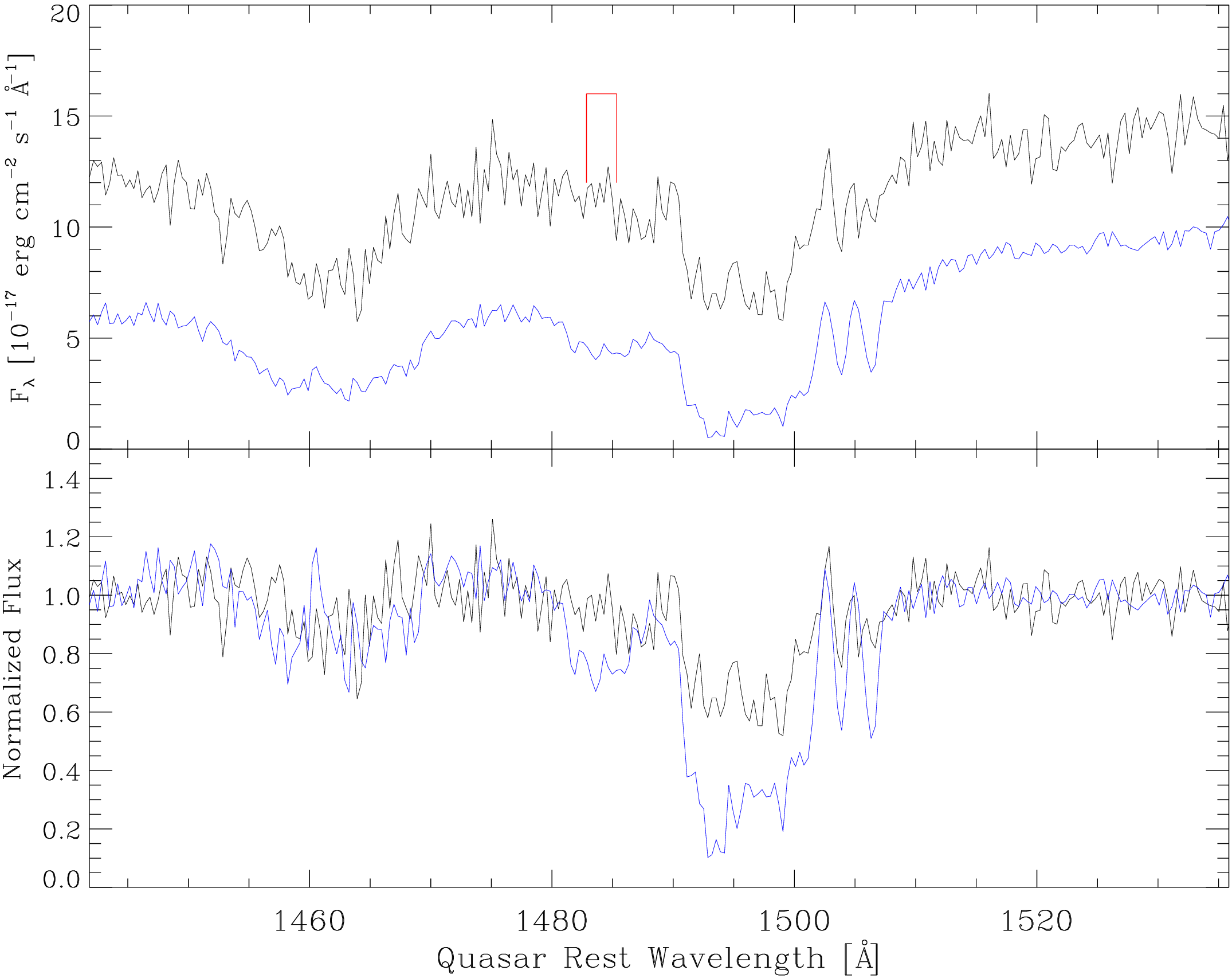}
\caption{The quasar J020629.33+004843.1 with $z_{\rm em}=2.4988$. See Figure A1 for the meanings of the color lines. The variable $\rm C~IV$ absorption system is located at $z_{\rm abs}=2.3624$, and has a relative velocity value of $\beta=0.0397$ with respect to the emission line redshift.}
\end{figure}

\begin{figure}
\centering
\includegraphics[width=8.cm,height=3.cm]{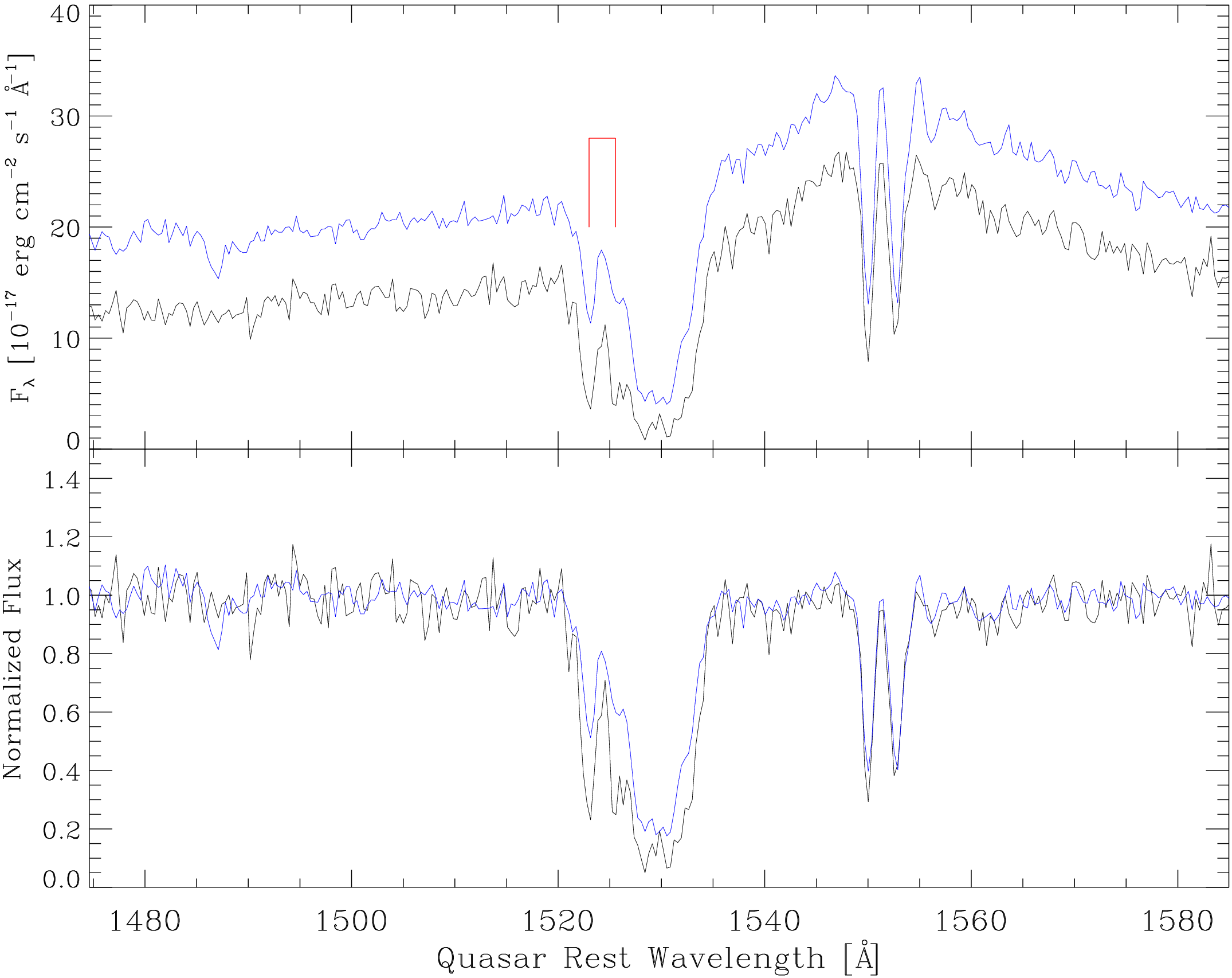}
\caption{The quasar J024304.68+000005.4 with $z_{\rm em}=2.0069$. See Figure A1 for the meanings of the color lines. The variable $\rm C~IV$ absorption system is located at $z_{\rm abs}=1.9426$, and has a relative velocity value of $\beta=0.0216$ with respect to the emission line redshift.}
\end{figure}

\begin{figure}
\centering
\includegraphics[width=8.cm,height=3.cm]{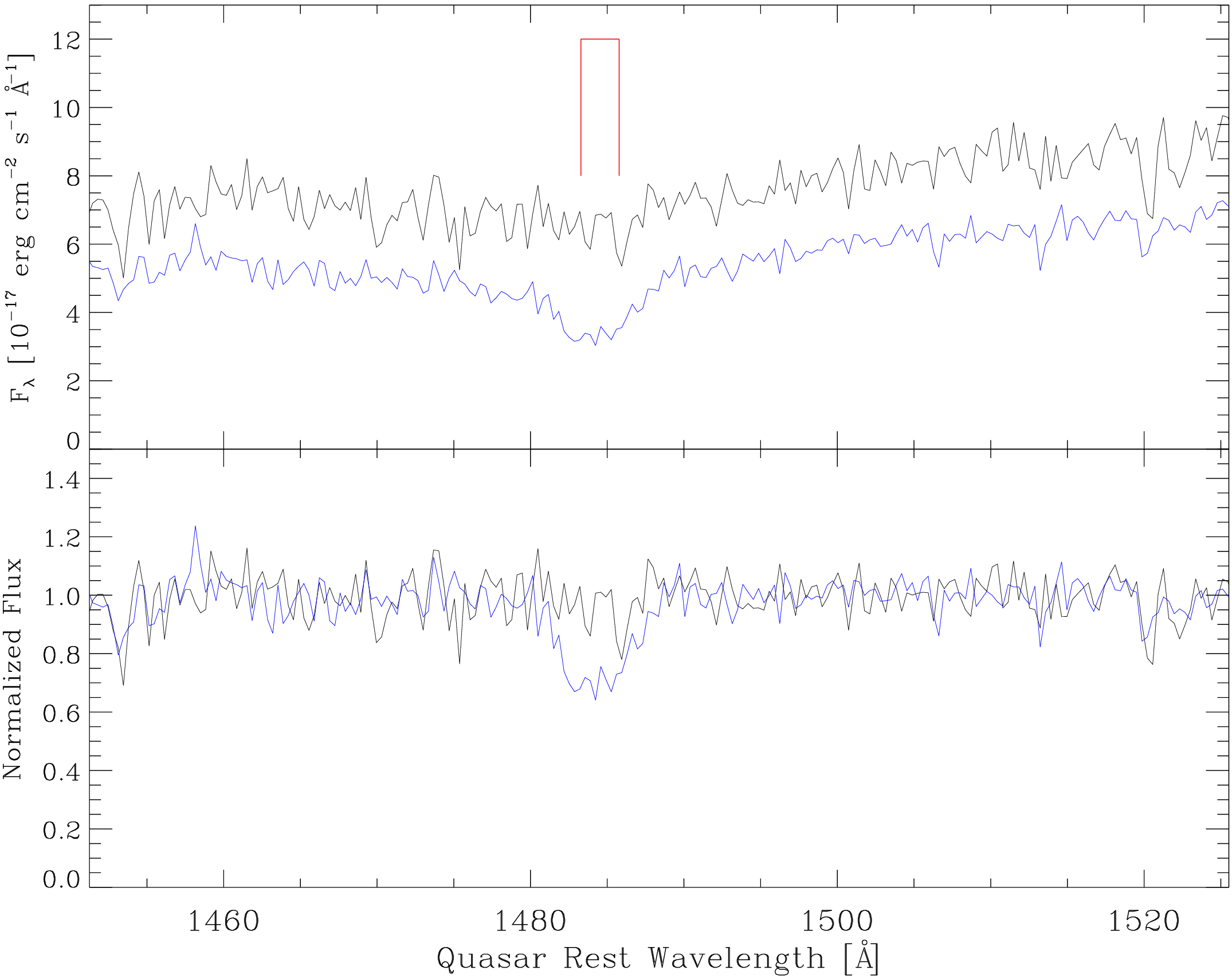}
\caption{The quasar J073232.79+435500.4 with $z_{\rm em}=3.4618$. See Figure A1 for the meanings of the color lines. The variable $\rm C~IV$ absorption system is located at $z_{\rm abs}=3.2569$, and has a relative velocity value of $\beta=0.0470$ with respect to the emission line redshift.}
\end{figure}
	
\begin{figure}
\centering
\includegraphics[width=8.cm,height=3.cm]{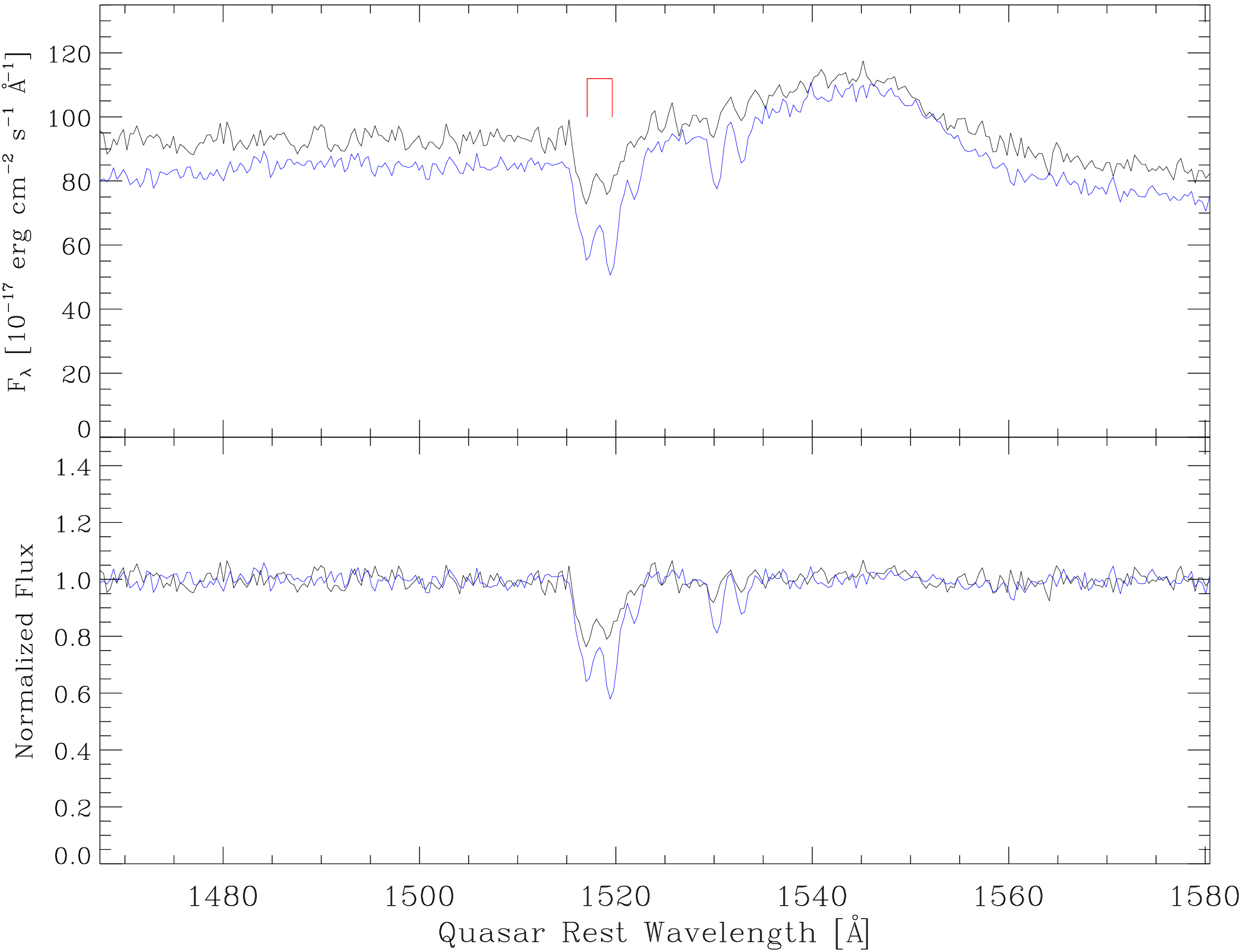}
\caption{The quasar J073406.75+273355.6 with $z_{\rm em}=1.9239$. See Figure A1 for the meanings of the color lines. The variable $\rm C~IV$ absorption system is located at $z_{\rm abs}=1.8609$, and has a relative velocity value of $\beta=0.0218$ with respect to the emission line redshift.}
\end{figure}

\begin{figure}
\centering
\includegraphics[width=8.cm,height=3.cm]{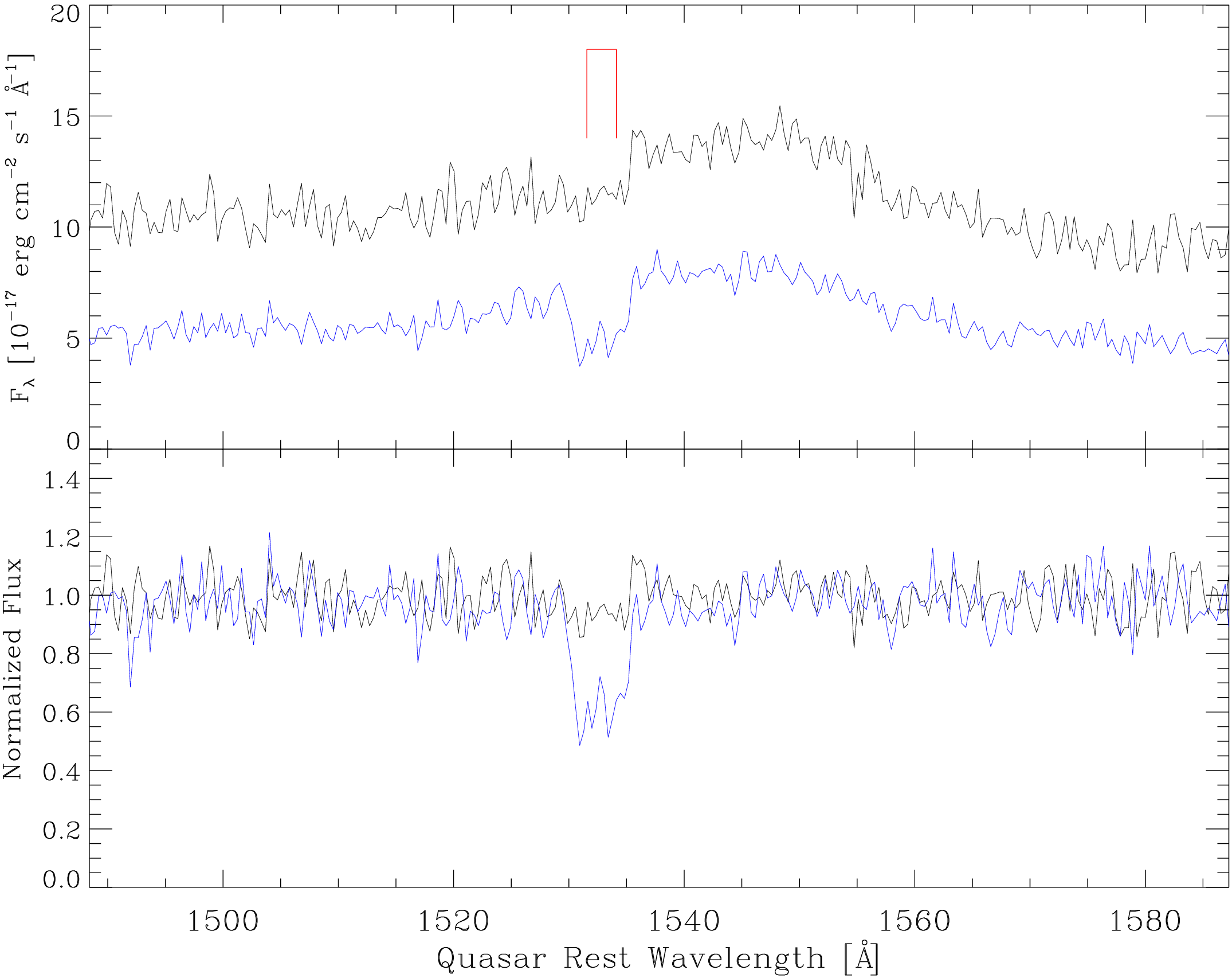}
\caption{The quasar J080006.59+265054.7 with $z_{\rm em}=2.3438$. See Figure A1 for the meanings of the color lines. The variable $\rm C~IV$ absorption system is located at $z_{\rm abs}=2.3033$, and has a relative velocity value of $\beta=0.0122$ with respect to the emission line redshift.}
\end{figure}

\begin{figure}
\centering
\includegraphics[width=8.cm,height=3.cm]{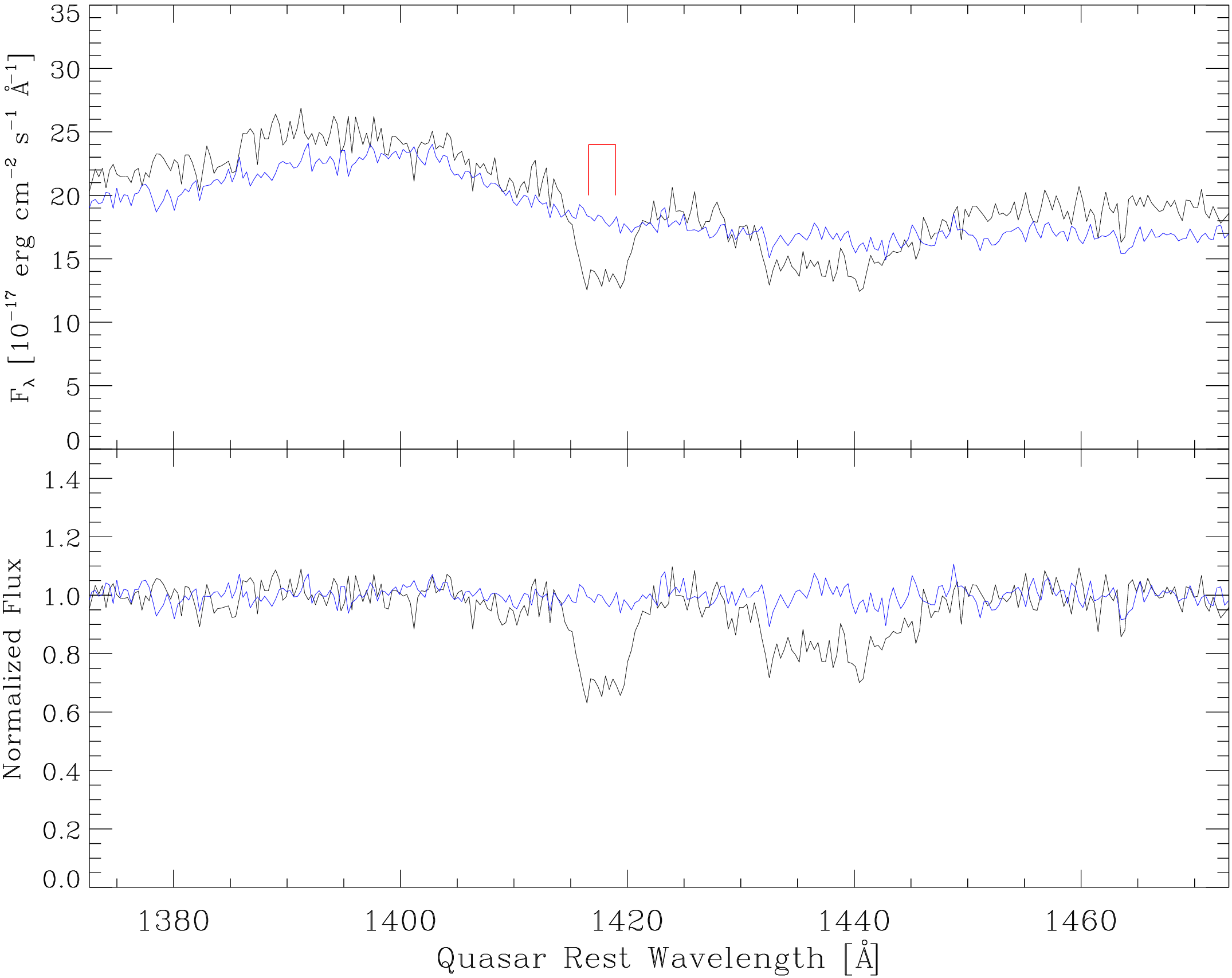}
\caption{The quasar J080609.24+141146.4 with $z_{\rm em}=2.2877$. See Figure A1 for the meanings of the color lines. The variable $\rm C~IV$ absorption system is located at $z_{\rm abs}=2.0062$, and has a relative velocity value of $\beta=0.0893$ with respect to the emission line redshift.}
\end{figure}

\clearpage
\begin{figure}
\centering
\includegraphics[width=8.cm,height=3.cm]{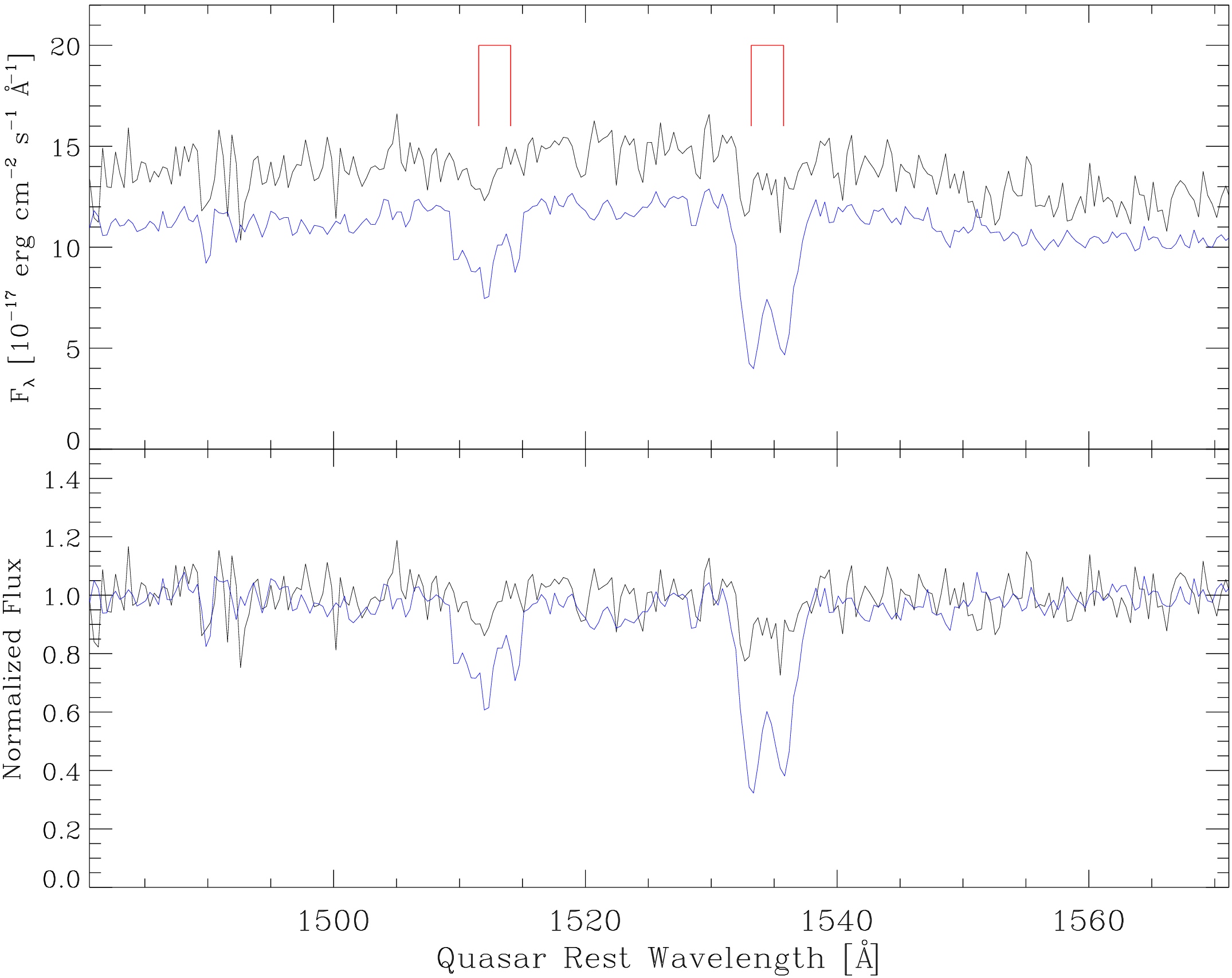}
\caption{The quasar J080906.88+172955.1 with $z_{\rm em}=2.9770$. See Figure A1 for the meanings of the color lines. The variable $\rm C~IV$ absorption systems are located at $z_{\rm abs}=2.8828$ and 2.9384, respectively, and have relative velocity values of $\beta=0.0240$ and 0.0098, respectively, with respect to the emission line redshift.}
\end{figure}

\begin{figure}
\centering
\includegraphics[width=8.cm,height=3.cm]{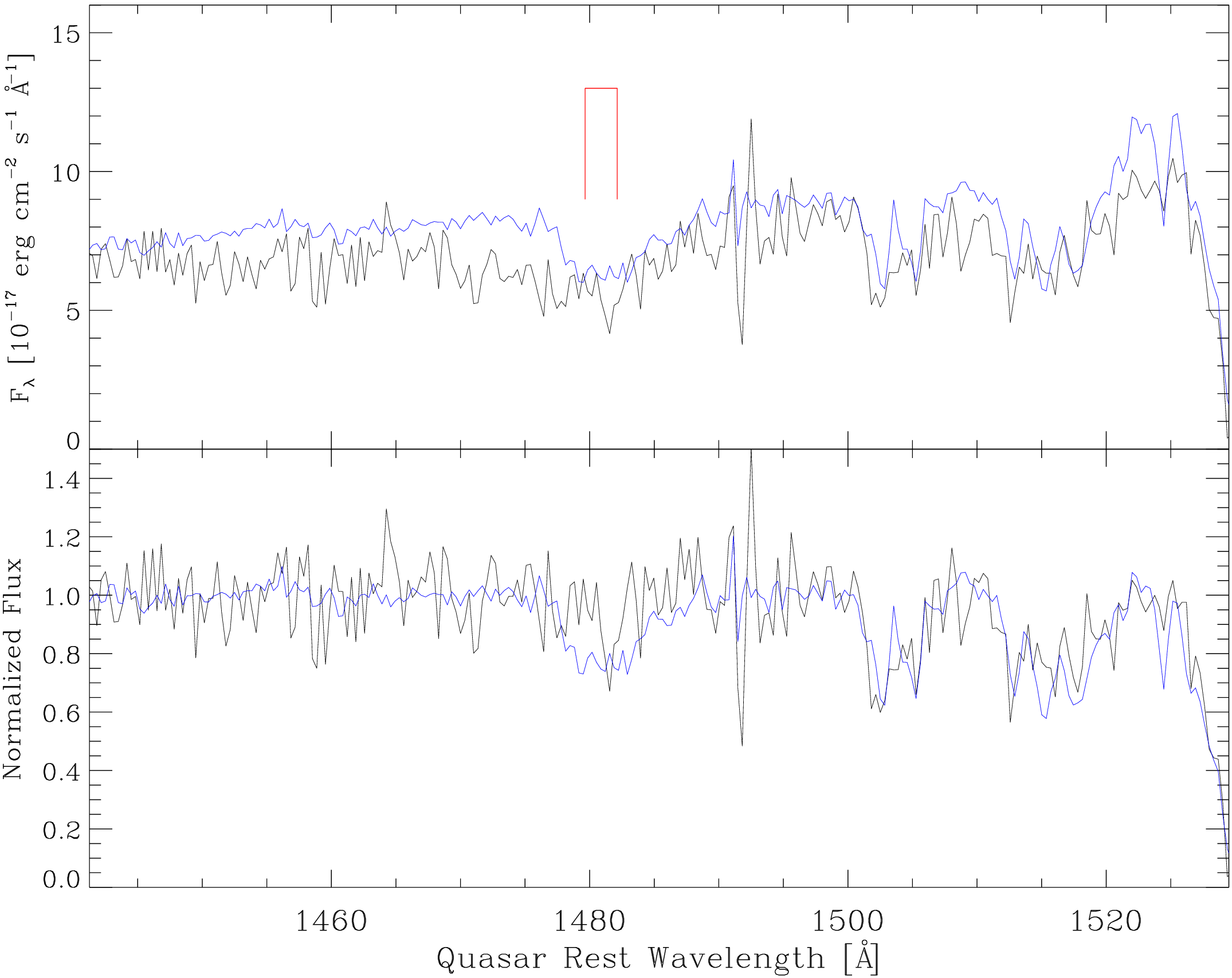}
\caption{The quasar J081655.49+455633.7 with $z_{\rm em}=2.7168$. See Figure A1 for the meanings of the color lines. The variable $\rm C~IV$ absorption system is located at $z_{\rm abs}=2.5745$, and has a relative velocity value of $\beta=0.0390$ with respect to the emission line redshift.}
\end{figure}	
	
\begin{figure}
\centering
\includegraphics[width=8.cm,height=3.cm]{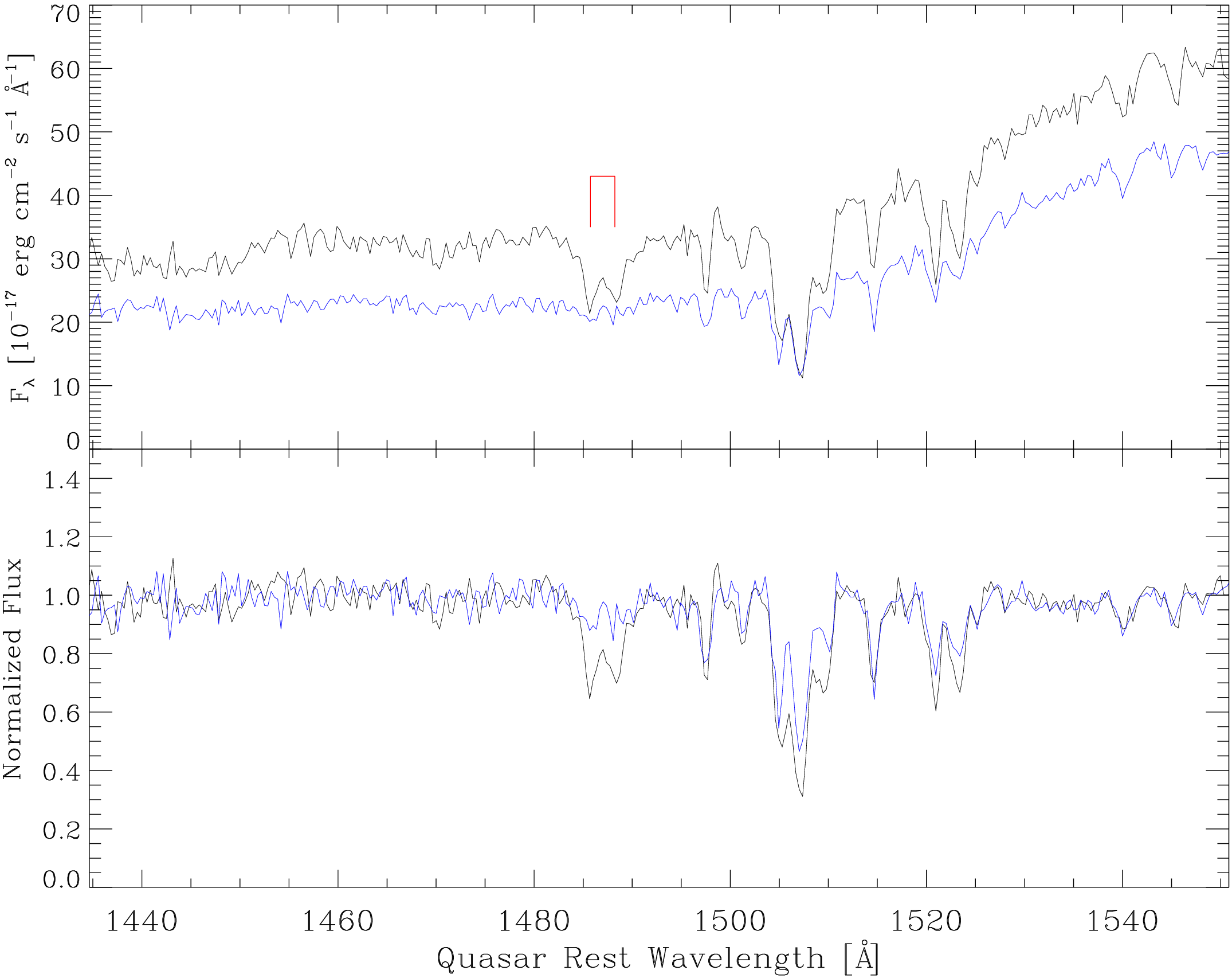}
\caption{The quasar J081929.59+232237.4 with $z_{\rm em}=1.8467$. See Figure A1 for the meanings of the color lines. The variable $\rm C~IV$ absorption system is located at $z_{\rm abs}= 1.7258$, and has a relative velocity value of $\beta=0.0434$ with respect to the emission line redshift.}
\end{figure}

\begin{figure}
\centering
\includegraphics[width=8.cm,height=3.cm]{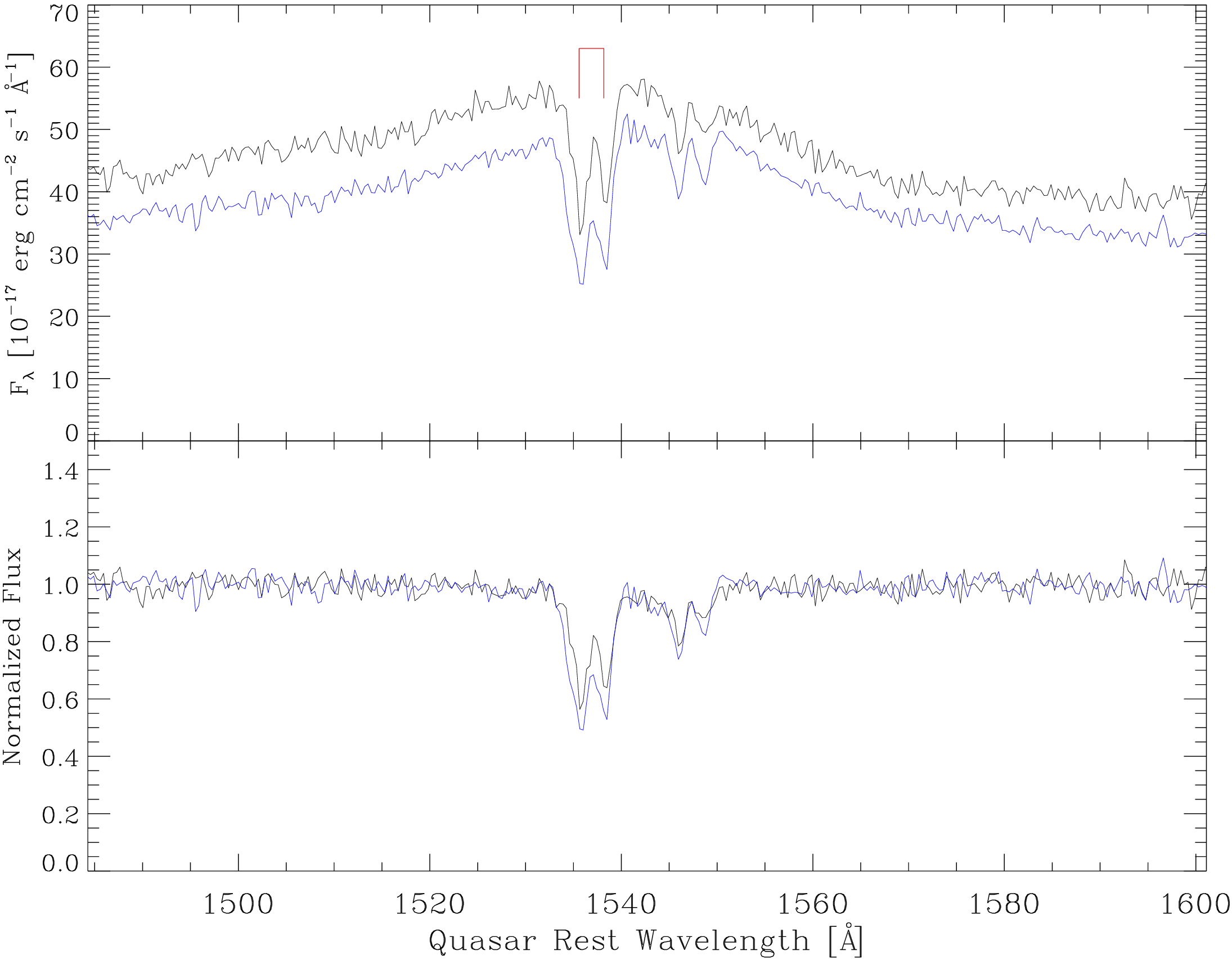}
\caption{The quasar J082751.78+132107.2 with $z_{\rm em}=1.8289$. See Figure A1 for the meanings of the color lines. The variable $\rm C~IV$ absorption system is located at $z_{\rm abs}= 1.8019$, and has a relative velocity value of $\beta=0.0096$ with respect to the emission line redshift.}
\end{figure}

\begin{figure}
\centering
\includegraphics[width=8.cm,height=3.cm]{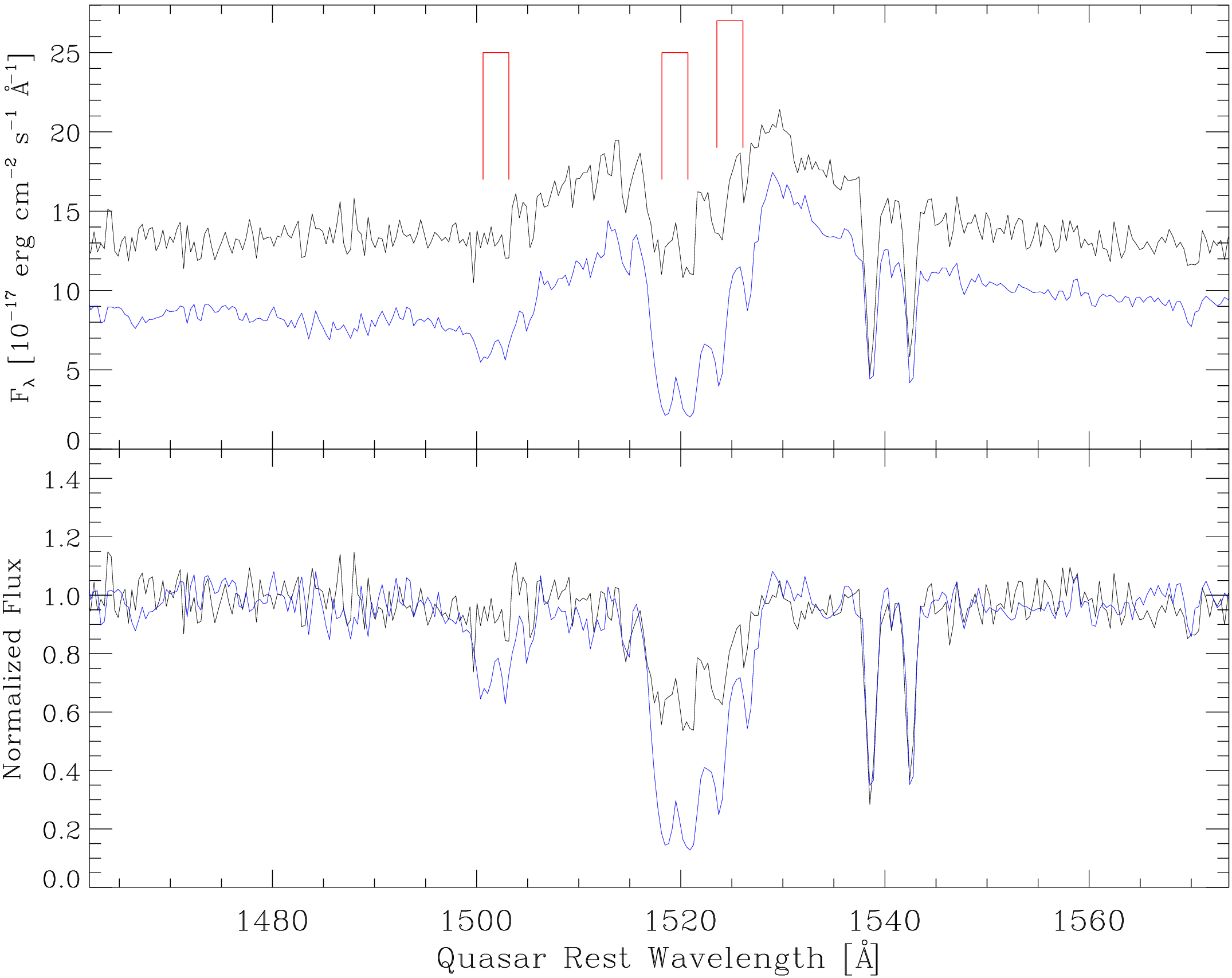}
\caption{The quasar J091621.46+010015.4 with $z_{\rm em}=2.2255$. See Figure A1 for the meanings of the color lines. The variable $\rm C~IV$ absorption systems are located at $z_{\rm abs}=2.1264$, 2.1629, and 2.1741, respectively, and have relative velocity values of $\beta=0.0312$, 0.0196, and 0.0161, respectively, with respect to the emission line redshift.}
\end{figure}	

\begin{figure}
\centering
\includegraphics[width=8.cm,height=3.cm]{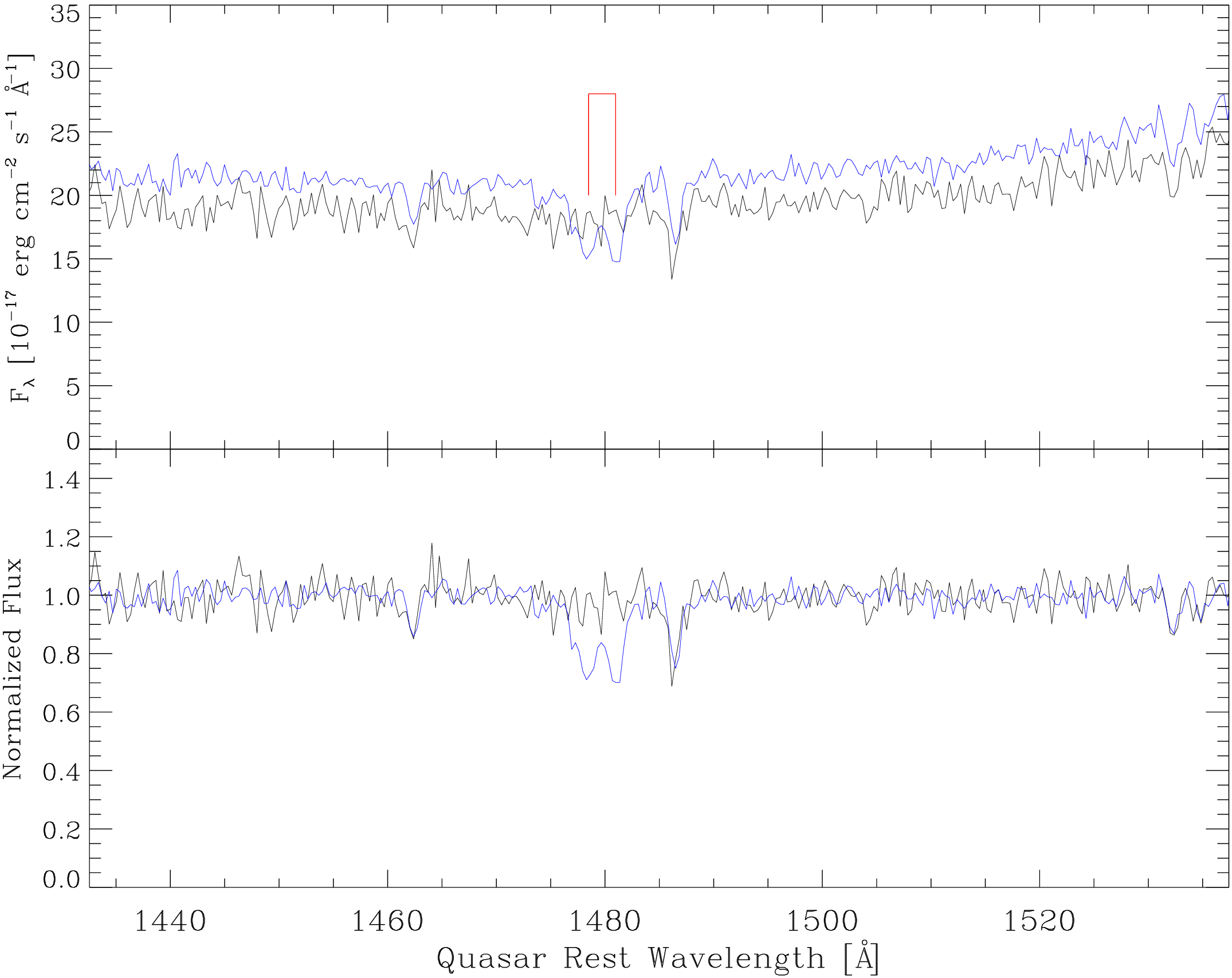}
\caption{The quasar J095254.10+021932.8 with $z_{\rm em}=2.1526$. See Figure A1 for the meanings of the color lines. The variable $\rm C~IV$ absorption system is located at $z_{\rm abs}= 2.0056$, and has a relative velocity value of $\beta=0.0477$ with respect to the emission line redshift.}
\end{figure}	

\begin{figure}
\centering
\includegraphics[width=8.cm,height=3.cm]{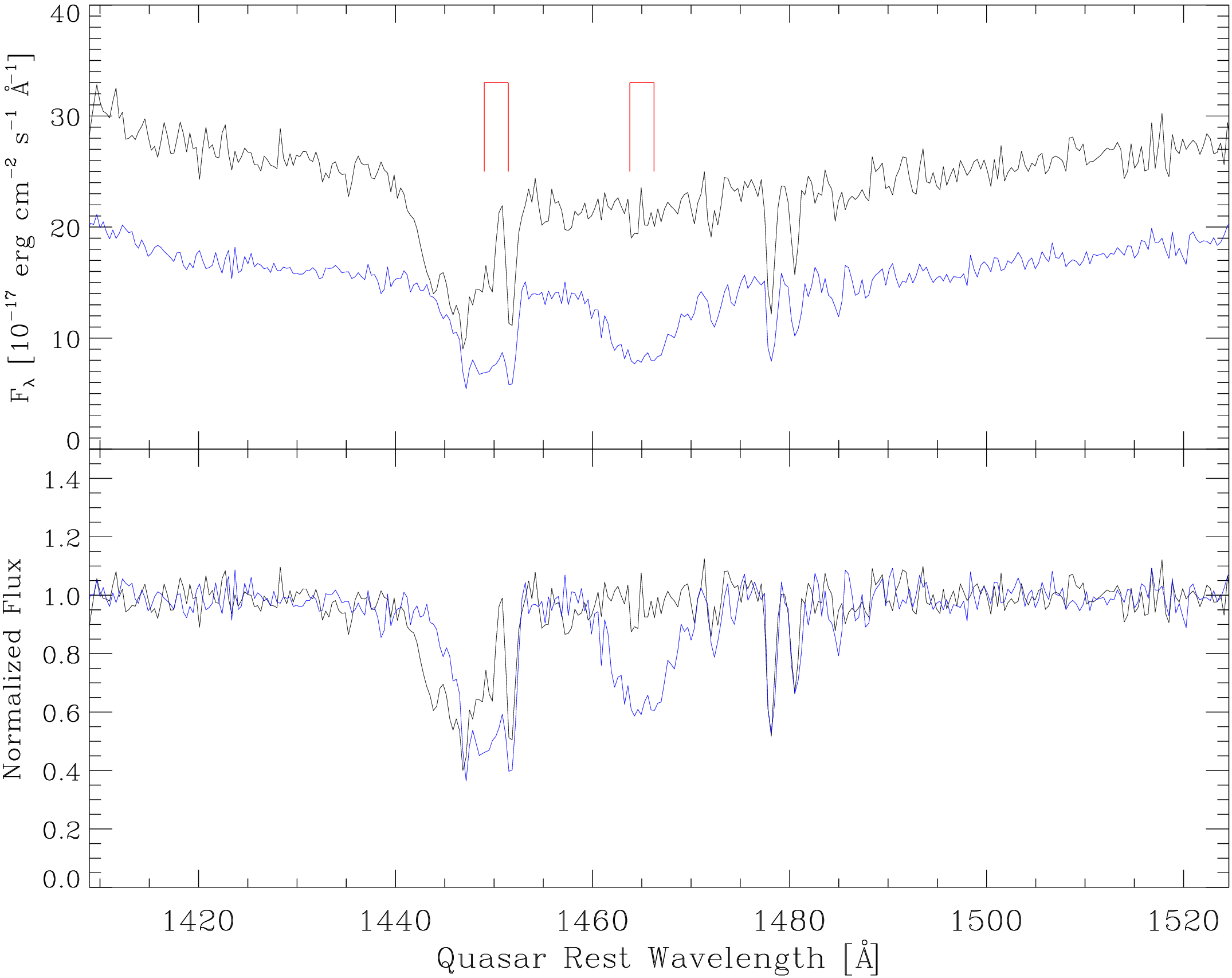}
\caption{The quasar J100716.69+030438.7 with $z_{\rm em}=2.1241$. See Figure A1 for the meanings of the color lines. The variable $\rm C~IV$ absorption systems are located at $z_{\rm abs}=1.9129$ and 1.9426, respectively, and have relative velocity values of $\beta=0.0699$ and 0.0598, respectively, with respect to the emission line redshift.}
\end{figure}

\begin{figure}
\centering
\includegraphics[width=8.cm,height=3.cm]{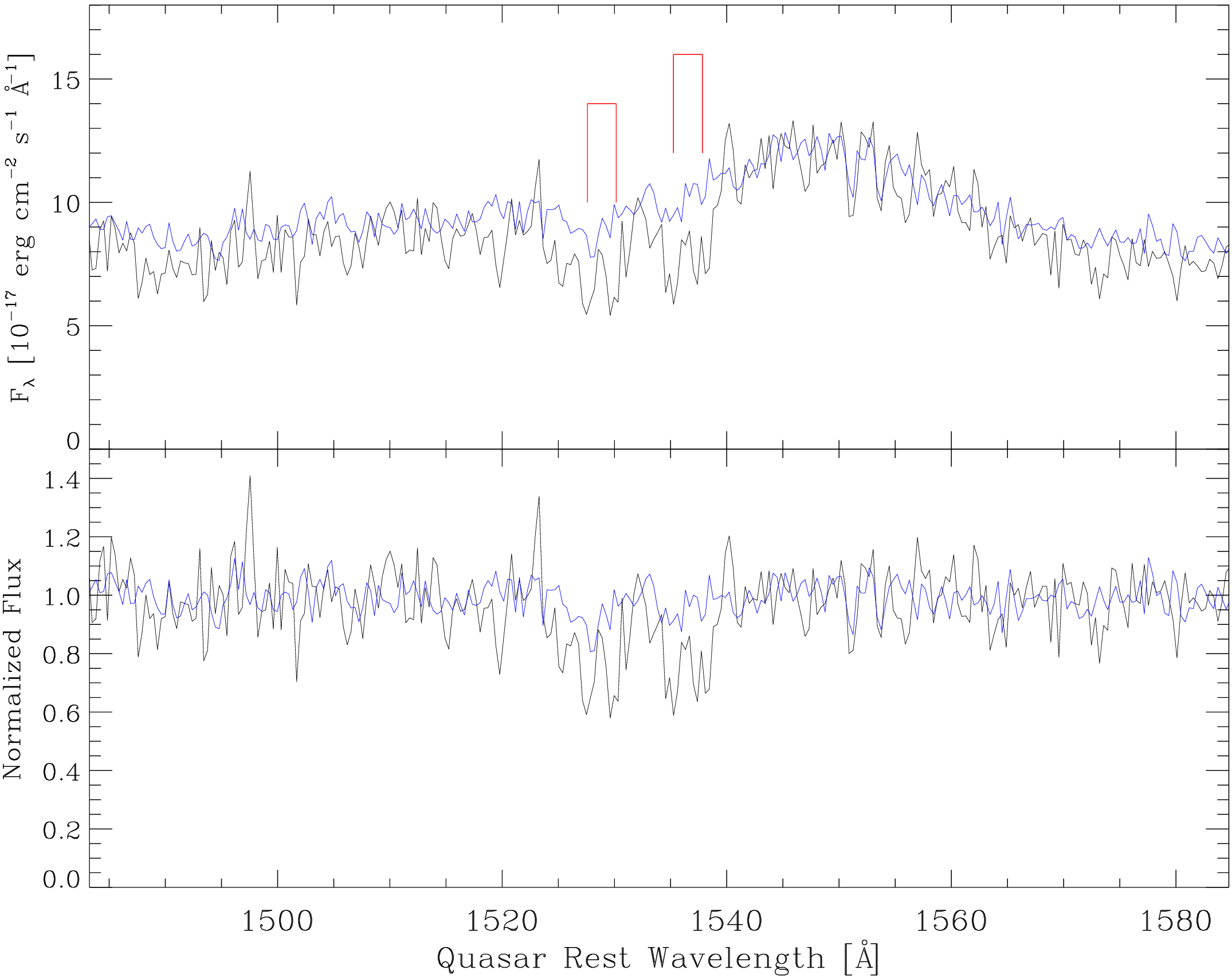}
\caption{The quasar J103115.69+374849.5 with $z_{\rm em}=2.2590$. See Figure A1 for the meanings of the color lines. The variable $\rm C~IV$ absorption systems are located at $z_{\rm abs}= 2.2092$ and 2.2253, respectively, and has a relative velocity value of $\beta=0.0154$ and 0.0104, respectively, with respect to the emission line redshift.}
\end{figure}	

\clearpage
\begin{figure}
\centering
\includegraphics[width=8.cm,height=3.cm]{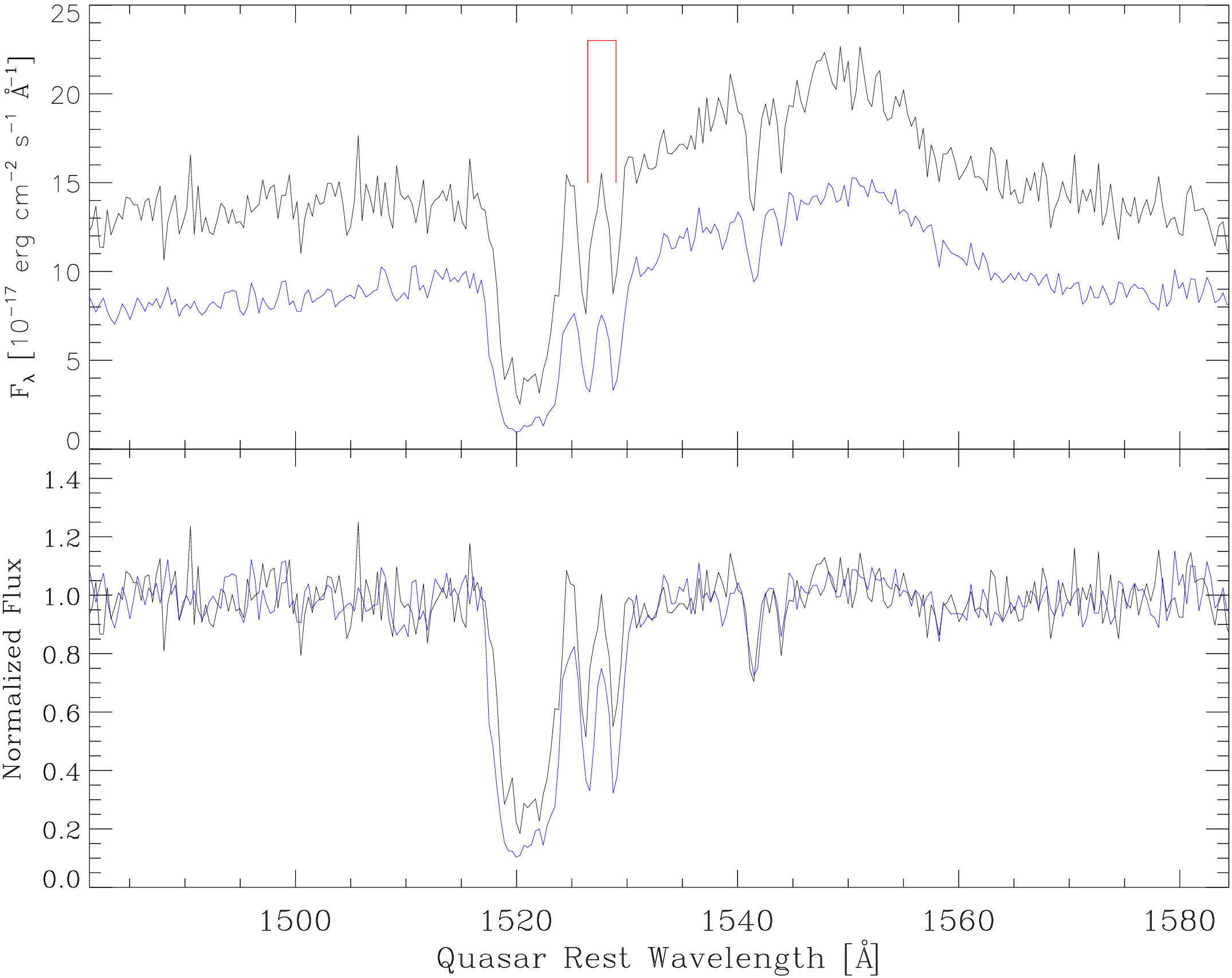}
\caption{The quasar J103842.14+350906.9 with $z_{\rm em}=2.2049$. See Figure A1 for the meanings of the color lines. The variable $\rm C~IV$ absorption system is located at $z_{\rm abs}=2.1563$, and has a relative velocity value of $\beta=0.0153$ with respect to the emission line redshift.}
\end{figure}	

\begin{figure}
\centering
\includegraphics[width=8.cm,height=3.cm]{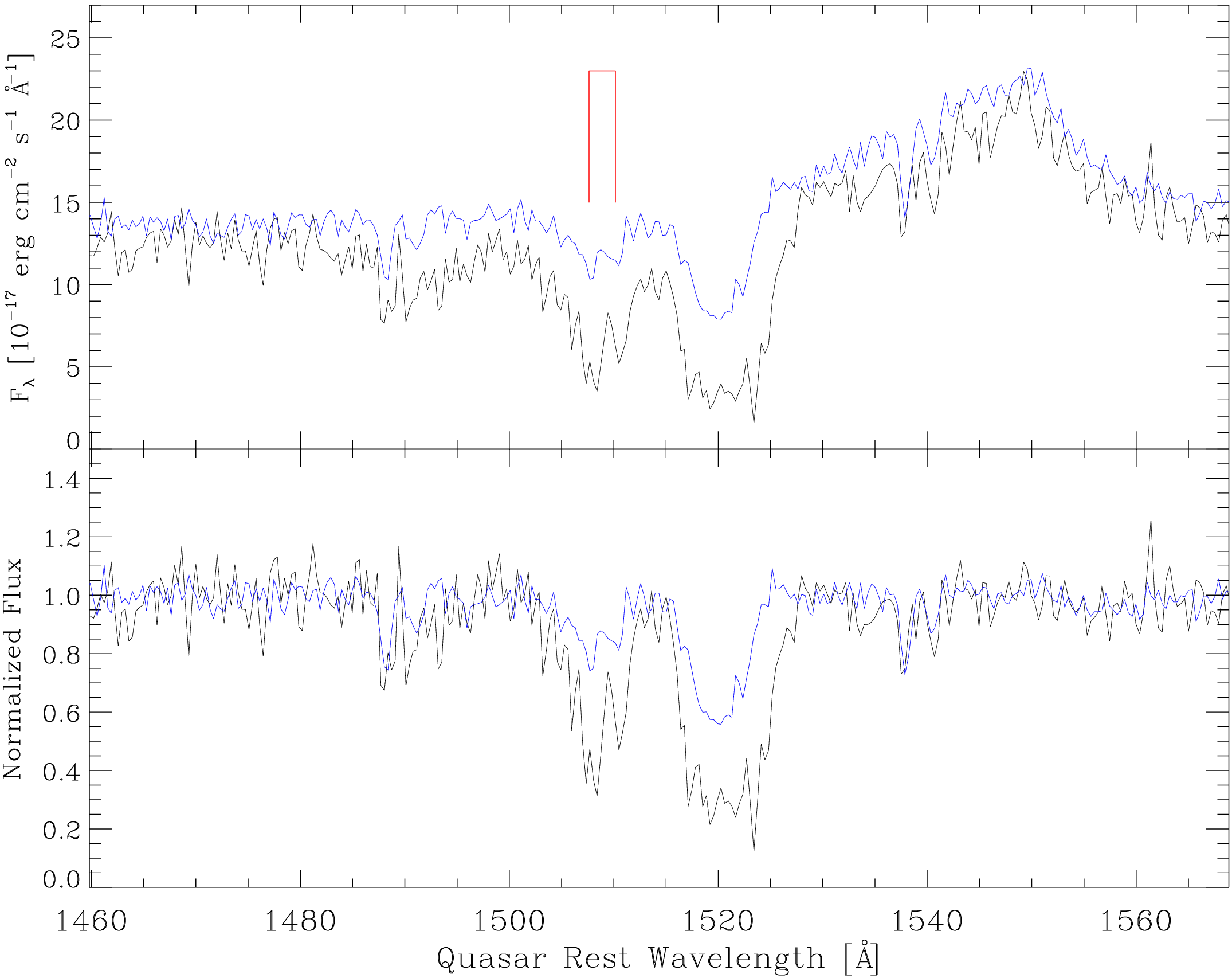}
\caption{The quasar J104841.02+000042.8 with $z_{\rm em}=2.0246$. See Figure A1 for the meanings of the color lines. The variable $\rm C~IV$ absorption system is located at $z_{\rm abs}=1.9468$, and has a relative velocity value of $\beta=0.0261$ with respect to the emission line redshift.}
\end{figure}

\begin{figure}
\centering
\includegraphics[width=8.cm,height=3.cm]{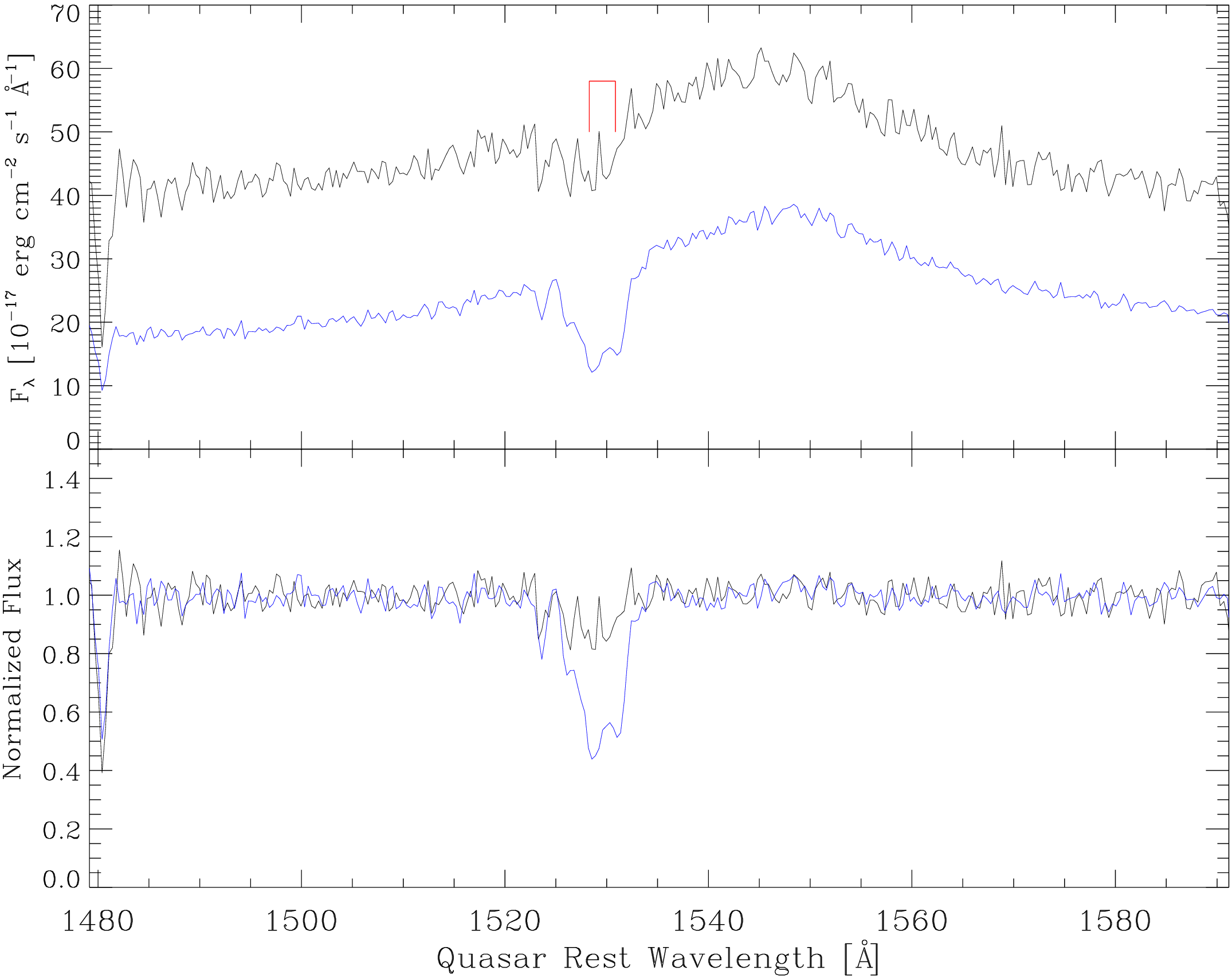}
\caption{The quasar J104923.94+012224.6 with $z_{\rm em}=1.9454$. See Figure A1 for the meanings of the color lines. The variable $\rm C~IV$ absorption system is located at $z_{\rm abs}=1.9087$, and has a relative velocity value of $\beta=0.0125$ with respect to the emission line redshift.  }
\end{figure}

\begin{figure}
\centering
\includegraphics[width=8.cm,height=3.cm]{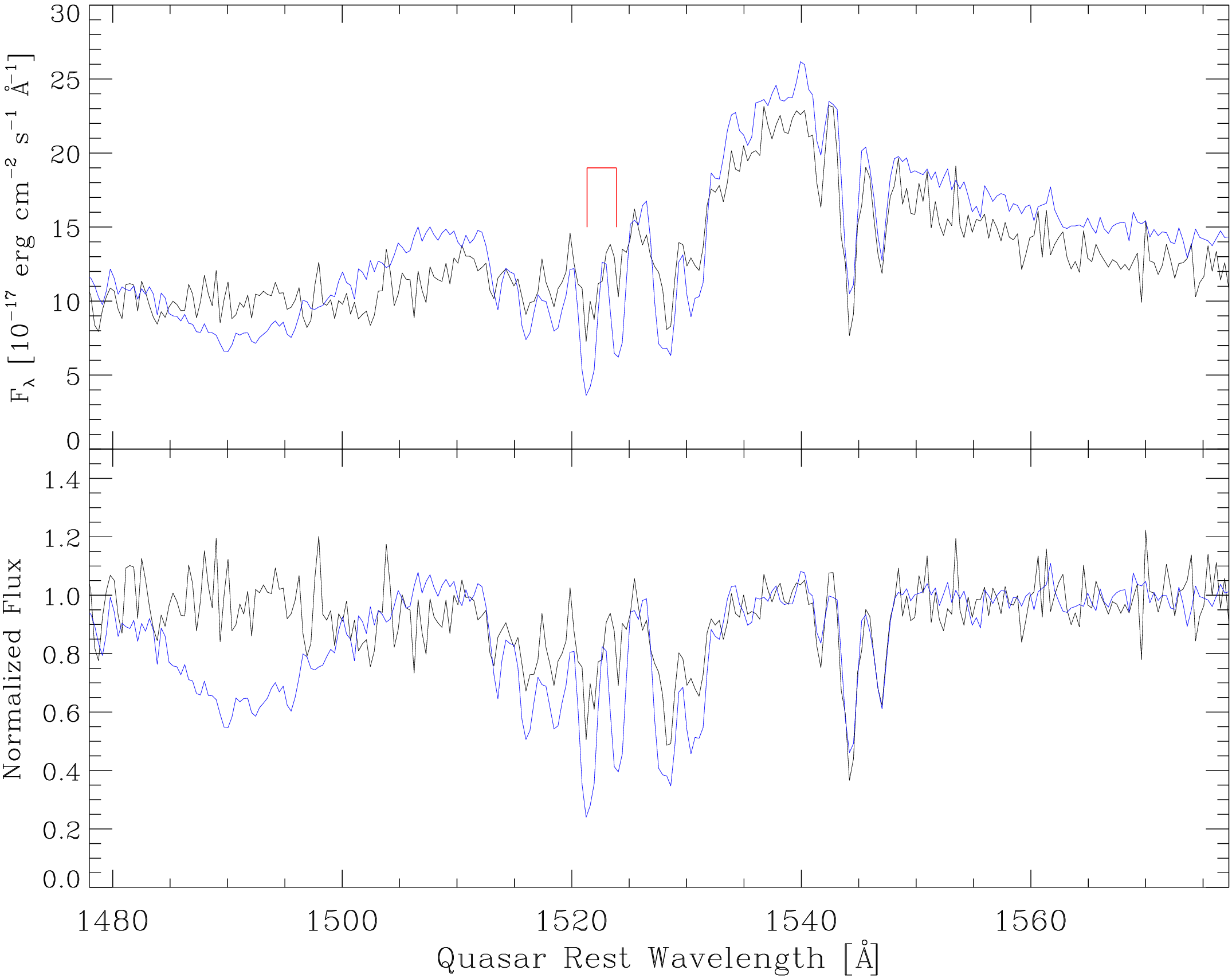}
\caption{The quasar J105207.90+362219.4 with $z_{\rm em}=2.3157$. See Figure A1 for the meanings of the color lines. The variable $\rm C~IV$ absorption system is located at $z_{\rm abs}= 2.2666$, and has a relative velocity value of $\beta=0.0149$ with respect to the emission line redshift.  }
\end{figure}	

\begin{figure}
\centering
\includegraphics[width=8.cm,height=3.cm]{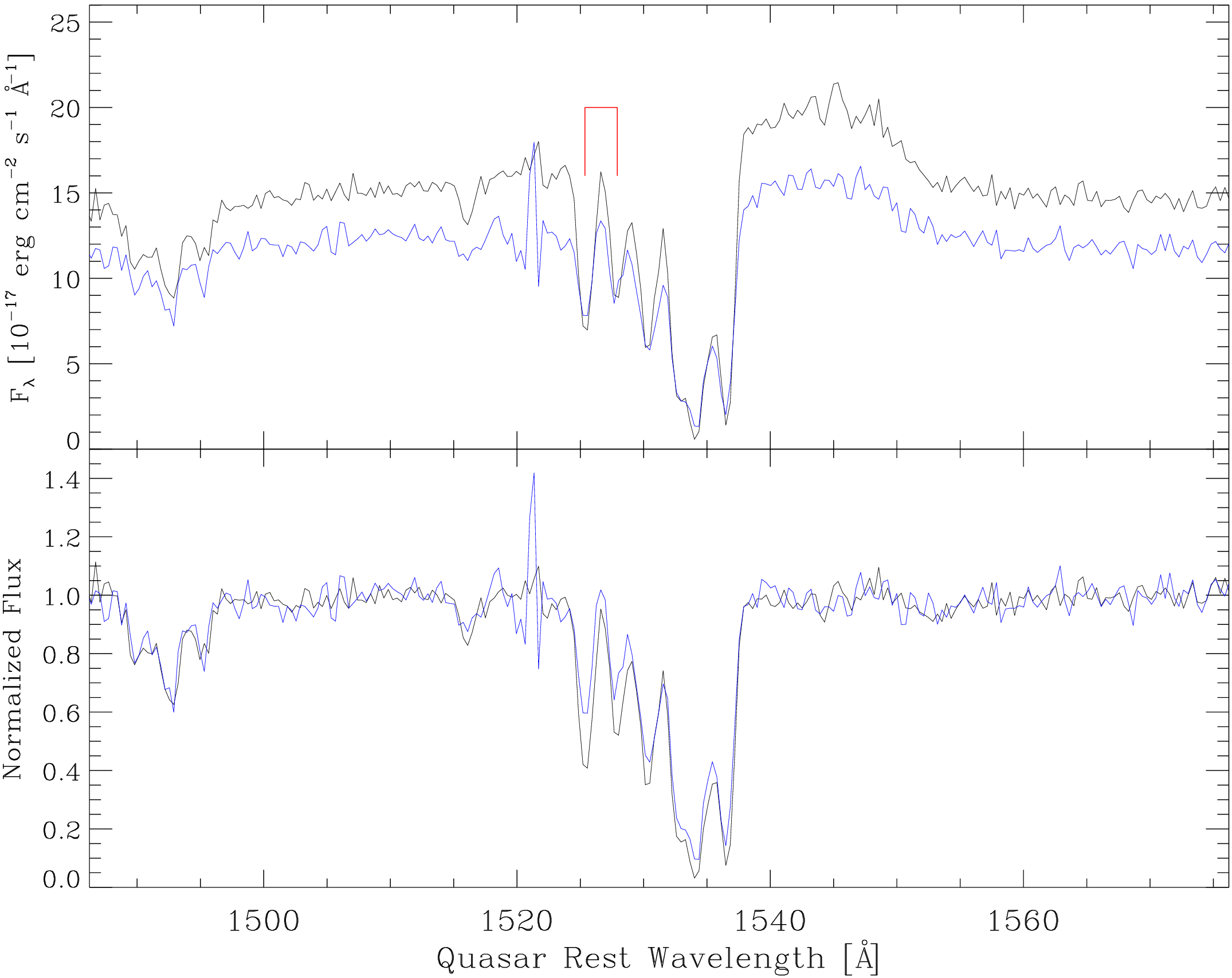}
\caption{The quasar J110726.04+385158.2 with $z_{\rm em}=2.6603$. See Figure A1 for the meanings of the color lines. The variable $\rm C~IV$ absorption system is located at $z_{\rm abs}=2.6134$, and has a relative velocity value of $\beta=0.0129$ with respect to the emission line redshift.  }
\end{figure}		

\begin{figure}
\centering
\includegraphics[width=8.cm,height=3.cm]{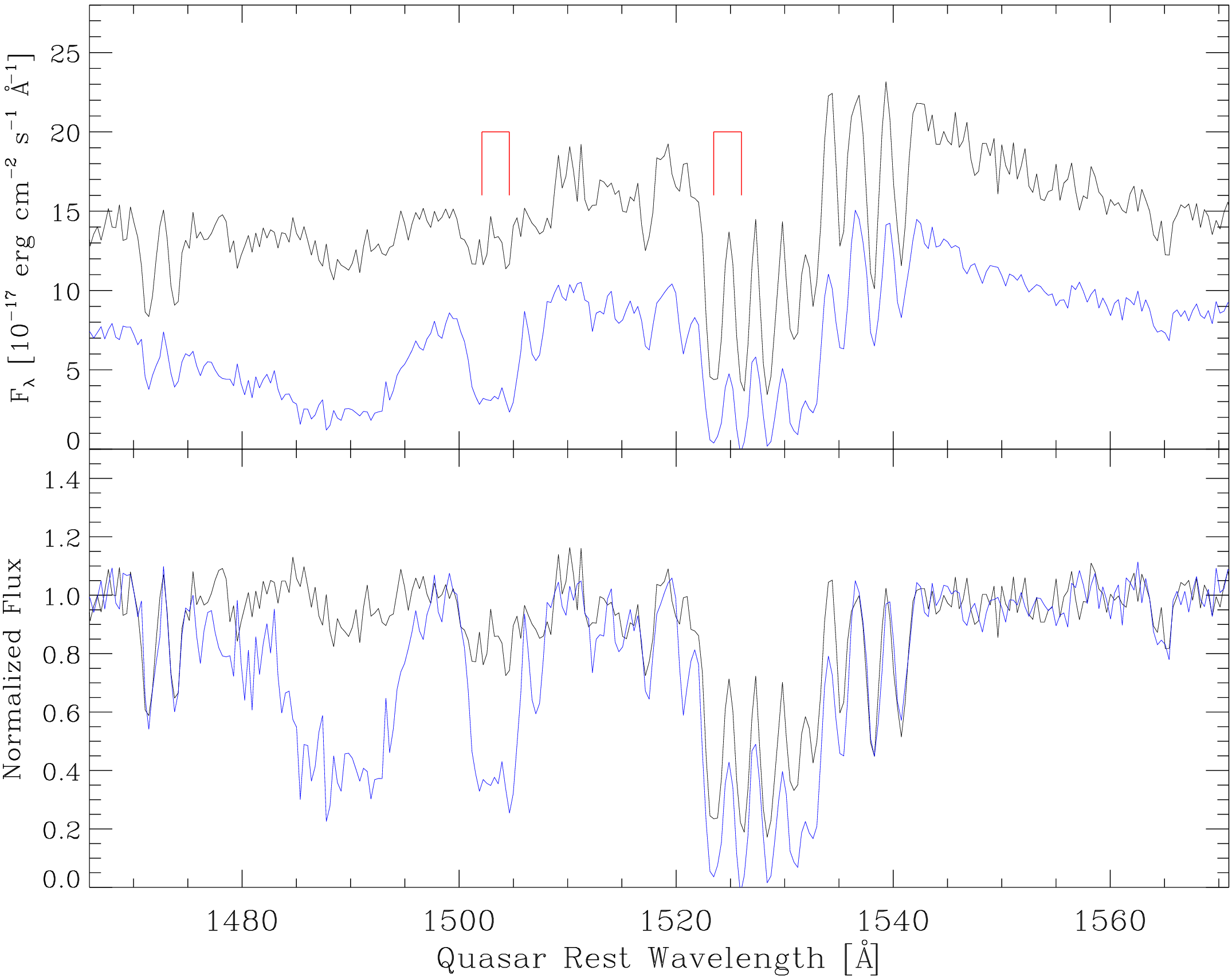}
\caption{The quasar J115122.14+020426.3 with $z_{\rm em}=2.4085$. See Figure A1 for the meanings of the color lines. The variable $\rm C~IV$ absorption systems are located at $z_{\rm abs}=2.3269$ and 2.3742, respectively, and have relative velocity values of $\beta=0.0242$ and 0.0101, respectively, with respect to the emission line redshift.  }
\end{figure}	

\begin{figure}
\centering
\includegraphics[width=8.cm,height=3.cm]{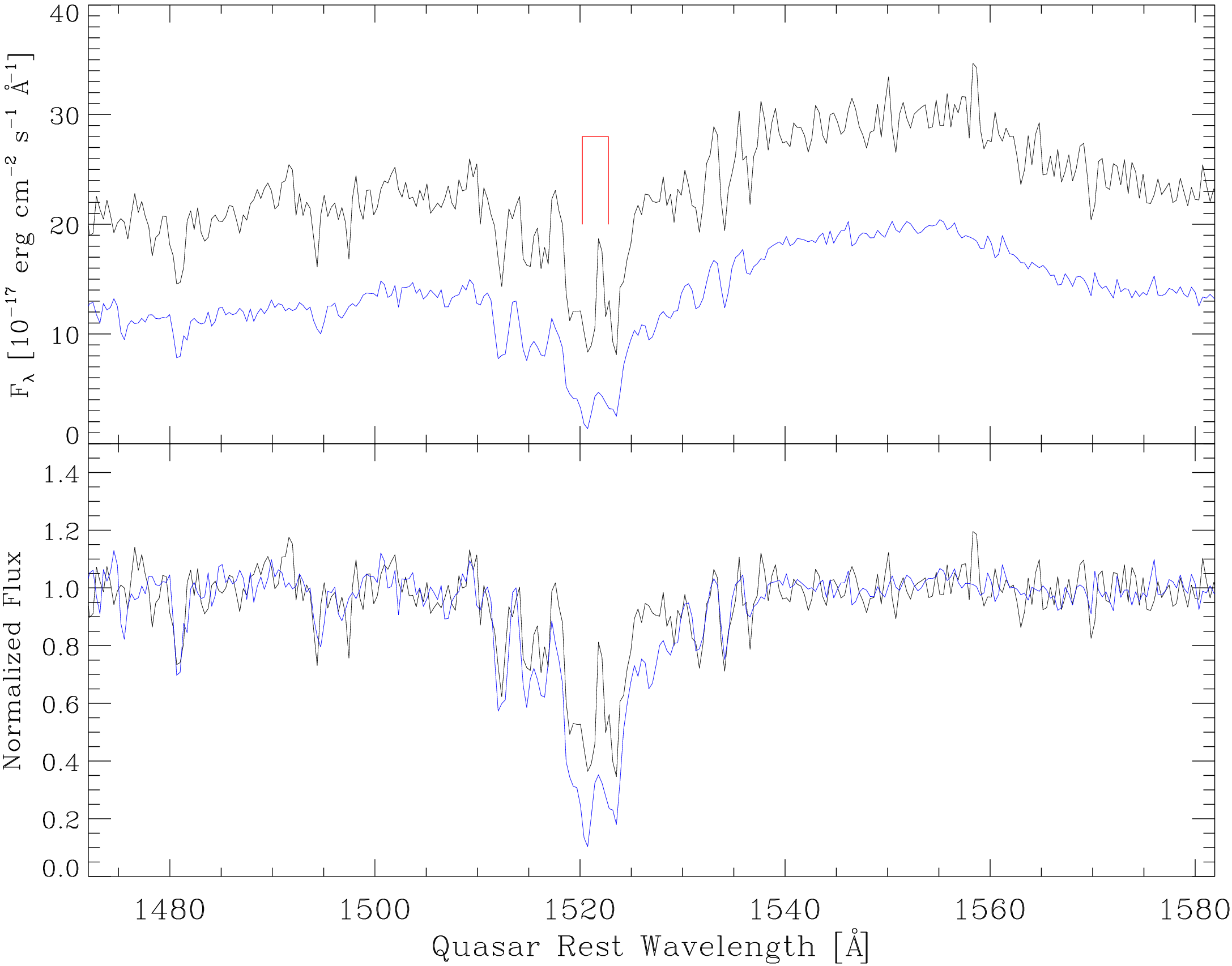}
\caption{The quasar J120819.29+035559.4 with $z_{\rm em}=2.0213$. See Figure A1 for the meanings of the color lines. The variable $\rm C~IV$ absorption system is located at $z_{\rm abs}= 1.9500$, and has a relative velocity value of $\beta=0.0239$ with respect to the emission line redshift.}
\end{figure}

\begin{figure}
\centering
\includegraphics[width=8.cm,height=3.cm]{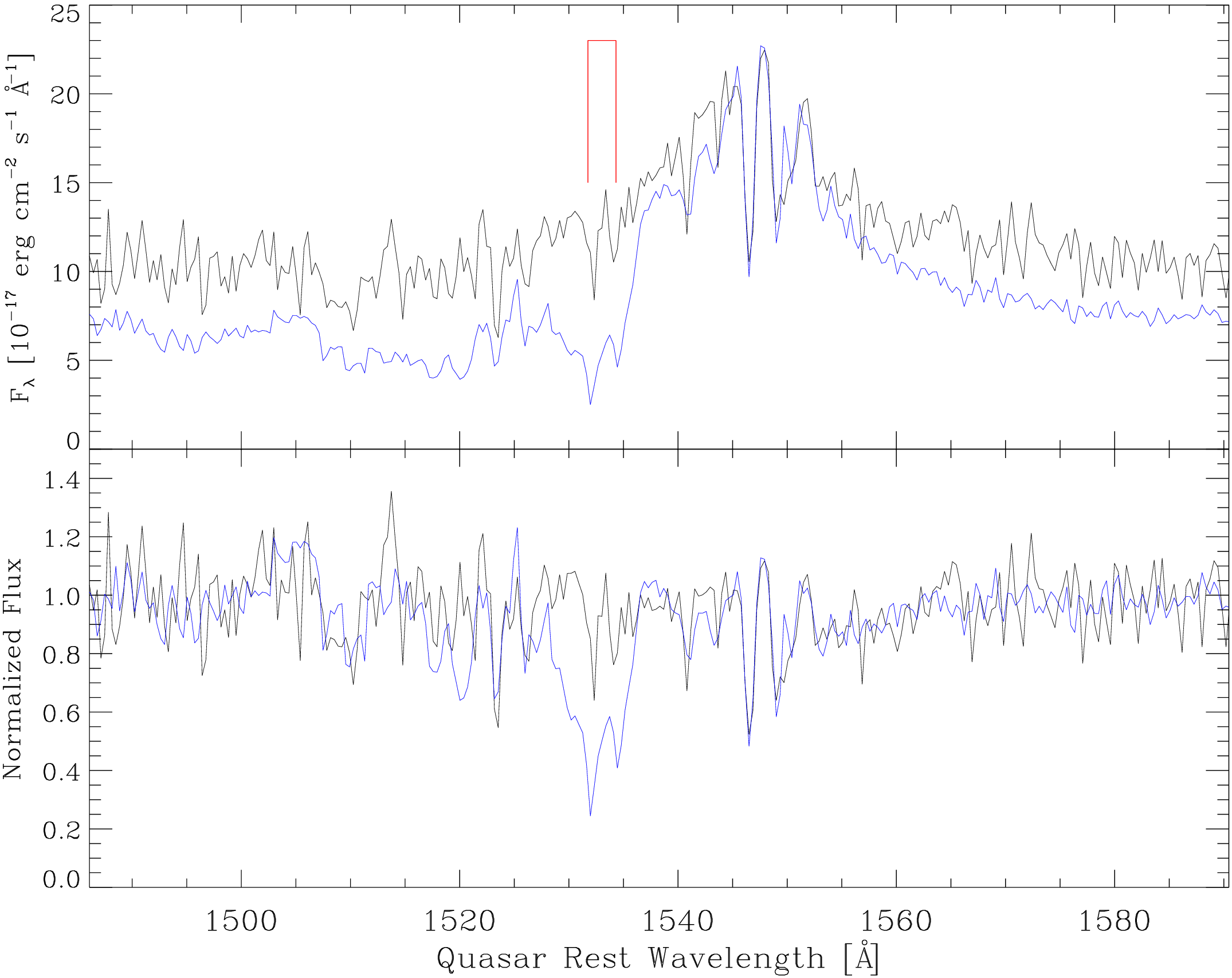}
\caption{The quasar J123720.85-011314.9 with $\rm z_{em}=2.1620$. See Figure A1 for the meanings of the color lines. The variable $\rm C~IV$ absorption system is located at $\rm z_{abs}= 2.1283$, and has a relative velocity value of $\rm \beta=0.0107$ with respect to the emission line redshift.}
\end{figure}

\clearpage
\begin{figure}
\centering
\includegraphics[width=8.cm,height=3.cm]{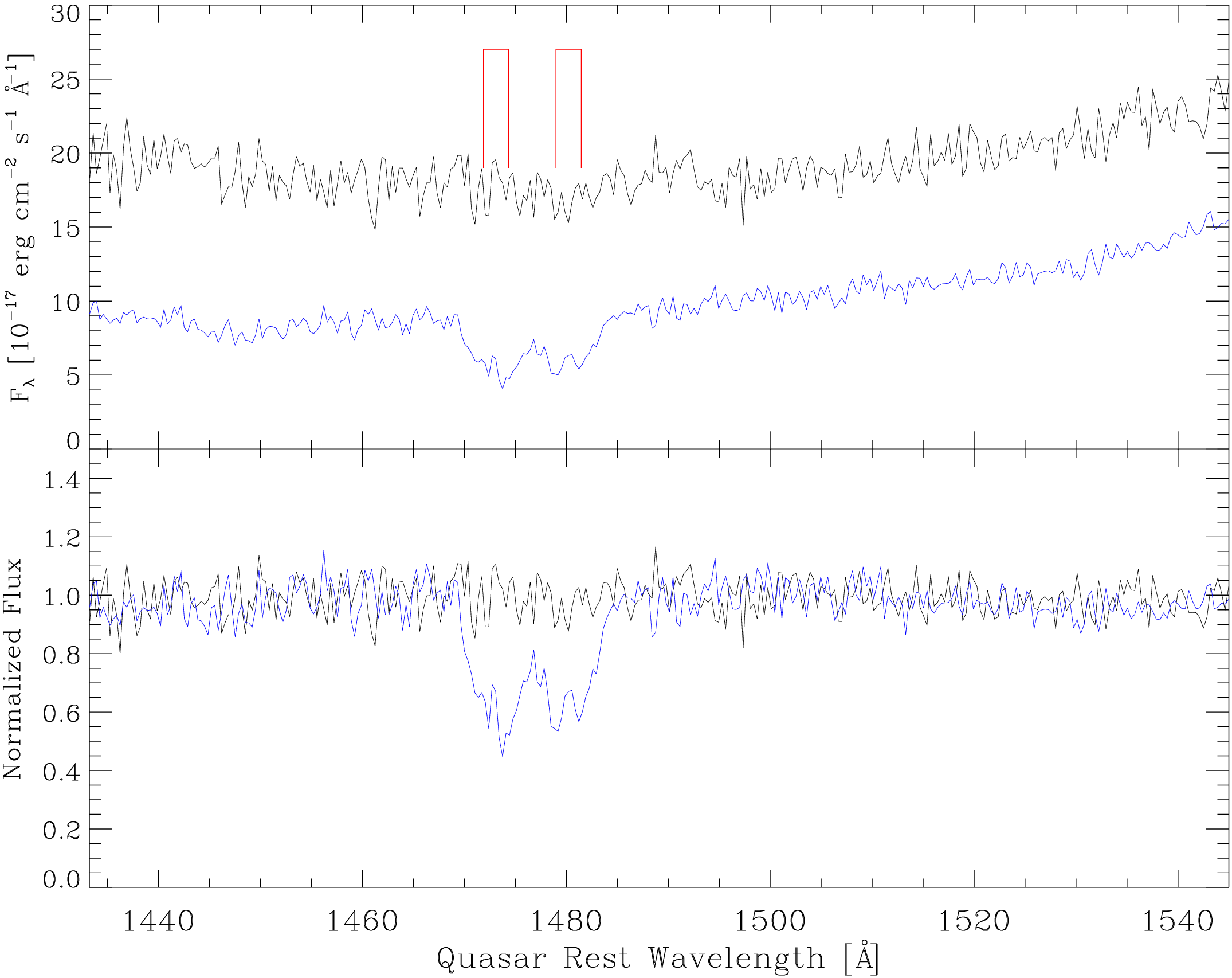}
\caption{The quasar J124829.46+341231.3 with $z_{\rm em}=2.2285$. See Figure A1 for the meanings of the color lines. The variable $\rm C~IV$ absorption systems are located at $z_{\rm abs}= 2.0621$ and 2.0769, respectively, and have relative velocity values of $\beta=0.0529$ and 0.0481, respectively, with respect to the emission line redshift.}
\end{figure}	

\begin{figure}
\centering
\includegraphics[width=8.cm,height=3.cm]{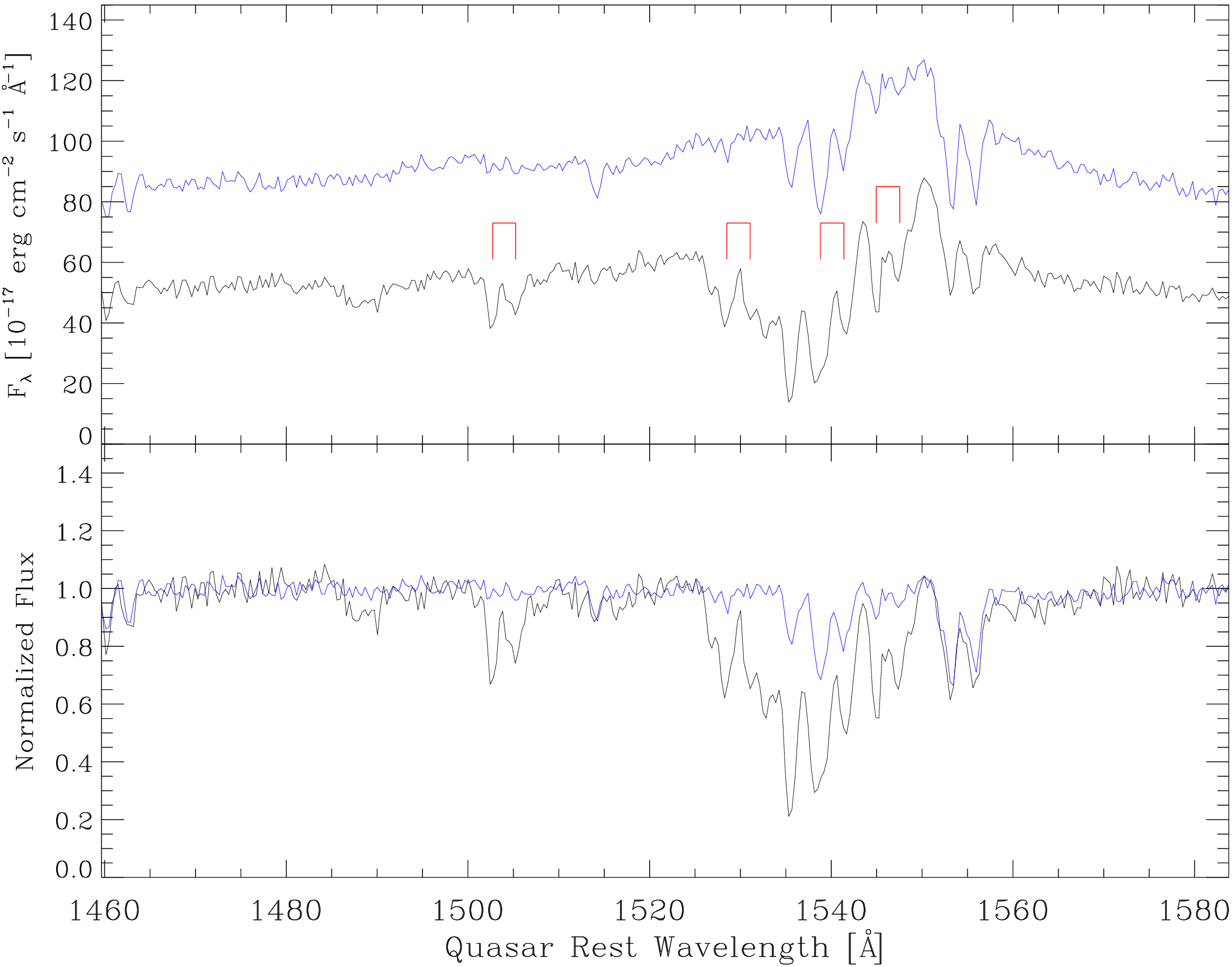}
\caption{The quasar J125216.58+052737.7 with $z_{\rm em}=1.9034$. See Figure A1 for the meanings of the color lines. The variable $\rm C~IV$ absorption systems are located at $z_{\rm abs}=1.8155$, 1.8638, 1.8831, and 1.8946, respectively, and have relative velocity values of $\beta=0.0307$, 0.0137, 0.0070, and 0.0030, respectively, with respect to the emission line redshift.}
\end{figure}	
	
\begin{figure}
\centering
\includegraphics[width=8.cm,height=3.cm]{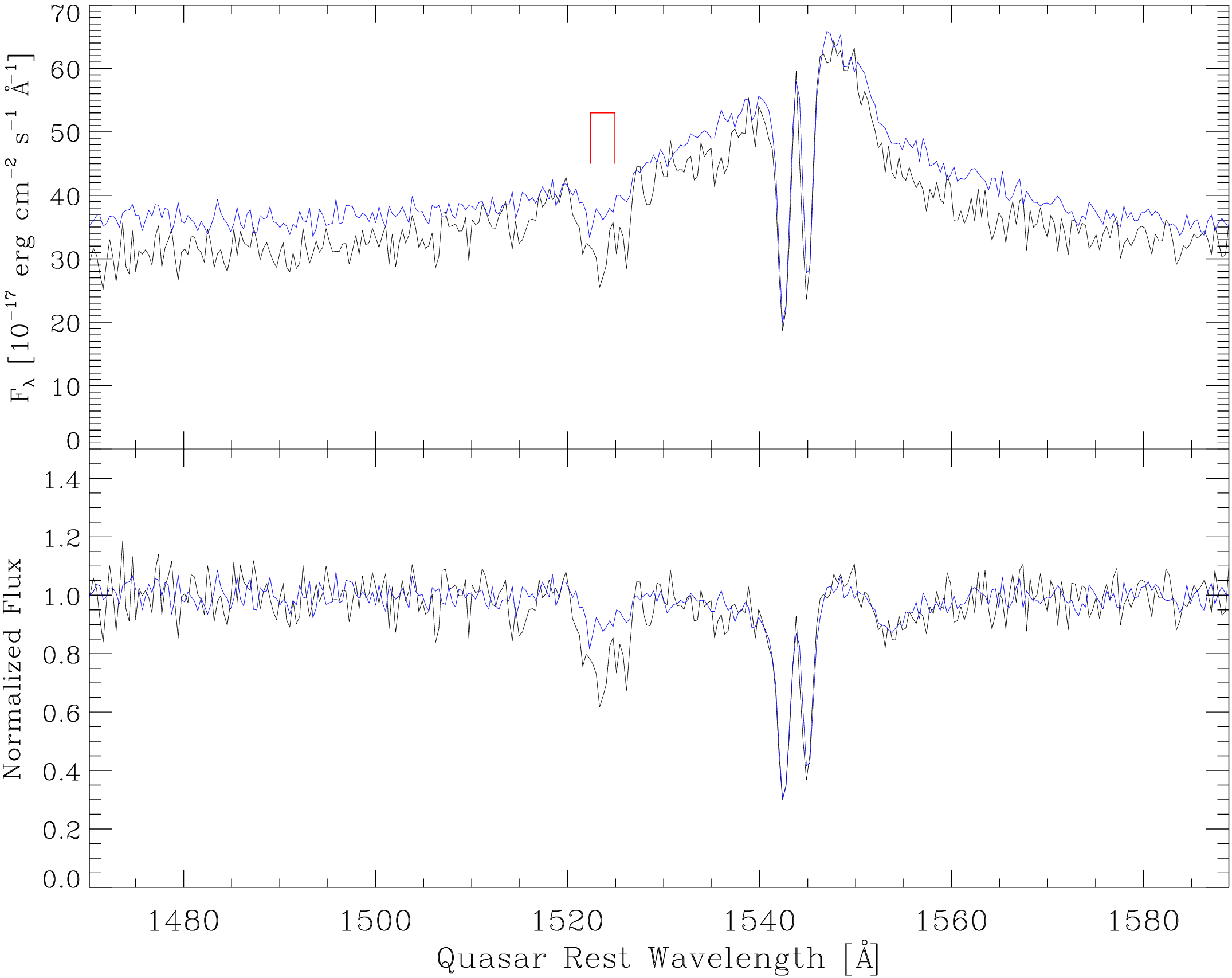}
\caption{The quasar J132333.03+004750.2 with $z_{\rm em}=1.7785$. See Figure A1 for the meanings of the color lines. The variable $\rm C~IV$ absorption system is located at $z_{\rm abs}= 1.7340$, and has a relative velocity value of $\beta=0.0161$ with respect to the emission line redshift.}
\end{figure}		

\begin{figure}
\centering
\includegraphics[width=8.cm,height=3.cm]{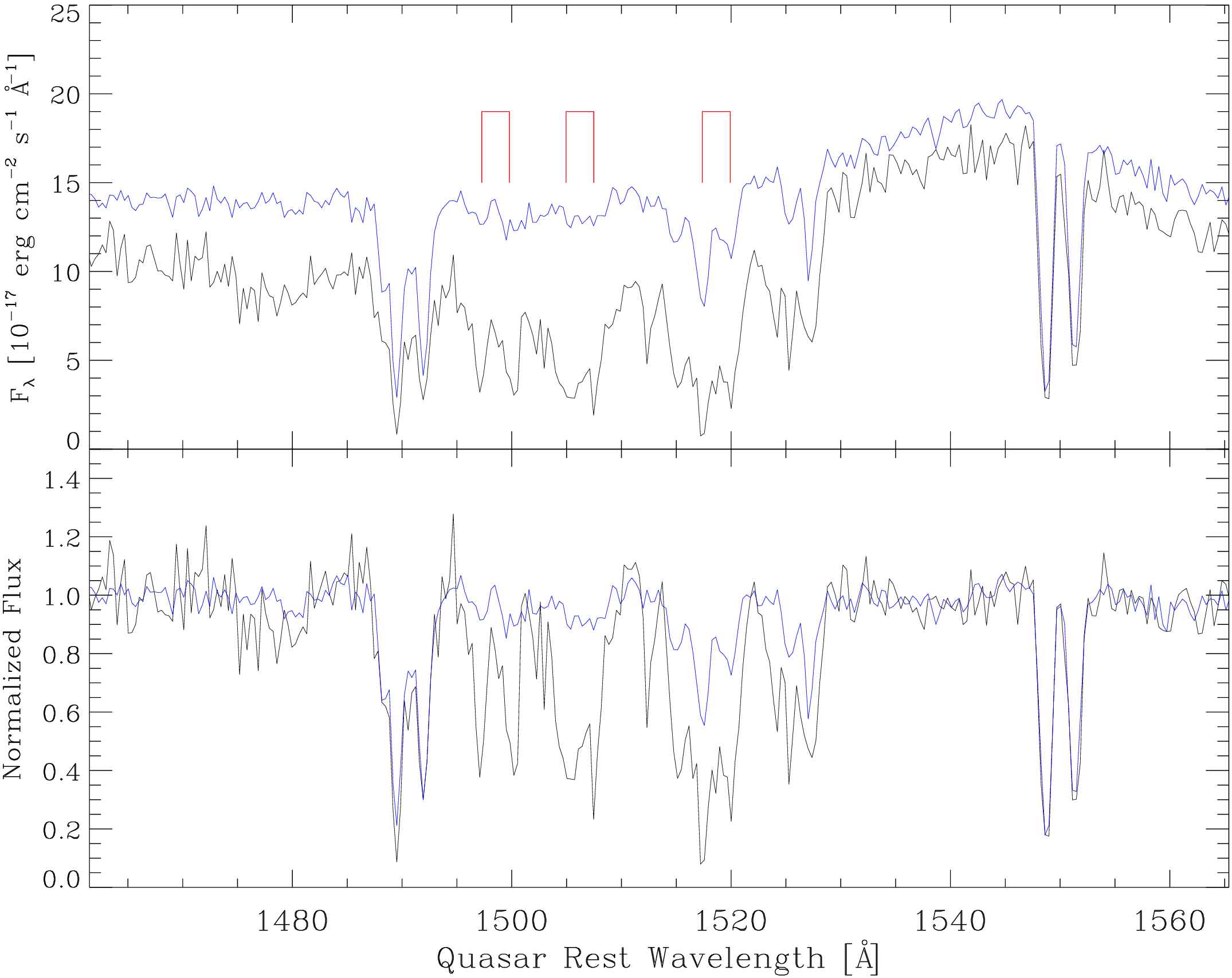}
\caption{The quasar J134544.55+002810.7 with $z_{\rm em}=2.4641$. See Figure A1 for the meanings of the color lines. The variable $\rm C~IV$ absorption systems are located at $z_{\rm abs}=2.3514$, 2.3686, and 2.3964, respectively, and have relative velocity values of $\beta=0.0331$, 0.0279, and 0.0197, respectively, with respect to the emission line redshift.}
\end{figure}	

\begin{figure}
\centering
\includegraphics[width=8.cm,height=3.cm]{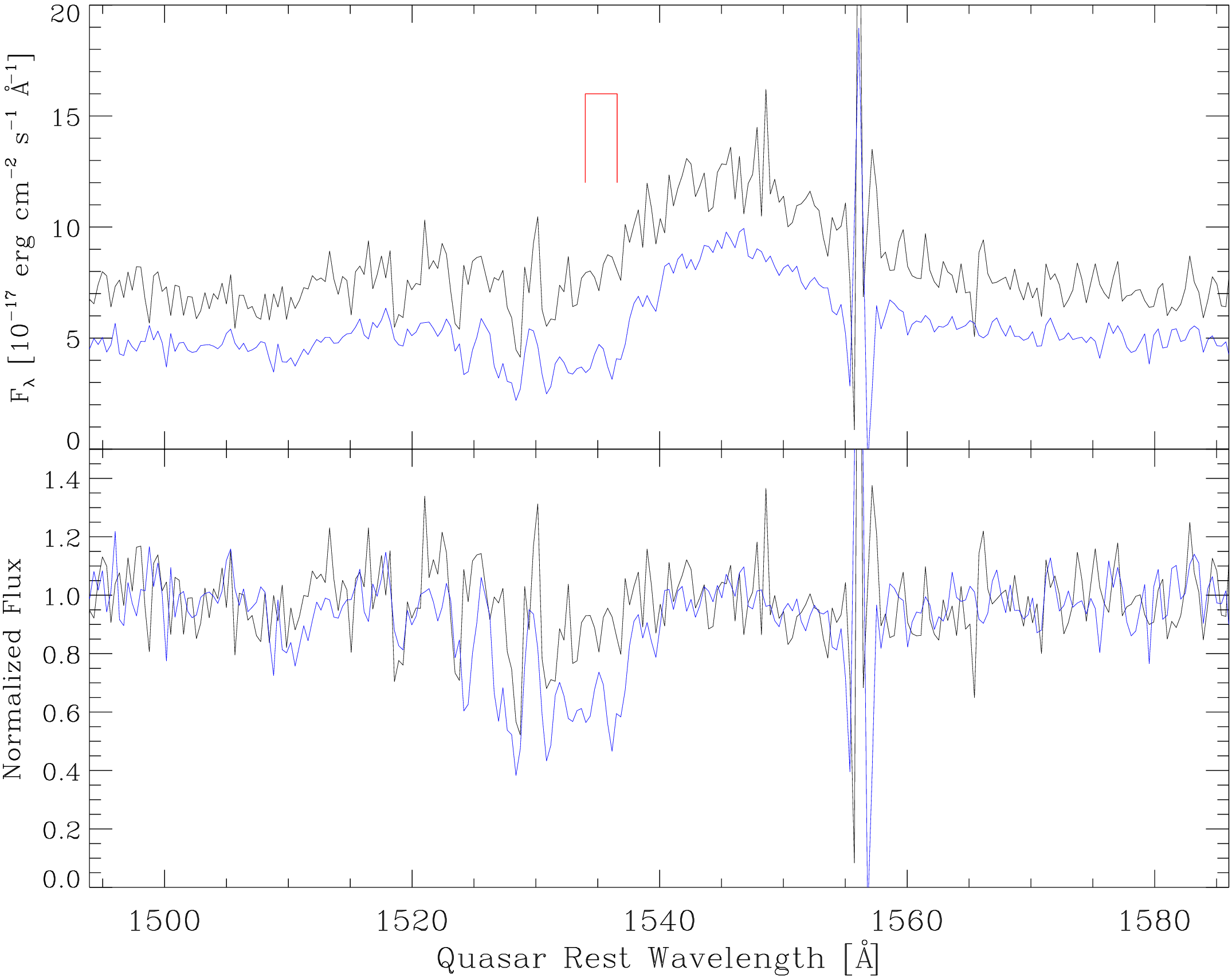}
\caption{The quasar J140815.58+060023.3 with $z_{\rm em}=2.5830$. See Figure A1 for the meanings of the color lines. The variable $\rm C~IV$ absorption system is located at $z_{\rm abs}=2.5519$, and has a relative velocity value of $\beta=0.0087$ with respect to the emission line redshift.}
\end{figure}	

\begin{figure}
\centering
\includegraphics[width=8.cm,height=3.cm]{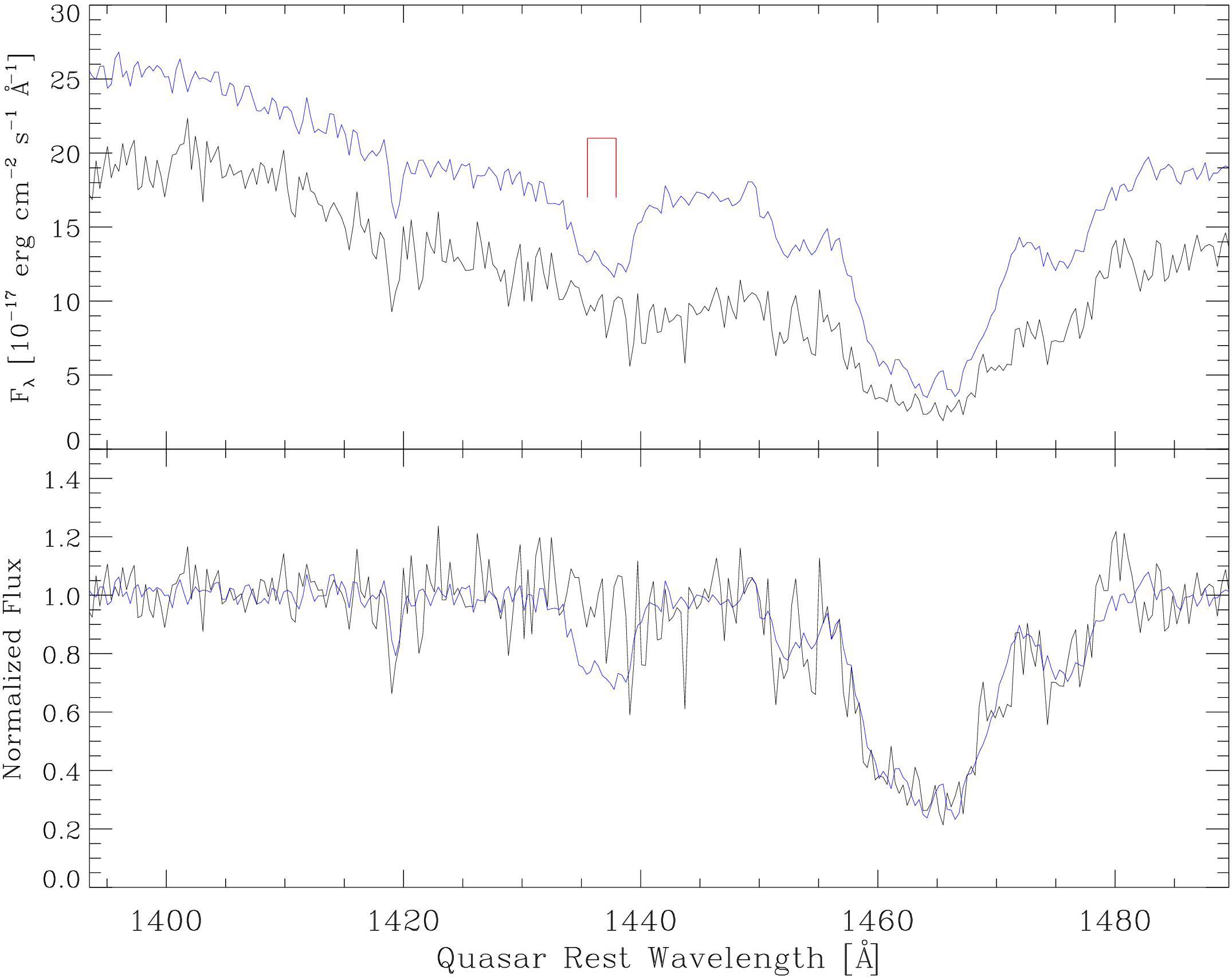}
\caption{The quasar J150033.53+003353.6 with $z_{\rm em}=2.4360$. See Figure A1 for the meanings of the color lines. The variable $\rm C~IV$ absorption system is located at $z_{\rm abs}=2.1849$, and has a relative velocity value of $\beta=0.0757$ with respect to the emission line redshift.}
\end{figure}	

\begin{figure}
\centering
\includegraphics[width=8.cm,height=3.cm]{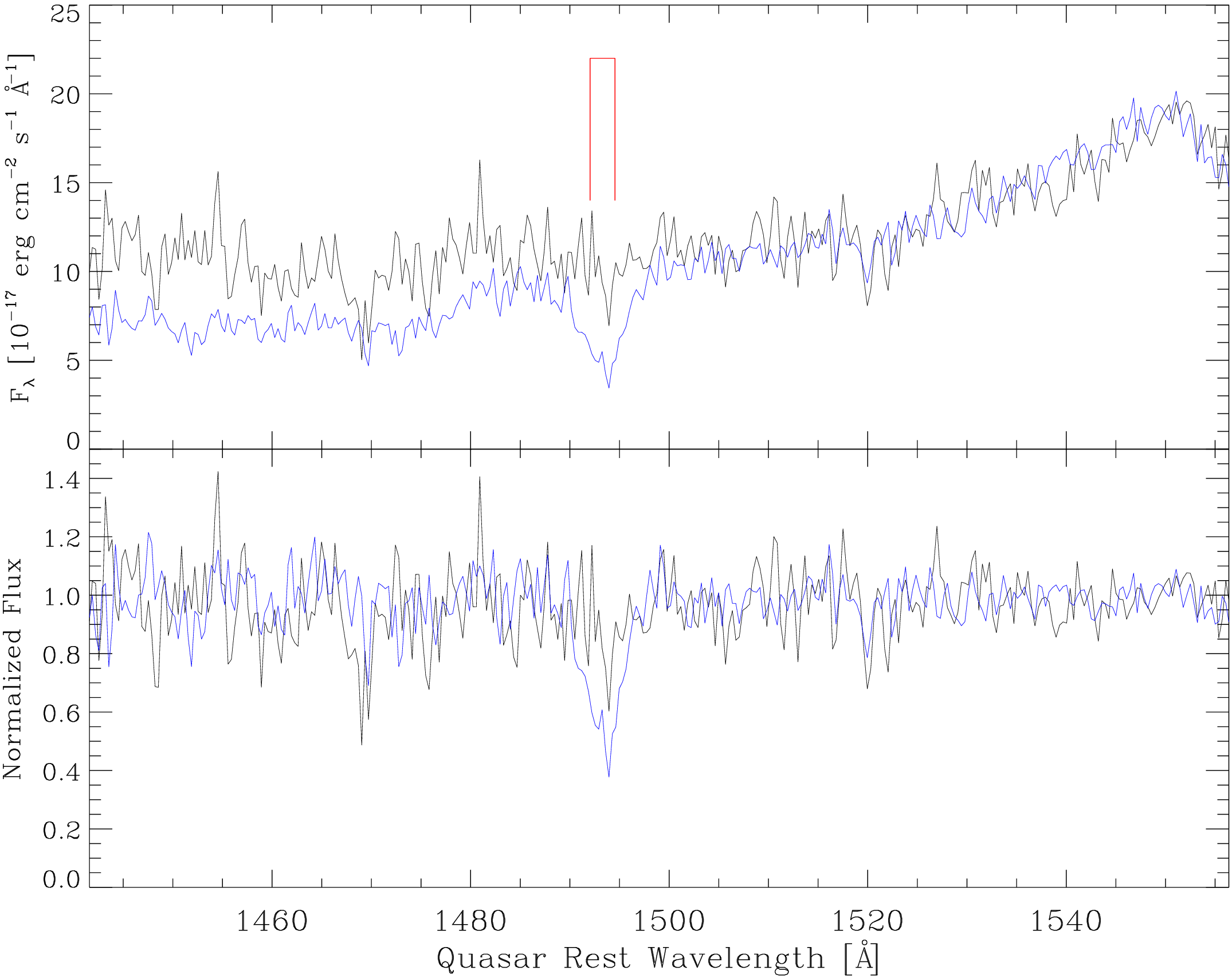}
\caption{The quasar J160445.92+335759.0 with $z_{\rm em}=1.8776$. See Figure A1 for the meanings of the color lines. The variable $\rm C~IV$ absorption system is located at $z_{\rm abs}=1.7709$, and has a relative velocity value of $\beta=0.0378$ with respect to the emission line redshift.}
\end{figure}		

\begin{figure}
\centering
\includegraphics[width=8.cm,height=3.cm]{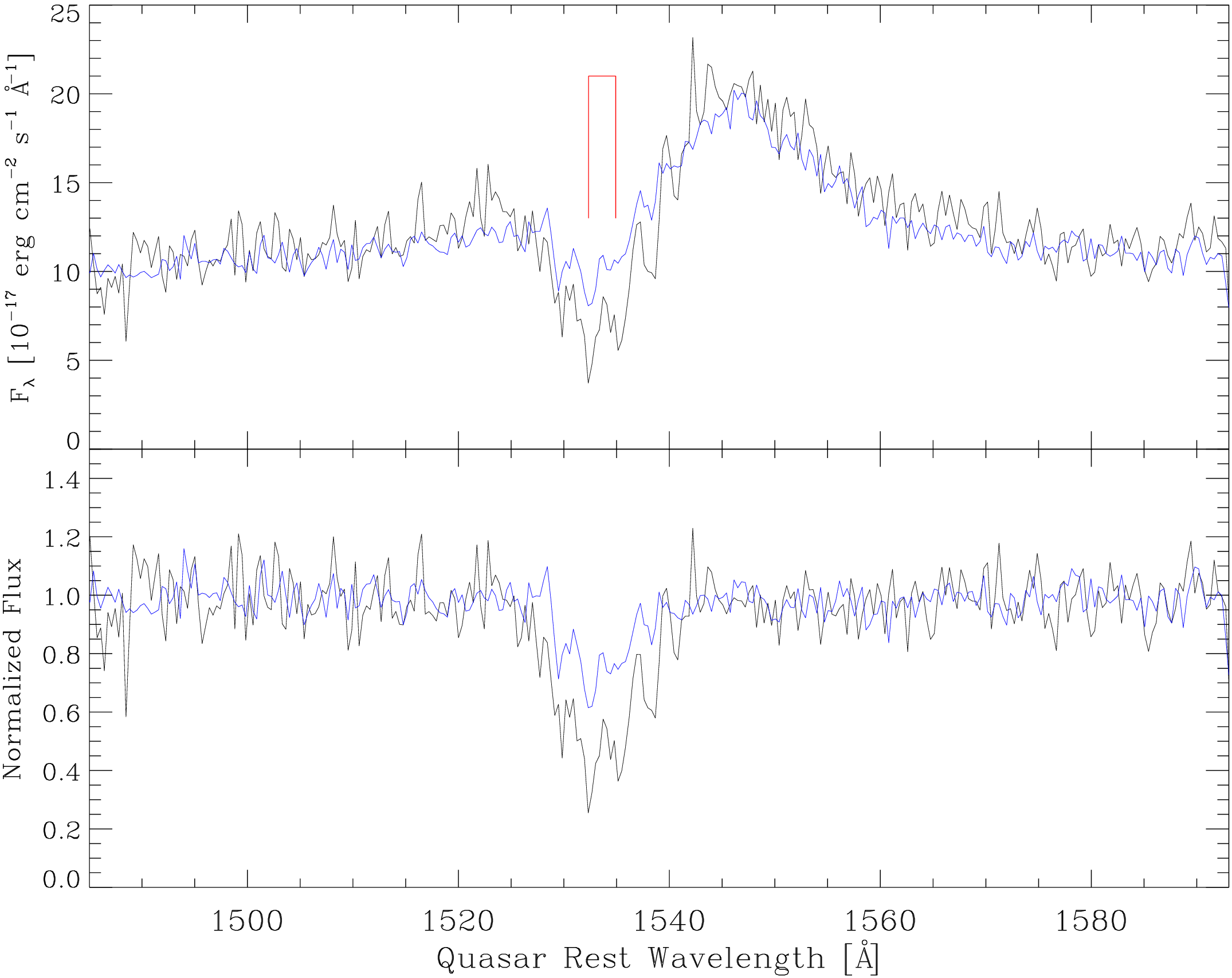}
\caption{The quasar J160613.99+314143.4 with $z_{\rm em}=2.0569$. See Figure A1 for the meanings of the color lines. The variable $\rm C~IV$ absorption system is located at $z_{\rm abs}=2.0240$, and has a relative velocity value of $\beta=0.0108$ with respect to the emission line redshift.}
\end{figure}	
	
\clearpage
\begin{figure}
\centering
\includegraphics[width=8.cm,height=3.cm]{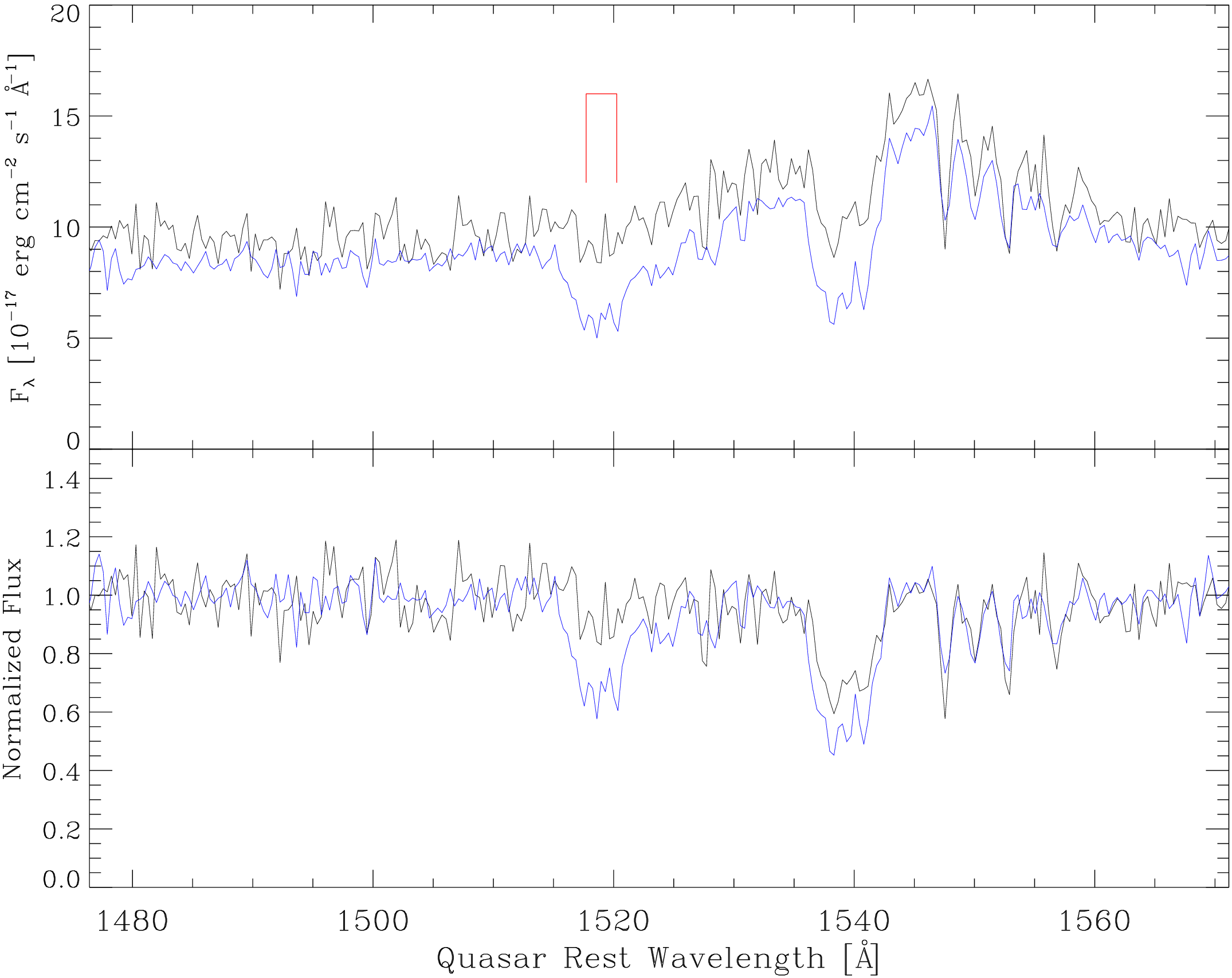}
\caption{The quasar J161336.81+054701.7 with $z_{\rm em}=2.4855$. See Figure A1 for the meanings of the color lines. The variable $\rm C~IV$ absorption system is located at $z_{\rm abs}= 2.4152$, and has a relative velocity value of $\beta=0.0204$ with respect to the emission line redshift.  }
\end{figure}	

\begin{figure}
\centering
\includegraphics[width=8.cm,height=3.cm]{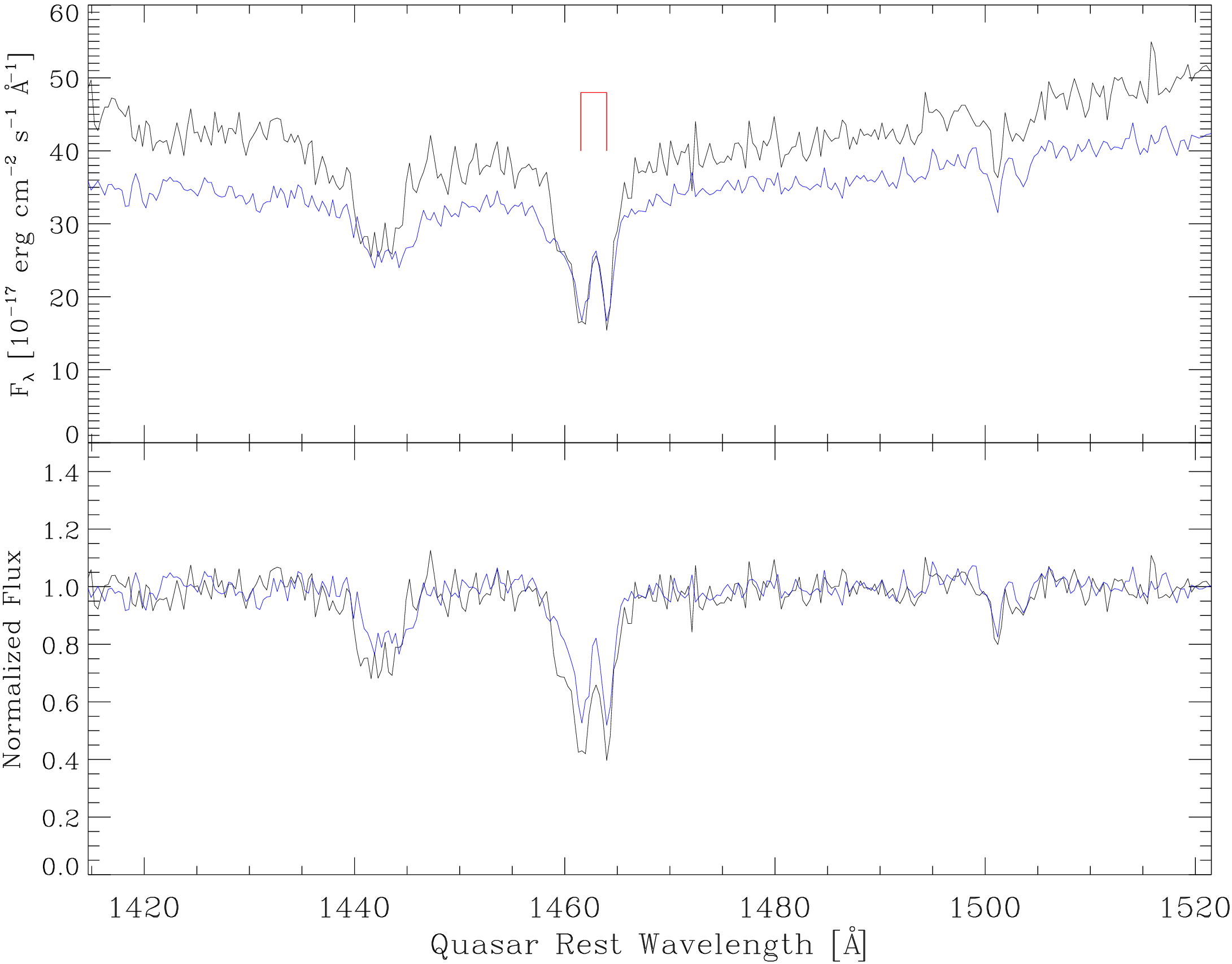}
\caption{The quasar J161511.35+314728.3 with $z_{\rm em}=2.0981$. See Figure A1 for the meanings of the color lines. The variable $\rm C~IV$ absorption system is located at $z_{\rm abs}=1.9157$, and has a relative velocity value of $\beta=0.0606$ with respect to the emission line redshift.}
\end{figure}		

\begin{figure}
\centering
\includegraphics[width=8.cm,height=3.cm]{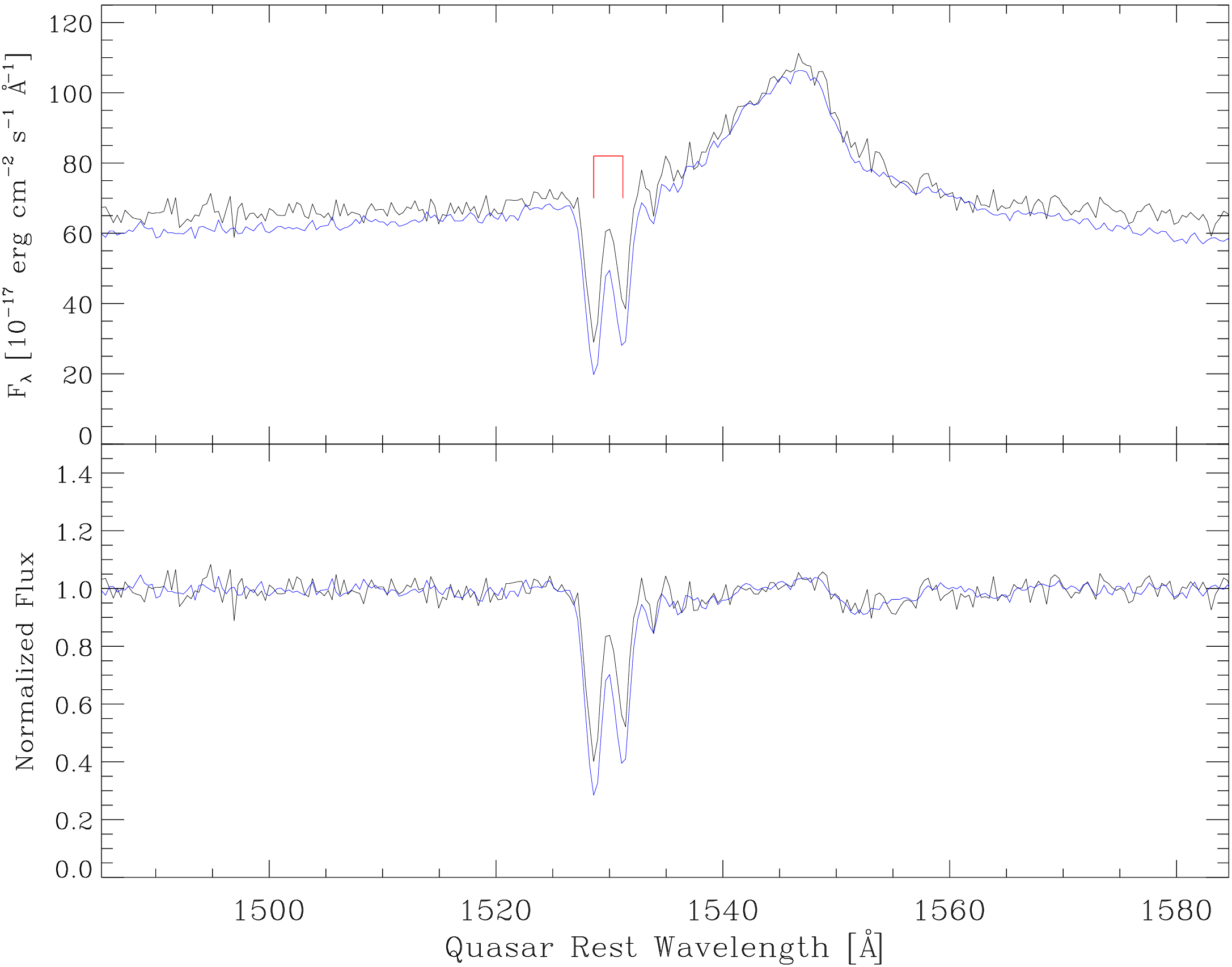}
\caption{The quasar J162701.94+313549.2 with $z_{\rm em}=2.3263$. See Figure A1 for the meanings of the color lines. The variable $\rm C~IV$ absorption system is located at $z_{\rm abs}=2.2785$, and has a relative velocity value of $\beta=0.0145$ with respect to the emission line redshift.}
\end{figure}	

\begin{figure}
\centering
\includegraphics[width=8.cm,height=3.cm]{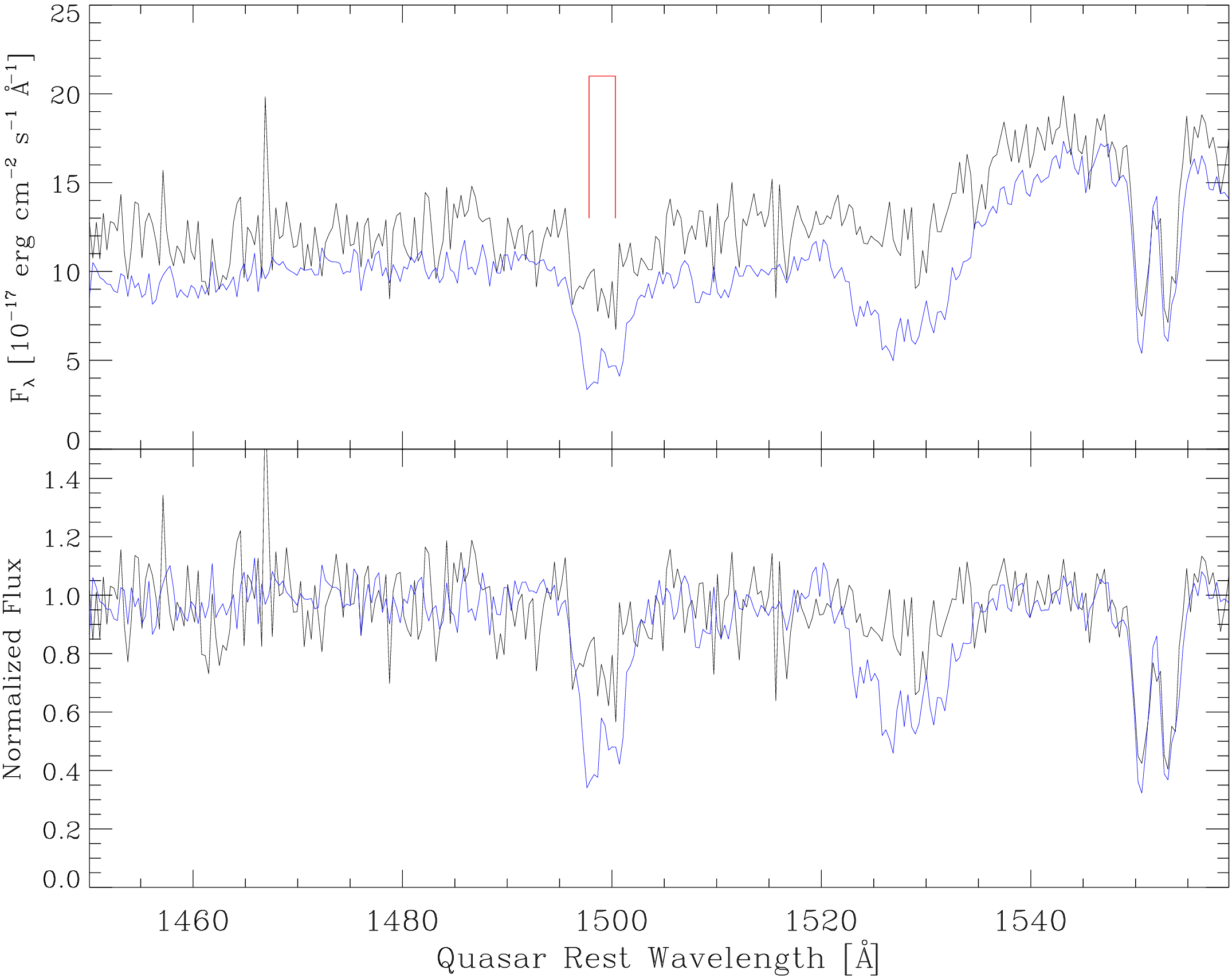}
\caption{The quasar J162935.68+321009.5 with $z_{\rm em}=2.0364$. See Figure A1 for the meanings of the color lines. The variable $\rm C~IV$ absorption system is located at $z_{\rm abs}= 1.9345$, and has a relative velocity value of $\beta=0.0341$ with respect to the emission line redshift.}
\end{figure}	

\begin{figure}
\centering
\includegraphics[width=8.cm,height=3.cm]{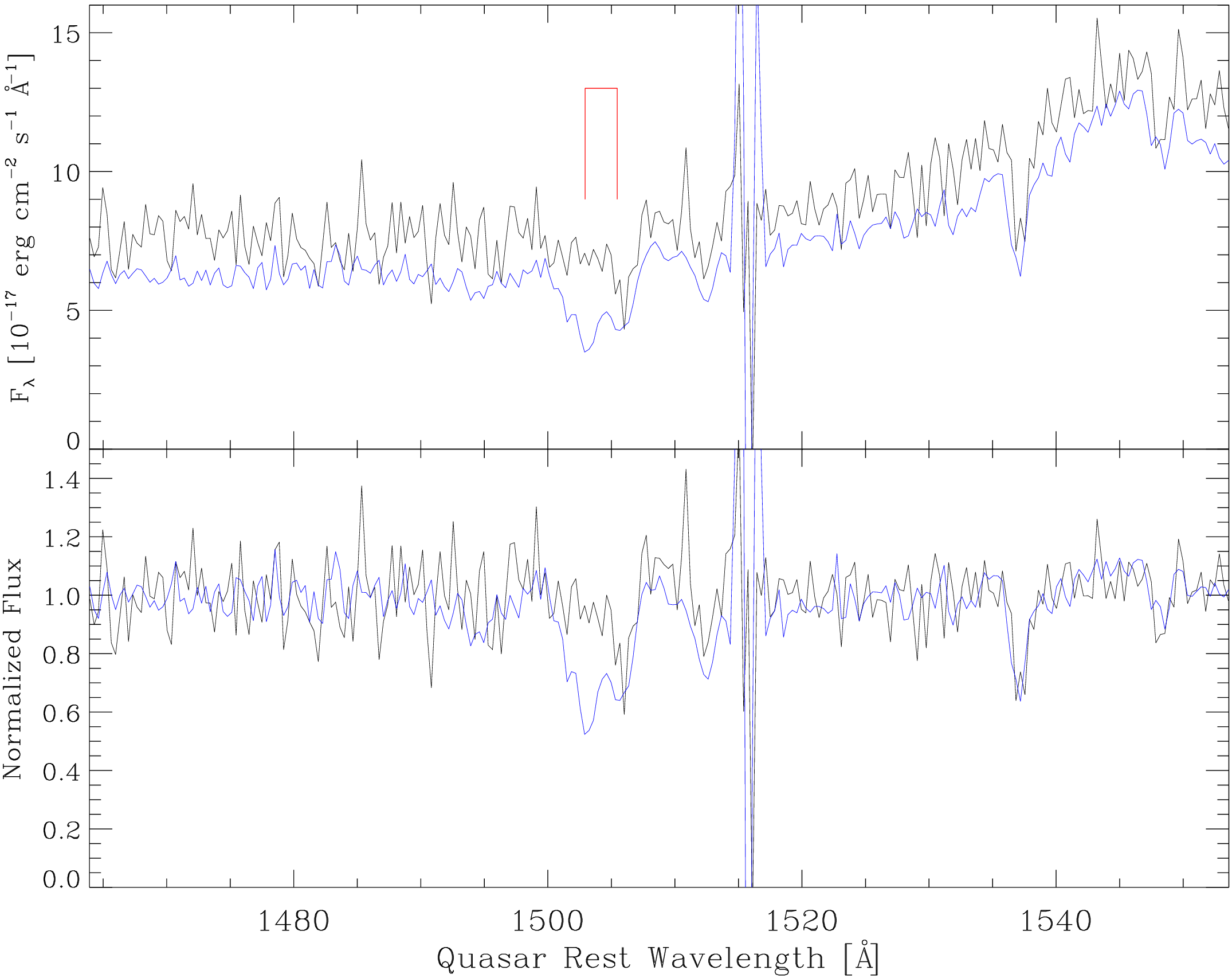}
\caption{The quasar J212943.25+003005.6 with $z_{\rm em}=2.6820$. See Figure A1 for the meanings of the color lines. The variable $\rm C~IV$ absorption system is located at $z_{\rm abs}= 2.5726$, and has a relative velocity value of $\beta=0.0297$ with respect to the emission line redshift.}
\end{figure}		

\begin{figure}
\centering
\includegraphics[width=8.cm,height=3.cm]{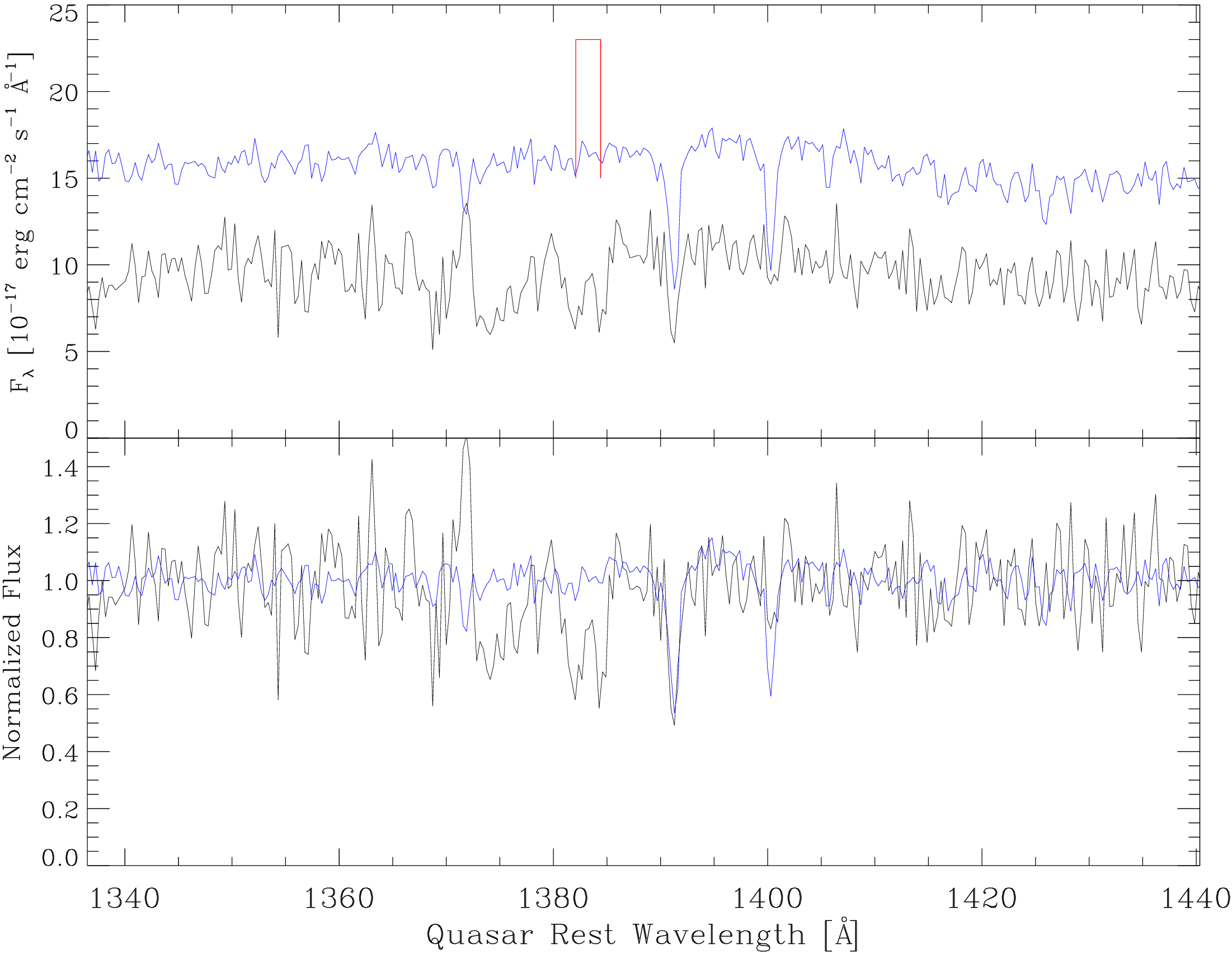}
\caption{The quasar J213648.17-001546.6 with $z_{\rm em}=2.1736$. See Figure A1 for the meanings of the color lines. The variable $\rm C~IV$ absorption system is located at $z_{\rm abs}= 1.8372$, and has a relative velocity value of $\beta=0.1116$ with respect to the emission line redshift.}
\end{figure}	

\begin{figure}
\centering
\includegraphics[width=8.cm,height=3.cm]{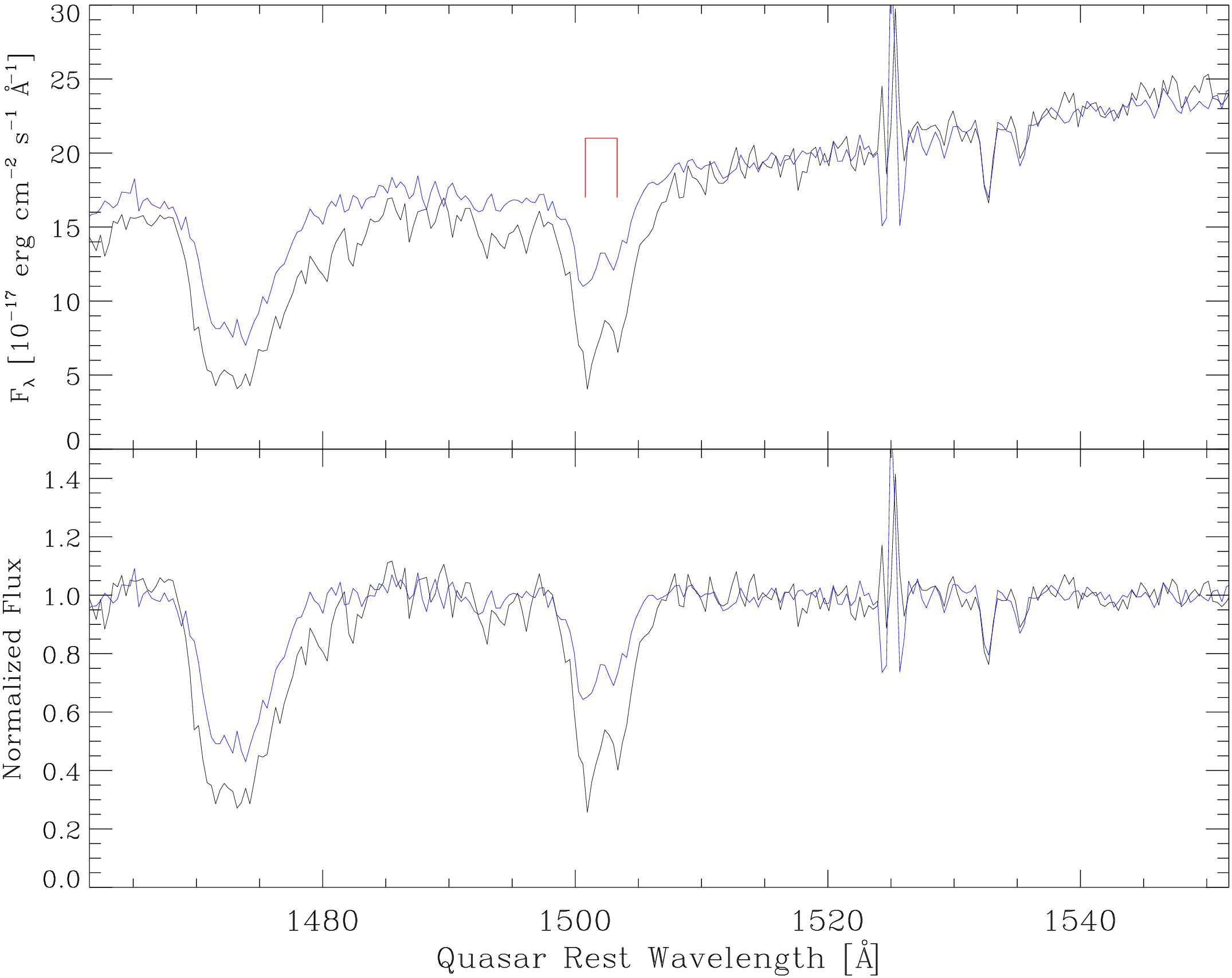}
\caption{The quasar J222157.97-010331.0 with $z_{\rm em}=2.6744$. See Figure A1 for the meanings of the color lines. The variable $\rm C~IV$ absorption system is located at $z_{\rm abs}=2.5459$, and has a relative velocity value of $\beta=0.0356$ with respect to the emission line redshift.}
\end{figure}	
	
\begin{figure}
\centering
\includegraphics[width=8.cm,height=3.cm]{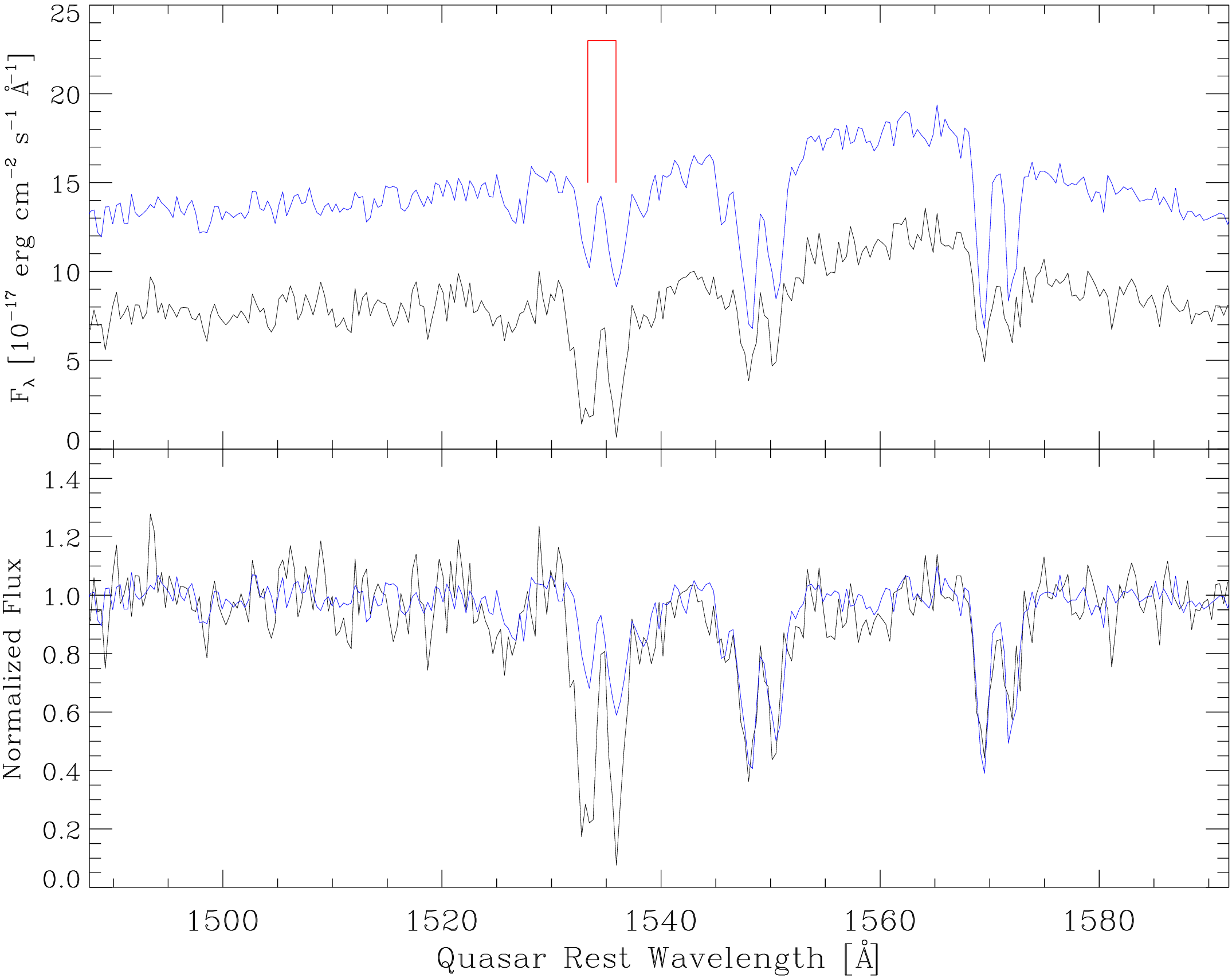}
\caption{The quasar J230034.04-004901.5 with $z_{\rm em}=2.2125$. See Figure A1 for the meanings of the color lines. The variable $\rm C~IV$ absorption system is located at $z_{\rm abs}= 2.1422$, and has a relative velocity value of $\beta=0.0221$ with respect to the emission line redshift.}
\end{figure}

\end{document}